\documentclass[11pt]{article}
\pdfoutput=1
\usepackage[utf8]{inputenc}
\usepackage{amsmath}
\usepackage{xfrac}
\usepackage[dvipsnames]{xcolor}
\usepackage{float}
\usepackage{graphicx}
\usepackage{caption}
\usepackage{subcaption}
\usepackage{array}
\usepackage{xr}
\makeatletter
\usepackage{url}
\usepackage{authblk}
\usepackage{cleveref}
\usepackage[backend=biber,style=chem-acs,]{biblatex}
\usepackage{geometry}
\usepackage{silence}
\usepackage{afterpage}
\usepackage{indentfirst}
\usepackage{amssymb}
\usepackage{setspace} 

\title{Biological free energy transduction is an Achilles heel of mean-field transport theory}

\begin{document}

\author[a]{Kiriko Terai}
\author[b,1]{Jonathon L. Yuly}
\author[a]{Peng Zhang}
\author[a,c,d,1]{David N. Beratan}
\affil[a]{Department of Chemistry, Duke University, Durham, NC, 27708}
\affil[b]{Lewis-Sigler Institute for Integrative Genomics, Princeton University, Princeton, NJ, 08540}
\affil[c]{Department of Physics, Duke University, Durham, NC, 27708}
\affil[d]{Department of Biochemistry, Duke University, Durham, NC, 27710}
\affil[1]{\small Corresponding authors. Email: jonathon.yuly@princeton.edu or david.beratan@duke.edu}

\maketitle

\begin{abstract}
Studies of nanoscale biological transport often use a mean-field approximation that is exact only when the system is  at equilibrium and there are no interactions between particles on different sites in the network. We explore the limitations of this approximation to describe many-particle transport in the context of enzyme function and biological transport networks. Our focus is on three  bioenergetic networks: a linear electron transfer chain (as found in bacterial nanowires), a redox-coupled proton pump (as in complex IV of respiration), and a near reversible electron bifurcation network (as  in complex III of respiration and other recently discovered structures). Away from equilibrium and with typical site-site interactions, we find that the mean-field approximation adequately describes linear transport chains. However, the mean-field approximation fails catastrophically to describe energy-transducing systems, as in the redox coupled proton pump and reversible electron bifurcation reactions. The mean-field approximation fails to capture the essential correlations that are needed to prevent slippage events and to produce efficient energy transduction.
\end{abstract}

\section*{Significance statement}
Nature boasts molecular nanomachines that catalyze reactions, turn motors, drag loads, transport particles, and interconvert energy among chemical, mechanical, and optical forms. Energy transduction in these machines couples  energetically downhill reactions to drive uphill reactions. Nanomachines are subject to thermal buffeting, so analyzing their function requires the consideration of statistical fluctuations that are not needed to describe macroscopic machines.  Yet, the full statistical analysis of many degrees of freedom is  costly. In the context of simple biological transport networks, an approximation that neglects all correlations is often used. We find that the mean-field approximation fails catastrophically to describe energy transduction at the nanoscale. Thus, including many-body correlations is essential to describe free energy transduction in molecular and biophysical systems.

\section{Introduction}  \label{Introduction}
\par
Classical machines harness a source of energy to perform work. For instance, rotation of one gear drives rotation of another, exploiting the rigidity of the gears' teeth. Wire coils convey the AC voltage in one circuit to another because both are coupled to the same magnetic flux. Pistons in combustion engines couple the chemical energy available from fuel oxidation to push a piston against a mechanical load. In each of these macroscopic machines, a core principle is the strong coupling between macroscopic components. Slippage (or loosening) of the coupling among components results in energy dissipation.  Examples of slippage in macroscopic machines include the motion of one gear without transmitting motion to its partner, the loss of energy to electromagnetic radiation in a wire coil, and leaks in a piston chamber of an engine that depressurizes the chamber without piston motion.
\par
Living systems use nanoscale molecular machines (in addition to macroscale machines) to accomplish wide ranging tasks~\cite{urry1997physical, hirokawa2009kinesin,brown2019theory}, including free energy transduction, whereby exergonic reactions (downhill particle flow, or $\Delta G < 0$) is leveraged to drive endergonic reactions (uphill particle flow, or $\Delta G > 0$)~\cite{hill2013free}. Often, these coupled downhill and uphill reactions involve transporting a species across a membrane, or the flow of electrons along redox chains~\cite{hill2013free, alberts2008molecular, jorgensen2003structure}. Energy transduction reactions must suppress slippage (or short-circuiting) reactions that direct all particles downhill, dissipating the available free energy as heat. Free energy transduction is intrinsically a non-equilibrium process, since the underpinning coupled reactions (with driving forces of opposite sign) are not in equilibrium.
\par
Biological transport networks, including those that perform energy transduction, involve the coordinated motion of multiple particles.   Many-particle transport theory describes free energy transduction, electron and ion transport, and mRNA translation~\cite{RN292,muneyuki2000properties,RevModPhys1997Modeling,zia2011modeling,hirokawa2009kinesin}.
The comprehensive descriptions of many-particle transport kinetics can require an astronomical number of microstates as the network scales to biologically-relevant sizes, making simulations computationally challenging. For example, bacterial nanowires involve redox hopping chains with as many as thousands of redox active sites~\cite{chong2018nature}, and some energy transducing redox active proteins can house $\sim 50$ iron sulfur clusters, cofactors that can transport electrons~\cite{huwiler2019one}. 
\par
To explore these biological systems, several approximations are commonly made.
Single-particle approximations assume that  transport of one particle through the network is characteristic of the full kinetics, and these approximations have been used to study many biological electron transfer  reactions~\cite{RN191,teo2019mapping,RN129}.  Energy transduction is an intrinsically multi-particle process, so single-particle descriptions provide a poor starting point for their description.  Energy transduction engages the flow of one particle or process downhill coupled to the motion of a second particle or process uphill.
Another approximation to describe many-particle transport kinetics, the totally asymmetric simple exclusion process (TASEP) model, is used to study molecular motors and protein synthesis~\cite{zia2011modeling,denisov2015totally}.  TASEP models assume irreversible transport and uniform transport rates between all neighboring sites~\cite{RN292,zia2011modeling,denisov2015totally}.
As such, TASEP models are not appropriate to describe systems with either multiple transport rates or reversible transport steps, circumsances that are typically encountered in biological transport networks  of bioenergetics~\cite{RN155,RN251,RN353,RN354,RN355,RN215}.
\par
The mean-field approximation for many-particle transport, which is our focus, neglects explicit statistical correlations between the occupancies of particle sites~\cite{RN152,RN250,RN290}. A similar approach that neglects correlations (but differs from the mean field approximation when interactions are present, see section \ref{Many-particle transport master equation for biological transport networks}) is implicitly used in many studies of biological electron transport networks~\cite{RN392,messelink2016site,RN155,RN36,RN219,RN222,RN289}.
The mean-field approximation fails when correlations between site occupancies are important.  Statistical dependence arises from thermodynamic disequilibrium or interactions between particles at different sites~\cite{mcquarrie2000statistical}. How the mean-field approximation of many particle transport may fail to describe biological function is poorly understood, especially in the context of molecular machines~\cite{jorgensen2003structure,nicholls2013bioenergetics,kim2007kinetic} and transport networks, such as bacterial nanowires and cable bacteria~\cite{pirbadian2014shewanella,RN241}.
\par
We assess the reliability of the mean-field approximation to describe the function of a linear electron transport chain and two free energy transducing nanomachines. 
The examples we study are: (1) a redox transport chain consisting of tetraheme proteins~\cite{RN222,tsapin2001identification,harada2002directional}, (2) a redox-coupled proton pump~\cite{kim2007kinetic,kim2012proton}, and (3) a near reversible electron bifurcation machine that transduces energy~\cite{RN197,RN296}.
Even with thermodynamic disequilibrium and interactions between particles at different sites, we find that the mean-field approximation is sufficient to describe linear transport chains. The mean-field treatment reproduces the electron transport fluxes in these models within an order of magnitude of those predicted by the fully correlated model. However, the mean-field approximation fails to describe  free energy transduction by models for a redox-coupled proton pump and for a nearly reversible electron bifurcating network. Section \ref{Discussion} discusses the nature of mean-field transport theory and its failure to describe energy transduction in nanoscale and biological machines.

\section{Many-particle transport master equation for biological transport networks}  \label{Many-particle transport master equation for biological transport networks}
Many-particle Markovian transport kinetics may be described using a classical master equation that captures all statistical correlations among site occupancies. Applications of this model to particle transport in ordered and disordered media are well known ~\cite{RN152,RN269,RN266,RN267,RN264,RN358,RN361,RN360}. The transport master equation may also be used to describe energy transduction in  biological systems ({\it vide infra}). A summary of this many-particle transport framework follows (see also section 1 of SI Appendix), including a description of how the approach may be generalized to include  degrees of freedom other than site occupancy  (e.g., conformational states) by labeling these additional states as fictitious particles and sites (see section 3 in SI Appendix). 
\par
Each microstate of a transport network with $N$ particle sites, is described using
\begin{equation}  \label{microstates}
\boldsymbol{\sigma} = [\sigma_{1} \hspace{0.8mm} \sigma_{2} ... \hspace{0.8mm} \sigma_{i} \hspace{0.8mm} ... \hspace{0.8mm} \sigma_{j} \hspace{0.8mm} ...\hspace{0.8mm} \sigma_{N}]
\end{equation}
where $\sigma_{i}$ and $\sigma_{j}$ are non-negative integers that indicate the number of particles on sites $i$ and $j$. Although many-particle transport problems are often modeled with sites that accommodate at most one particle, biological transport may support multiple particles on a single site, such as electron transfer through cofactors that accommodate multiple electrons (examples include flavins, quinones, the H-cluster in hydrogenase, and the P-cluster in  nitrogenase~\cite{kar2022understanding,zhang2008quinone,lubitz2014hydrogenases,peters1995involvement}). The probability $P_{\boldsymbol{\sigma}}$ that the network is in microstate $\boldsymbol{\sigma}$ at time $t$ is described by the master equation~\cite{RN152}:
 \begin{equation}  \label{master eq}
 \frac{dP_{\boldsymbol{\sigma}}(t)}{dt} = -\sum_{\boldsymbol{\sigma'}}W_{\boldsymbol{\sigma}\boldsymbol{\sigma'}}P_{\boldsymbol{\sigma'}}(t)
 \end{equation}
where $W_{\boldsymbol{\sigma}\boldsymbol{\sigma'}}=\delta_{\boldsymbol{\sigma}\boldsymbol{\sigma'}} \sum_{\boldsymbol{\sigma'}} K_{\boldsymbol{\sigma}\boldsymbol{\sigma''}}-K_{\boldsymbol{\sigma'}\boldsymbol{\sigma}}$, $K_{\boldsymbol{\sigma}\boldsymbol{\sigma'}}$ is the rate constant for the transition from microstate $\boldsymbol{\sigma}$ to $\boldsymbol{\sigma'}$, and $\delta$ is the Kronecker delta.
Quantities of interest include the probability of finding site $i$ in occupation state $q$ (denoted by $p_{iq}$), calculated using
 \begin{equation}    \label{probability}
 p_{iq} = \sum_{\boldsymbol{\sigma}} \delta_{\sigma_{i},q} P_{\boldsymbol{\sigma}}
 \end{equation}
 The microstate probabilities $P_{\boldsymbol{\sigma}}$ may also be used to calculate $X$-point correlation functions (denoted by $p_{i_{1}q_{1},i_{2}q_{2},...,i_{X}q_{X}}$, $X$ must be $\leq N$) that are the joint probabilities of finding sites $i_1$, $i_2$, ..., and $i_X$ in occupation states $q_1$, $q_2$, ..., and $q_X$, respectively. The $X$-point correlation functions are calculated using
  \begin{equation}   \label{X-point correlation function}
 p_{i_{1}q_{1},i_{2}q_{2},...,i_{X}q_{X}} = \sum_{\boldsymbol{\sigma}} \delta_{\sigma_{i_1},q_1} \delta_{\sigma_{i_2},q_2}...\delta_{\sigma_{i_X},q_X} P_{\boldsymbol{\sigma}}.
 \end{equation}
 \par
Site occupancy constraints are enforced by the values of the rate constants $K_{\boldsymbol{\sigma}\boldsymbol{\sigma'}}$ and the rate constant $K_{\boldsymbol{\sigma}\boldsymbol{\sigma'}}$ is zero if either microstate $\boldsymbol{\sigma}$ or $\boldsymbol{\sigma'}$ violates the occupancy constraint.  $n$-particle transport  from site $i$ to site $j$ is denoted:
 \begin{equation}    \label{site-site}
[\sigma_{1} \hspace{0.8mm} \sigma_{2} ... \hspace{0.8mm} \sigma_{i} \hspace{0.8mm} ... \hspace{0.8mm} \sigma_{j} \hspace{0.8mm} ...] \rightleftharpoons [\sigma_{1} \hspace{0.8mm} \sigma_{2} ... \hspace{0.8mm} (\sigma_{i}-n) \hspace{0.8mm} ... \hspace{0.8mm} (\sigma_{j}+n) \hspace{0.8mm} ...]
 \end{equation}  
 In addition to transport between sites (Eq. \ref{site-site}), particles can enter or leave the system through interactions with reservoirs (Eq. \ref{site-reservoir}). For example, bulk ions, pools of oxidants or reductants, and external electrodes may serve as reservoirs. A typical $n$-particle transfer between site $i$ and  a reservoir is:
 \begin{equation}       \label{site-reservoir}
 \hspace{1cm}[\sigma_{1} \hspace{0.8mm} \sigma_{2} ... \hspace{0.8mm} \sigma_{i} \hspace{0.8mm} ... \hspace{0.8mm} \sigma_{j} \hspace{0.8mm} ...] \rightleftharpoons [\sigma_{1} \hspace{0.8mm} \sigma_{2} ... \hspace{0.8mm} (\sigma_{i}+n) \hspace{0.8mm} ... \hspace{0.8mm} \sigma_{j} \hspace{0.8mm} ...]
 \end{equation}
 Conformational or chemical changes (provided they are Markovian) may be incorporated in a many-particle transport master equation by mapping conformational and chemical degrees of freedom to fictitious sites and particles. An example of this mapping is given in section 3 in SI Appendix using a classic model of biological energy transduction framed by T.L. Hill \cite{hill2013free}.
 \par
The number of possible microstates scales exponentially with the number of particle sites. For example, if $N$ sites can each accommodate one particle, the number of microstates is $2^{N}$. This exponential increase in complexity makes exact simulation of large systems computationally demanding. To reduce the computational complexity of the exact master equation (Eq. \ref{master eq}), a mean-field approximation is sometimes used.

\subsection{The mean-field approximation to many-particle transport}    \label{The mean-field approximation to many-particle transport}
The site occupancies of a transport network with $N$ particle sites are statistically uncorrelated if any $X$-point correlation function (Eq. \ref{X-point correlation function}) can be factored into a product of $X$ probabilities~\cite{RN152,RN250,RN290}. Eq. \ref{mean-field approximation} defines the mean-field approximation:
\begin{equation}    \label{mean-field approximation}
    p_{i_{1}q_{1},i_{2}q_{2},...,i_{X}q_{X}} \approx \prod_{x=1}^{X}p_{i_{x}q_{x}}   \hspace{1cm}
\end{equation}
Eq. \ref{mean-field approximation} is exact for systems (1) at thermodynamic equilibrium and (2) lacking site-site interactions between particles at different sites (particles on the same site may interact and Eq. \ref{mean-field approximation} remains exact at equilibrium in that case)~\cite{richards1977theory,ambegaokar1971hopping}. This equivalence of the mean-field approximation and the exact master equation under conditions (1) and (2) can be shown by exploiting a mapping between the transport chain (Eq. \ref{microstates}) and a spin chain (see SI Appendix section 2 for a derivation).
\par
The exact time derivative of $p_{iq}$ in the absence of site-site interactions is described by the exact master equation (Eq. \ref{master eq}), is~\cite{RN152}:
\begin{equation}   \label{exact rate eq - no interaction} 
    \frac{dp_{iq}}{dt} = -\sum_{\substack{j\neq i\\n,s}} \sum_{r, u} w_{i,j, n, s}^{r,u} \hspace{3pt} p_{ir,ju}
\end{equation}
where $w_{i,j, n, s}^{r,u}$ is defined in Eq. 11 of SI Appendix, and contains information about sites $i$ and $j$ and the particle transport rate constants between them (see SI Appendix section 1.2 for a derivation).
The corresponding equation in the case with site-site interactions involves an $N$-point correlation function (Eq. 9 in SI Appendix).
Applying the mean-field approximation (Eq. \ref{mean-field approximation}) to Eq. \ref{exact rate eq - no interaction} leads to a closed set of equations that describes the mean-field approximated probabilities $p_{iq}$ in the absence of site interactions~\cite{RN152}:
\begin{equation}   \label{mean-field rate eq - no interaction} 
    \frac{dp_{iq}}{dt} = -\sum_{\substack{j\neq i \\n,s}} \sum_{r,u} w^{r,u}_{i,j,n,s} p_{ir}p_{ju}
\end{equation}
In the case with site-site interactions, the rate of change of $p_{iq}$ is
\begin{equation}   \label{mean-field rate eq - interaction} 
    \frac{dp_{iq}}{dt} = -\sum_{\substack{j\neq i \\n,s}} \sum_{r,u} w^{r,u(\text{MF})}_{i,j,n,s} (\{ p_{kv} \}) p_{ir}p_{ju}
\end{equation}
where $w^{r,s(\text{MF})}_{i,j} (\{ p_{kv} \})$ is a function of the site probabilities $p_{kv}$ for all $k$ and $v$ (see SI Appendix section 1.3 and 1.4 for a derivation).
Many previous studies of mean-field transport kinetics focus on systems without site-site interactions, greatly simplifying the mean-field kinetics to Eq. \ref{mean-field rate eq - no interaction}. The dependence of the mean-field rates $w^{r,s(\text{MF})}_{i,j} (\{ p_{kv} \})$ on the site probabilities $p_{kv}$ increases the order of the polynomials used to calculate fluxes from $2$ (Eq. \ref{mean-field rate eq - no interaction}) to $N$ (Eq. \ref{mean-field rate eq - interaction}). Solving Eq. \ref{mean-field rate eq - interaction} for steady-state solutions involves finding the roots of these high-order polynomials that are not all degenerate. In our studies, we always find one physically relevant solution (i.e., fluxes of expected orders of magnitude and probabilities between zero and one).
\par
Many approaches that neglect correlations between site occupancies may be used to estimate the  fluxes through transport networks (see section \ref{Introduction}). An example is the approach of Jiang $et$ $al.$~\cite{RN222} (see section 5 in SI Appendix) for electron transport through the STC protein in bacterial nanowires.  That treatment neglects correlations among redox site occupancies. The approach of Jiang {\it et al.} is not equivalent to the mean-field kinetic treatment in the presence of site-site interactions discussed here (Eq. \ref{mean-field rate eq - interaction}), because the approach of Jiang {\it et al.} does not apply Eq. \ref{mean-field approximation} to the exact master equation (Eq. \ref{master eq}). Instead, the approach of Jiang {\it et al.} results from first calculating the mean-field driving force for electron transfer from site $i$ to site $j$, then using that driving force to estimate the electron-transfer rate. In the absence of site-site interactions, the mean-field kinetics (Eq. \ref{mean-field rate eq - interaction}) and the approach of Jiang $et$ $al.$ are equivalent, both reducing to Eq. \ref{mean-field rate eq - no interaction}. The conclusion of our study (namely that the explicit treatment of correlations between sites is required to explain energy transducing networks) is also observed using the approach of Jiang {\it et al.} All of the simulations described in the next section using the mean-field approximation were repeated using the approach of Jiang {\it et al.}, with qualitatively similar results as compared to the mean-field approximation (see SI appendix for details).

\section{Models and results}
\subsection{Linear transport chains}
\begin{figure}[bt!]
    \centering
    \includegraphics[width=0.25\textwidth]{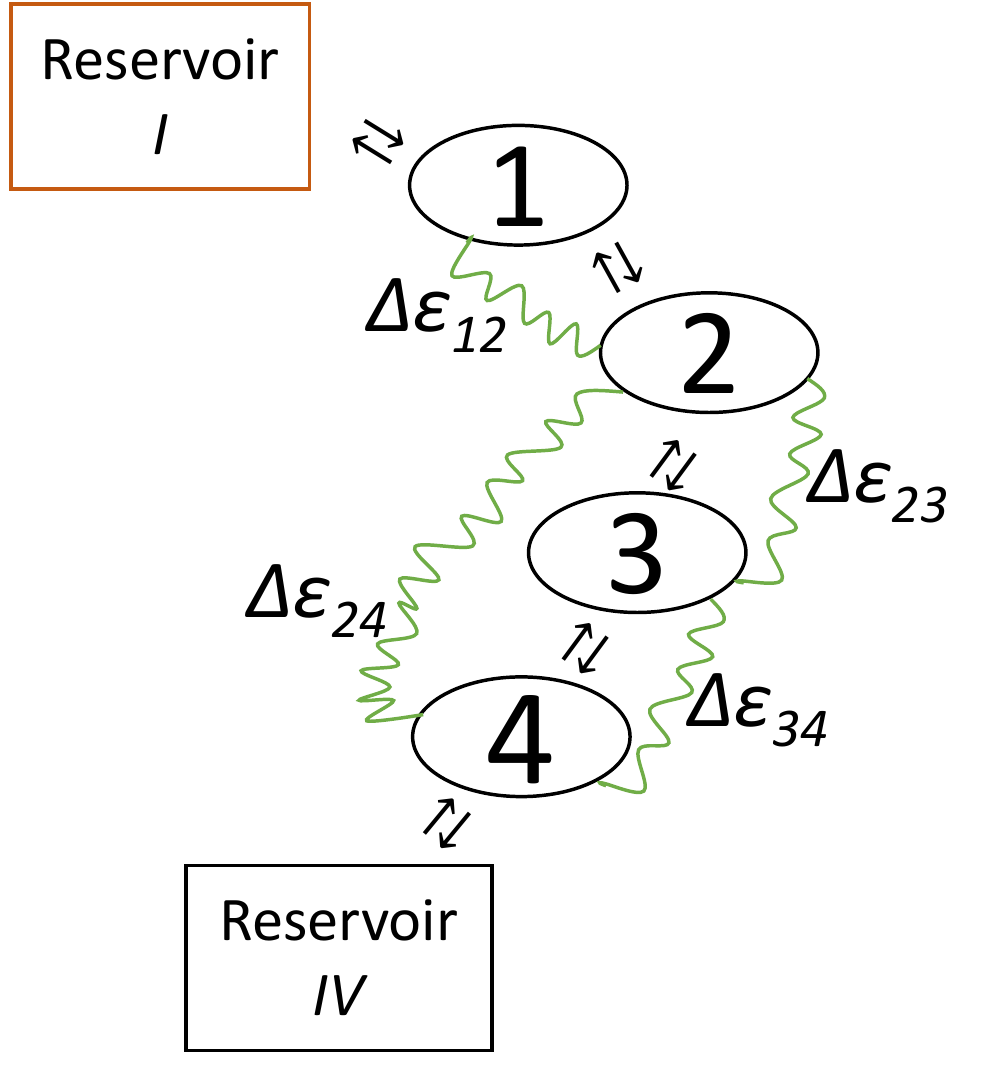}
    \caption{Model of a linear  transport chain inspired by the STC protein~\cite{RN222}. The redox cofactors can each have an occupancy of 0 or 1. The terminal cofactors $1$ and $4$ are each coupled to a reservoir. The transport chain is driven by $\Delta G$ (meV), the free energy to transport electrons from reservoir $I$ to $IV$. $\Delta\epsilon_{ij}$ is the change of redox potential (meV) of cofactor $i(j)$ when $j(i)$ is unoccupied (see SI Appendix for parameters used for computation)~\cite{RN222,fonseca2009tetraheme}.}  
    \label{fig:psuedoSTC_model}
\end{figure}

Linear transport chains are found throughout biology. Extracellular appendages known as bacterial nanowires, for example, conduct electrons from metabolic electron donors inside the cell to extracellular electron sinks - such as iron oxide~\cite{malvankar2014microbial,ucar2017overview,denisov2015totally,zia2011modeling,RevModPhys1997Modeling,hirokawa2009kinesin}. The tetra-heme STC protein is a multi-heme protein that mediates extracellular electron transport in $S.$ $oneidensis$ nanowires~\cite{RN222,tsapin2001identification,harada2002directional}. Fig. \ref{fig:psuedoSTC_model} shows our model of a simple linear transport chain, inspired by the STC protein. The modeled network has a strong driving force between the electron reservoirs at the termini, and the model also includes interactions between particles at different sites in the transport chain~\cite{RN222}. The model is parameterized based on studies of Jiang $et$ $al.$~\cite{RN222} and Fonseca $et$ $al.$~\cite{fonseca2009tetraheme}.

\subsubsection*{The mean-field approximation adequately describes kinetics of simple linear transport, as occurs in the tetra-heme STC protein}  
We explore the reliability of the mean-field approximation to describe the electron transport kinetics of a linear transport chain inspired by the STC protein~\cite{RN222,tsapin2001identification,harada2002directional}. We simulate the kinetics (Fig. \ref{fig:psuedoSTC_model}) using the exact master equation (Eq. \ref{master eq}) and the mean-field approximation (Eq. \ref{mean-field rate eq - interaction}) (see SI Appendix for parameters used in the computations, which are drawn from previous  studies of the STC protein~\cite{RN222,fonseca2009tetraheme}). We compared the steady-state electron fluxes to reservoir IV computed using the exact master equation with the mean-field results. The approach of Jiang {\it et al.} is not equivalent to the mean-field approximation (see section \ref{The mean-field approximation to many-particle transport}). 
\par
\begin{figure}[bt!]
    \centering
    \includegraphics[width=0.5\textwidth]{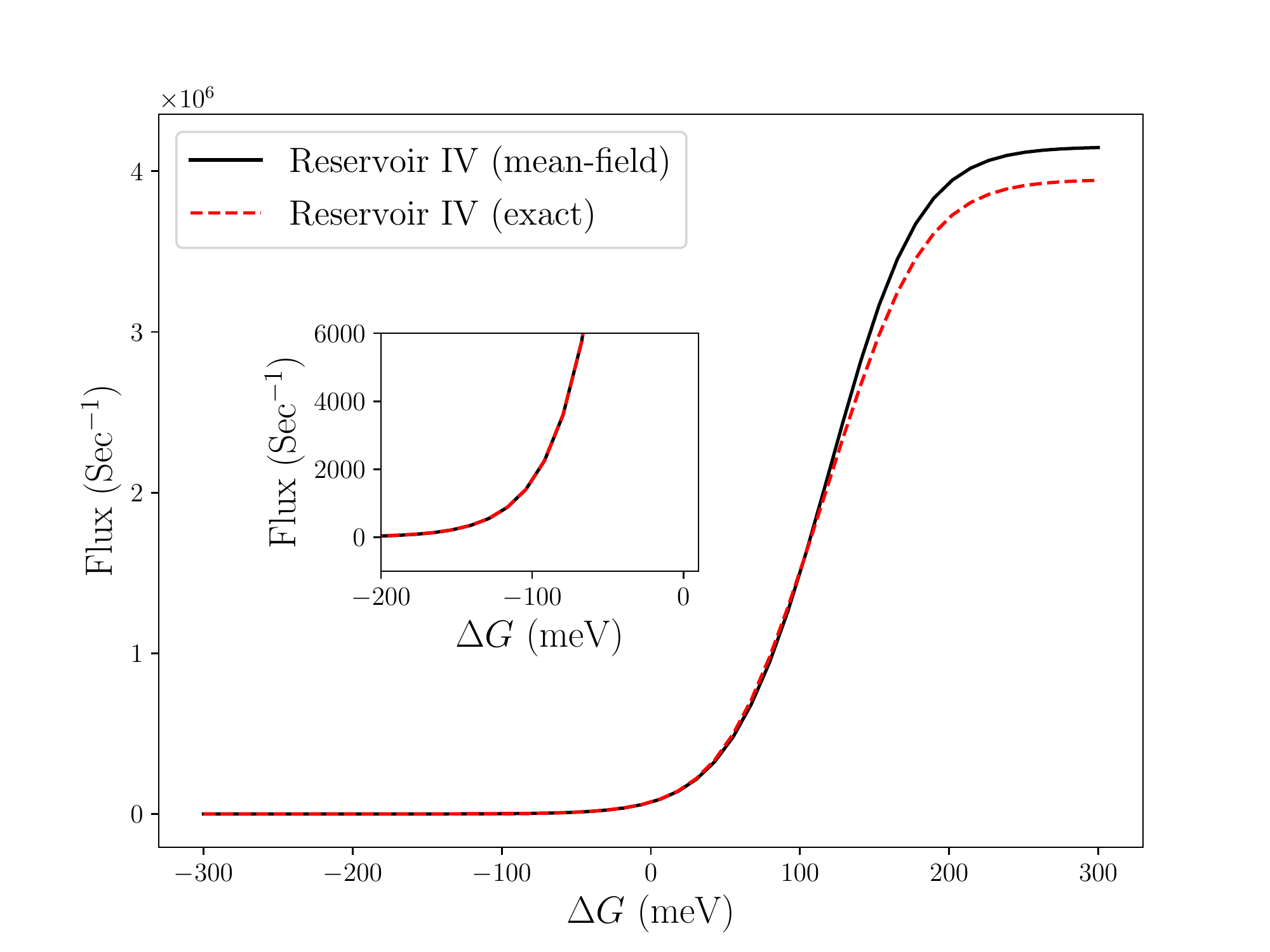}
    \caption{Kinetics of a linear transport chain inspired by the STC protein. The plot shows the steady-state net electron fluxes into coupled reservoirs as a function of $\Delta G$, varied from -300 to 300 meV. The inset shows the results for $\Delta G$ ranging from -200 to 10 meV. The fluxes computed with the mean-field model are within a factor of $\sim$ 2 of the exact computed fluxes. The exact and mean-field models predict similar steady-state fluxes with reservoir IV.  The fluxes are $\sim 0$ sec$^{-1}$ at negative $\Delta G$, and converge to $\sim 10^{6}$ sec$^{-1}$ at large $\Delta G$. As our focus is qualitative differences between the exact and mean-field treatments, we find that the mean-field approximation describes the electron transport kinetics of a linear transport chain well, regardless of thermodynamic disequilibrium and presence of site-site interactions. The plot in Fig. S2 in SI Appendix shows the steady-state net electron fluxes into coupled reservoirs when site-site interactions are absent. Comparison to fluxes computed by Jiang {\it et al.}~\cite{RN222} is also discussed in Fig. S2 in SI Appendix.}
    \label{fig:flux_psuedoSTC_MEK_trueMF}
\end{figure}

Fig. \ref{fig:flux_psuedoSTC_MEK_trueMF} shows that the fluxes computed at the mean-field level are within an order of magnitude of the exact computed fluxes for the linear chain. Our aim is to  identifying qualitative differences - where they exits - between the exact and mean-field results.  The mean-field approximation describes the electron transport kinetics of this linear transport chain, regardless of thermodynamic disequilibrium and the presence of site-site interactions.
\par
Since the mean-field approximation is often used to treat linear transport chains~\cite{RN392,messelink2016site,RN155,RN36,RN219,RN222,RN289}, it is comforting to see that the approximation works well in these systems. In bioenergetics, however, linear transport chains often exist between energy transducing catalytic sites, and transport through the chains may be coupled to other events, such as ATP hydrolysis, ion translocation across membranes, or protonation of amino acid residues or cofactors~\cite{alberts2008molecular,jorgensen2003structure}. The validity of the mean-field approximation for transport chains with energy input from coupled reactions in the interior of the chains is poorly understood. Below, we explore the validity of the mean-field approximation to describe such transport networks with coupled reactions, specifically those that transduce energy (electron transfer driven proton pumping and electron bifurcation, for example).

\subsection{Free energy transduction networks}
Free energy transduction occurs when one or more exergonic reactions ($\Delta G < 0$ or ``downhill'')  are leveraged to drive one or more endergonic reactions ($\Delta G > 0$ or ``uphill'')~\cite{hill2013free}.
Reaction cycles that only perform exergonic processes without driving endergonic cycles are known as slippage events, dissipating free energy as heat~\cite{hill2013free}. Slippage  reduces the turnover rate of the endergonic reaction; efficient free energy transduction occurs when slippage is suppressed.

\subsubsection{Redox-coupled proton pump}
\begin{figure}[bt!]
    \centering
     \begin{subfigure}[b]{0.35\textwidth}
         \subcaption{}
         \includegraphics[width=\textwidth]{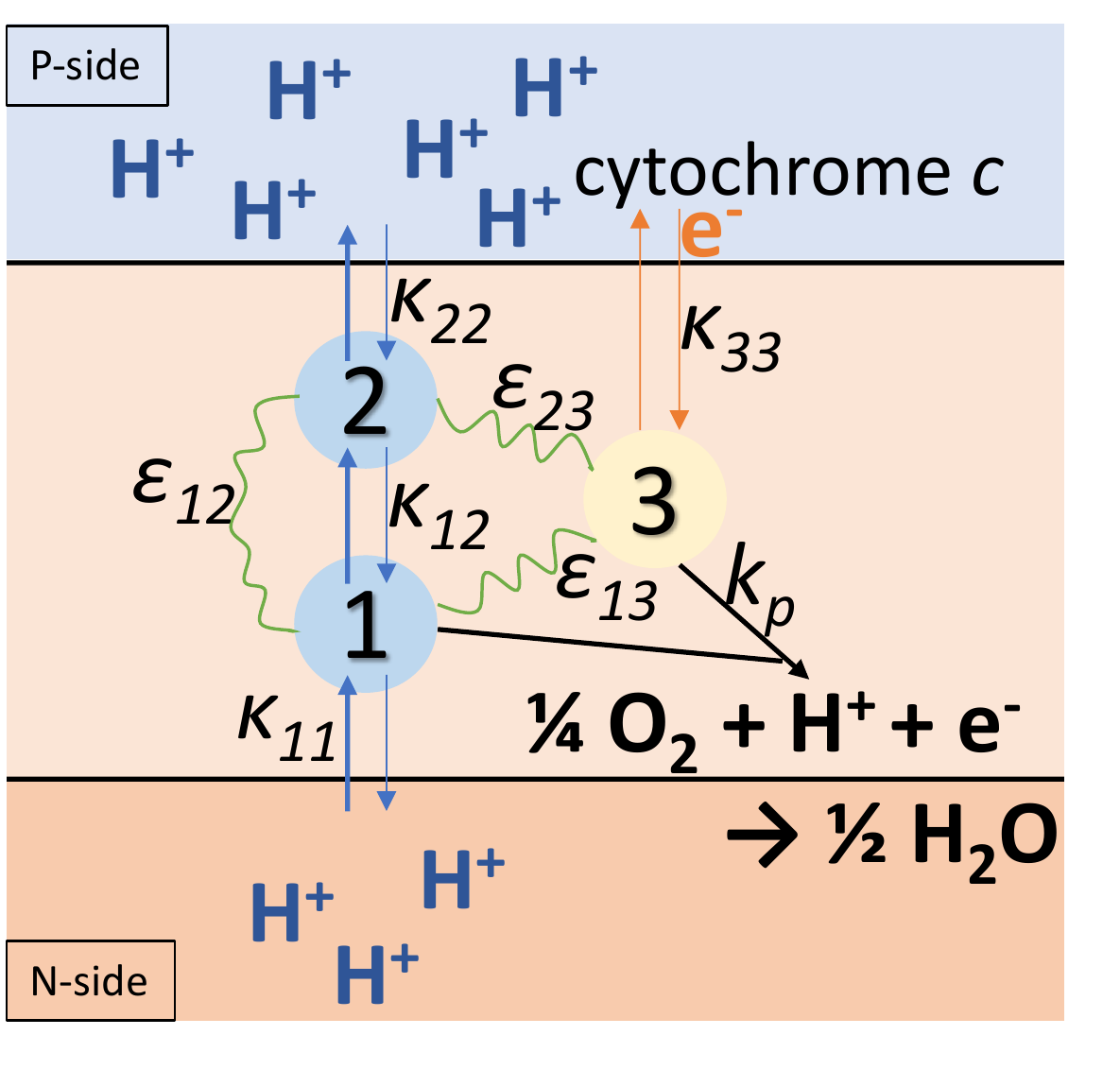}
         \label{fig:Proton pump model in sites}
     \end{subfigure}
     \begin{subfigure}[b]{0.45\textwidth}
         \subcaption{}
         \includegraphics[width=\textwidth]{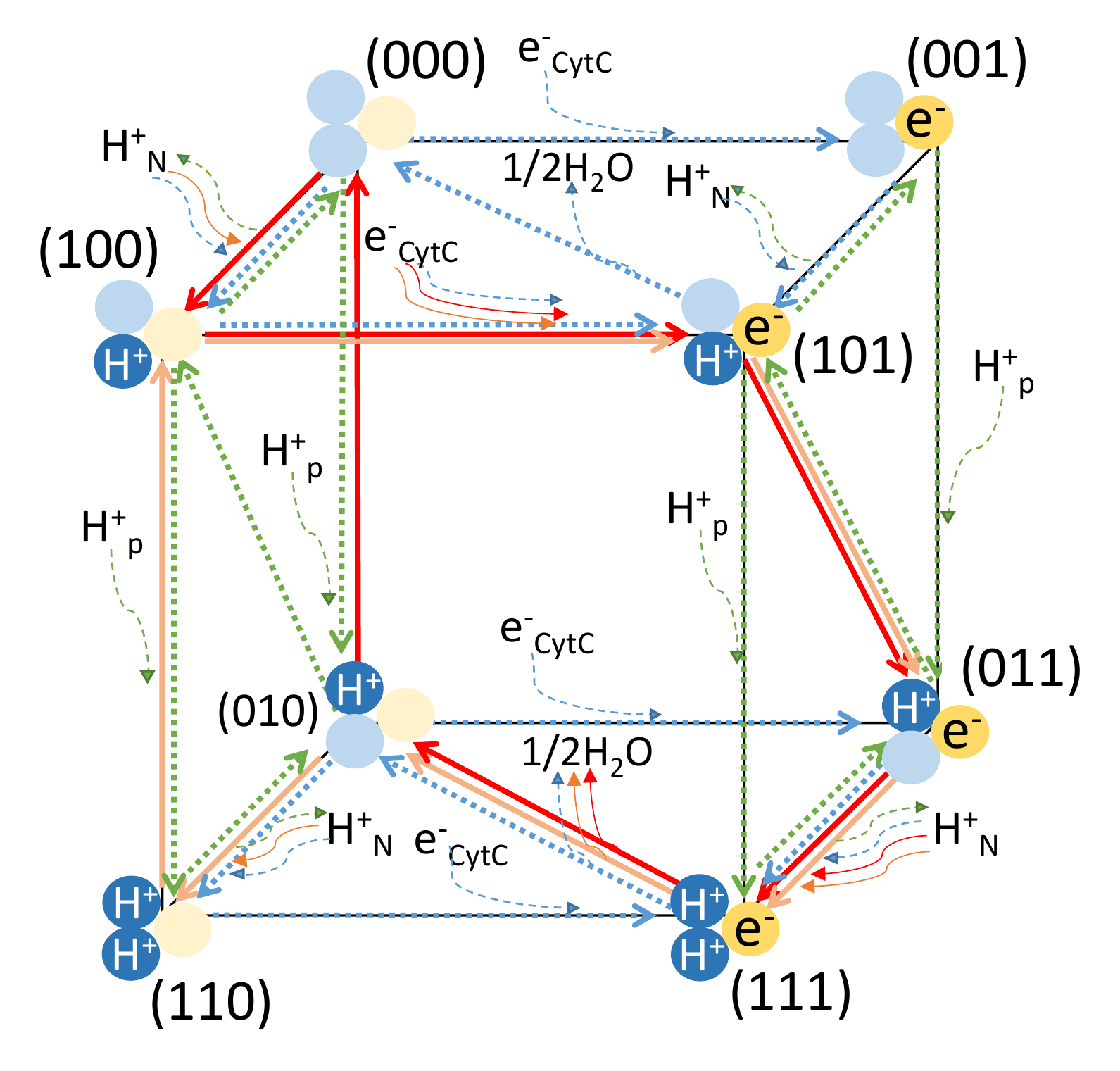}
         \label{fig:Proton pump model in microstates}
     \end{subfigure}
\caption{A model for redox-coupled proton pumping in CcO~\cite{kim2012proton,kim2007kinetic,kim2009kinetic}. (a) Sites 1 and 2 are proton binding sites and site 3 is an electron site (redox cofactor). The blue and orange arrows represent proton and electron transfer pathways, respectively, associated with the intrinsic rate constants $\kappa_{ij} (s^{-1})$. The black arrow represents $H_{2}O$ formation associated with the rate constant $k_{p} (s^{-1})$. $\epsilon_{ij}$ is the electrostatic interaction between sites (see SI Appendix for
parameters used for computation~\cite{kim2012proton,kim2007kinetic,kim2009kinetic}). Redox-coupled proton pumping in CcO is deemed irreversible because the free energy of $O_{2}$ reduction is large and negative. (b) Occupancy microstates and transitions within the simple redox-coupled proton pump~\cite{kim2009kinetic}. The microstates are labeled ($\sigma_{1}$, $\sigma_{2}$, $\sigma_{3}$), where $\sigma_{i}$ indicates the site occupancies of site $i = 1,2,$ or $3$). The colored circles are populated with either an electron or proton, with colors corresponding to the sites in (a). The species $H^{+}_{N}$ and $H^{+}_{P}$ represent reservoirs of protons on the N-side and P-side, respectively, with a difference in chemical potential (proton motive force across the membrane). The electrons from the cyt $c$ reservoir are represented as $e^{-1}_{CytC}$. For free energy transduction to occur, the cycles indicated by the red or orange arrows must occur, both with the net effect of transporting one proton from the N-side to the P-side, and generation of 1/2$H_{2}O$~\cite{kim2012proton}. The slippage cycles lead to $H_{2}O$ formation without proton pumping (blue dashed arrows), or proton flow from the P-side to the N-side of the membrane (green dashed arrow).}
\label{fig:Proton pump model}
\end{figure}

Cytochrome $c$ oxidase (CcO) in the mitochondrial electron transport chain performs free energy transduction as a redox-coupled proton pump. Harnessing the downhill reduction of oxygen by oxidation of cyt $c$, CcO pushes protons from the inside (N-side) to the outside (P-side) of the mitochondrial inner membrane against a proton motive force~\cite{kim2007kinetic,kim2012proton,nicholls2013bioenergetics}. Dioxygen reduction and proton pumping correspond to the M and L cycles, respectively, when mapping onto Hill's model~\cite{hill2013free} (see section 3 in SI Appendix). Recently, a simple model of redox-coupled proton pumping was used to study the role of electrostatic interactions in CcO~\cite{kim2007kinetic, kim2012proton}. The model includes one redox site (redox cofactor) and two proton sites; all sites interact.  The redox-coupled proton pump model, and its mapping onto the occupancy representation, is described in Fig. \ref{fig:Proton pump model}.
\par
The model of proton pumping has eight microstates, and four reservoirs: two proton reservoirs corresponding to each side of the membrane, one electron reservoir for the cyt \textit{c} pool, and one combined electron-proton reservoir representing the oxygen electron sink (flow of electrons and protons into this reservoir is assumed irreversible~\cite{kim2007kinetic}). Dioxygen reduction irreversibly draws a proton and an electron from sites 1 and 3, respectively, at a constant rate when the two sites are both occupied~\cite{kim2007kinetic,kim2012proton}. If energy transduction occurs, protons will flow uphill from the N-side of the membrane (at site 1) to the P-side of the membrane (at site 2). Fig. \ref{fig:Proton pump model in microstates} illustrates the flux through the microstates associated with the energy transducing and slippage cycles~\cite{kim2009kinetic}. Site-site interactions are required for free energy transduction in the proton-pump model~\cite{kim2007kinetic,kim2012proton}, and we use the parameters of previous studies~\cite{kim2012proton,kim2007kinetic,kim2009kinetic}.

\subsubsection*{The mean-field approximation does not describe free energy transduction by a redox-coupled proton pump}
We simulated the kinetics of this redox-coupled proton pump model using the exact master equation (Eq. \ref{master eq}).  
Our results are compared with simulations that make the mean-field approximation (Eq. \ref{mean-field approximation}) (see  SI Appendix for the parameters~\cite{kim2012proton,kim2007kinetic,kim2009kinetic}). 
\par
\begin{figure}[bt!]
    \centering
     \begin{subfigure}[b]{0.37\textwidth}
         \includegraphics[width=\textwidth]{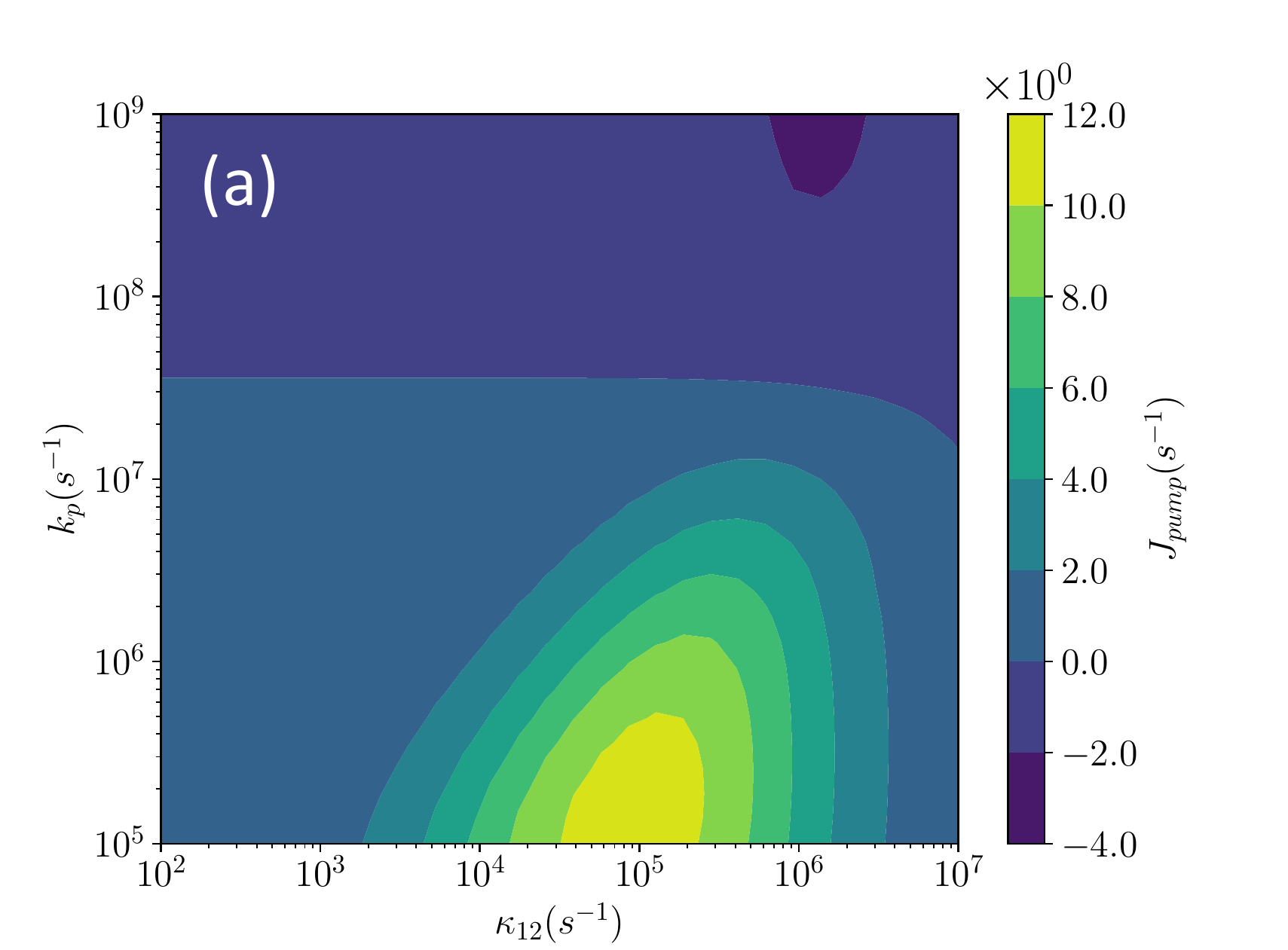}
         \label{fig:Exact pump flux}
     \end{subfigure}
     \begin{subfigure}[b]{0.37\textwidth}
         \includegraphics[width=\textwidth]{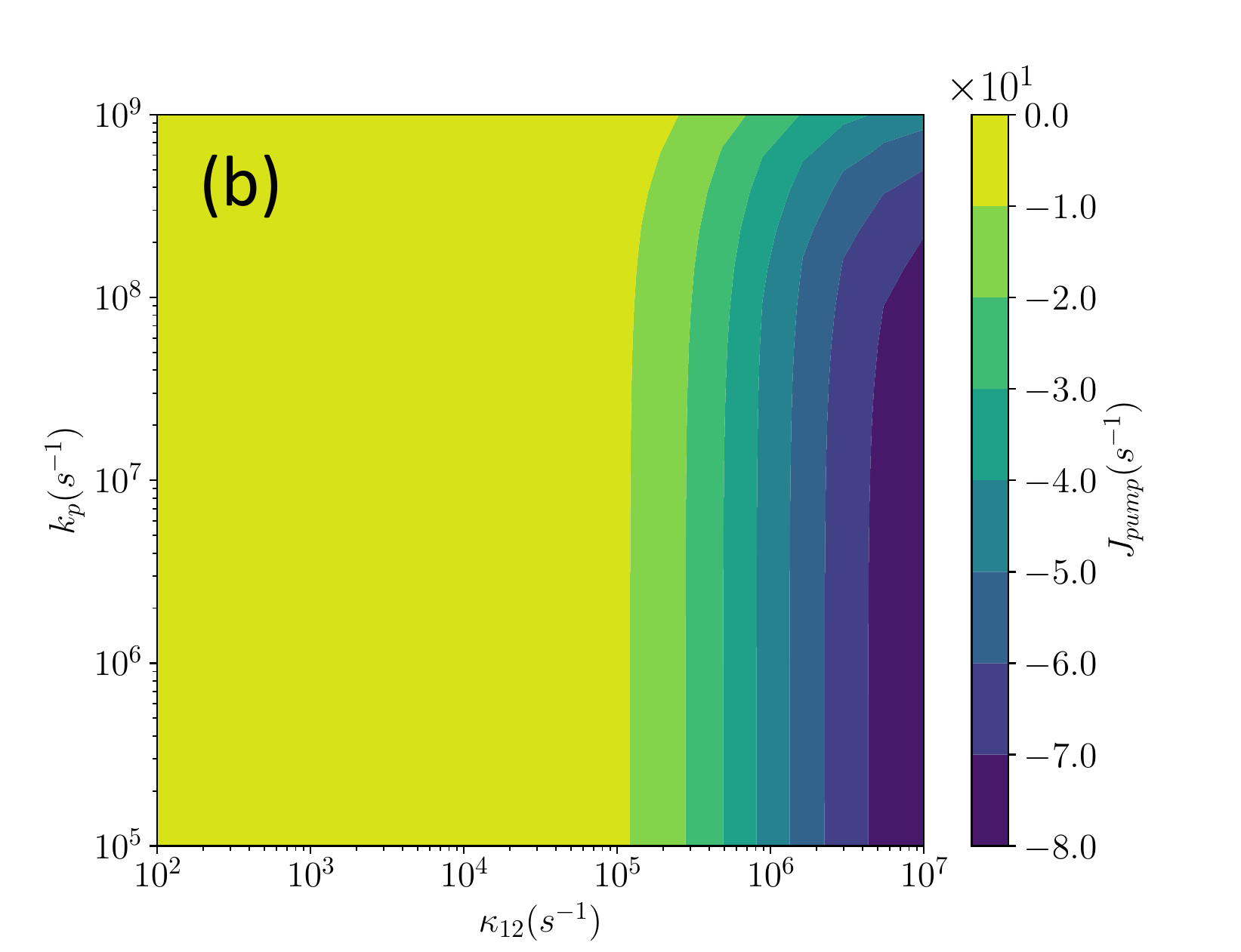}
         \label{fig:Mean-field pump flux}
     \end{subfigure}
\caption{Steady-state flux of protons pumped to the P-side ($J_{pump} (s^{-1})$), calculated using (a) the exact master equation including site-site interactions and (b) the mean-field model including site-site interactions (Fig. S3 in SI Appendix for $J_{pump} (s^{-1})$ shows the case with site-site interactions absent, and includes a comparison to the modeling approach of Jiang $et$ $al.$~\cite{RN222}). $\kappa_{12}$ is varied from $10^{2}$ to $10^{7}$ sec$^{-1}$ and $k_{p}$ is varied from $10^{5}$ to $10^{9}$ sec$^{-1}$. Positive $J_{pump}$ values indicate energy transduction, because protons spontaneously flow uphill from the N- to the P-side of the membrane (the membrane potential, which is the proton motive force, is $V_{m} =$ 100 mV). Negative $J_{pump}$ values indicate that slippage reactions  dominate, and protons  leak from the P-side. Comparing panels (a) and (b), in regimes  where the exact master equation predicts no free energy transduction ($k_{p} \gtrsim  10^{8}$ sec$^{-1}$), the mean-field model also predicts no free energy transduction. However, when the exact master equation predicts free energy transduction (in the case $k_{p} \lesssim 10^{8}$ sec$^{-1}$), the mean-field model  does not predict free energy transduction. Thus, the mean-field approximation is unable to describe energy transduction by this simple redox-coupled proton pump.}
\label{fig:Pump flux}
\end{figure}

We compute the steady-state proton flux pumped to the P-side ($J_{pump} (s^{-1})$) in the mean-field approximation.  We find that the mean-field approximation  fails to describe energy transduction, which is predicted correctly in simulations that use the exact master equation (as found in previous studies~\cite{kim2007kinetic}). In the mean-field treatment, the protons slip past the oxygen catalytic site and flow downhill from the P to the  N side of the membrane.
The full master equation treatment predicts $J_{pump} > 0$ (energy transduction) in most regimes of $\kappa_{12}$ (the proton transport rate constant from site 1 to site 2) varied from $10^{2}$ to $10^{7}$ sec$^{-1}$ and $k_{p}$ (the rate constant of $H_{2}O$ reduction) varied from $10^{5}$ to $10^{9}$ sec$^{-1}$, except for $k_{p} \gtrsim 10^{8}$ sec$^{-1}$ where slippage reactions dominate and $J_{pump} < 0$.  These data are shown in Fig. \ref{fig:Pump flux}a.
The mean-field model for the proton pump, in contrast, predicts $J_{pump} < 0$ in all regimes of $\kappa_{12}$ varied from $10^{2}$ to $10^{7}$ sec$^{-1}$ and $k_{p}$ varied from $10^{5}$ to $10^{9}$ sec$^{-1}$ (Fig. \ref{fig:Pump flux}b).  The mean-field treatment never predicts energy transduction (positive $J_{pump}$) for this model. 

\subsubsection{Near reversible electron bifurcation}
Electron bifurcation reactions transduce energy. The reactions play a vital role in bieenergetics~\cite{RN296}, including the Q cycle of respiration~\cite{RN366}, nitrogen fixation~\cite{RN257}, hydrogen reduction~\cite{RN307}, and methanogenesis~\cite{RN308}. A minimal model for reversible electron bifurcation was  described recently~\cite{RN197}.  Electron bifurcation occurs when a pool of two-electron donors is oxidized, and the  electrons thus produced reduce two separate acceptor species. The downhill flow of electrons to a high potential substrate is leveraged to drive the other electrons uphill to another substrate at lower potential~\cite{RN197,RN296,RN306,RN189,RN310,RN305,RN311,RN309}. Electron bifurcating enzymes typically contain a two-electron cofactor binding site, which serves as the site of bifurcation. Remarkably, electron bifurcation is typically nearly reversible, so  energy transduction occurs even when $\Delta G \approx 0$ for the bifurcation process~\cite{RN197}.
\par
\begin{figure}[bt!]
    \centering
    \includegraphics[width=0.5\textwidth]{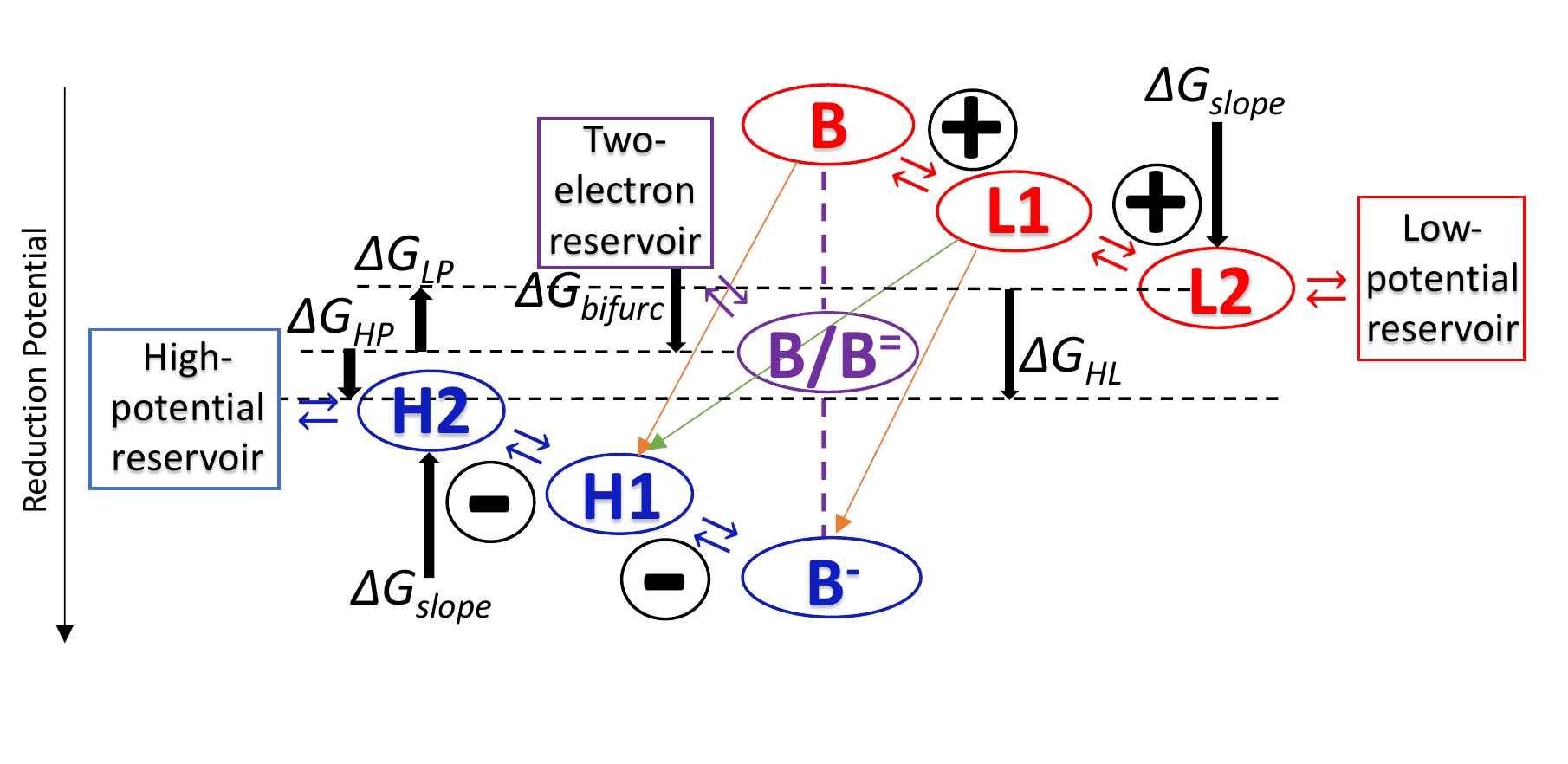}
\caption{Model for near-reversible electron bifurcation. Low-potential (higher energy) branch (red) and high-potential (lower energy) branch (blue). $B$ is a two-electron cofactor and $L1, L2, H1$, and $H2$ are one-electron cofactors in the low and high-potential branches. Two electrons are transferred to $B$ from the two-electron reservoir.  Then, one electron proceeds thermodynamically downhill through the high-potential branch, while the other moves uphill along the low-potential branch. There are substrates to be reduced (low and high-potential reservoirs) at the termini of the two branches. The redox potentials of the low and high-potential reservoirs are equal to those of the terminal cofactors $L2$ and $H2$. The redox potential of $L1$ ($H1$) is the average of $B$ and $L2$ ($B^{-}$ and $H2$). $\Delta G_{bifurc}$ is the energy to transfer two electrons from the two-electron reservoir to the two-electron cofactor and is also the overall free energy of the electron bifurcation system. $\Delta G_{slope}$  is the free energy for transfer of an electron from the two-electron cofactor to the terminal cofactor. $\Delta G_{HP}$ ($\Delta G_{LP}$) is the free energy of transporting an electron downhill (uphill) from the two-electron donor to the high-potential (low-potential) reservoir, which is the difference between the redox potential of the high-potential (low-potential) reservoir and the midpoint of the first and second redox potentials of $B$. $\Delta G_{HL}$ is the chemical potential difference between the low and high-potential reservoirs. The orange and green arrows represent the dominant short-circuiting channels. The plus and minus signs represent the buildup of holes and electrons near the bifurcating cofactor, suppressing microstates that can readily short-circuit~\cite{RN197}.}
\label{fig:EB scheme}
\end{figure}

Theoretical studies found that a privileged free-energy landscape (see Fig. \ref{fig:EB scheme}) can efficiently, reversely, and robustly perform energy transduction via electron bifurcation, minimizing energy wasting short-circuiting~\cite{RN197,RN371}. For strongly irreversible energy transduction, as occurs in the early events of photosynthesis, the short distances between nearest-neighbor redox cofactors and the Marcus invertedness of electron-hole recombination disfavors back electron transfer.  These factors minimize short-circuiting, but a large free energy cost is paid to do so~\cite{blankenship2021molecular,imahori2001modulating}. In contrast, nearly reversible electron-transfer reactions cannot exploit energy dissipation to eliminate short-circuiting because the overall free energy of the two-electron  bifurcation process occurs at close to zero overall driving force~\cite{RN306,osyczka2004reversible}. An earlier study found that steep free energy slopes ($\Delta G_{slope}$ in Fig. \ref{fig:EB scheme}) in the two redox branches produce occupancy blockade effects that minimize short-circuiting (this framework defines the EB scheme: Fig. \ref{fig:EB scheme})~\cite{RN197}. Electrons and holes form blockades in the high- and low-potential branches, minimizing short-circuiting electron transfer~\cite{RN197}. We examine whether the mean-field approximation captures the efficacy of this occupancy blockade effect.
\par
Site-site interactions are not required to produce efficient energy transduction via electron bifurcation~\cite{RN197}, so they are not included in the electron bifurcation model used here. Thus, electron bifurcation contrasts with many other energy transducing systems (such as the the redox-coupled proton pump model above~\cite{kim2007kinetic, kim2012proton}) that require site-site interactions for efficient energy transduction. In near reversible electron bifurcation, site occupancies in each of the two electron hopping branches are well approximated using Boltzmann weights that reflect the reduction potentials of the sites and their connected reservoirs~\cite{RN197, RN371}. Recall that the mean-field approximation only fails when site-site interactions or thermodynamic disequilibrium arise (see section \ref{The mean-field approximation to many-particle transport}); yet, some disequilibrium is required to drive energy transducing reactions (see section 3 in SI Appendix). Since the electron bifurcation model does not include site-site interactions, and each branch is approximately in local equilibrium (although out of equilibrium with the other branch~\cite{RN197}), reversible electron bifurcation seems to be a prime candidate for energy transduction that might be described well by the mean-field approximation (Eq. \ref{mean-field approximation}).  As such, we explore the viability of the mean-field approximation to describe free-energy transduction via electron bifurcation.

\subsubsection*{The mean-field approximation does not describe free-energy transduction by near reversible electron bifurcation}
\begin{figure}
    \centering
     \begin{subfigure}[b]{0.49\textwidth}
         \subcaption{}
         \includegraphics[width=\textwidth]{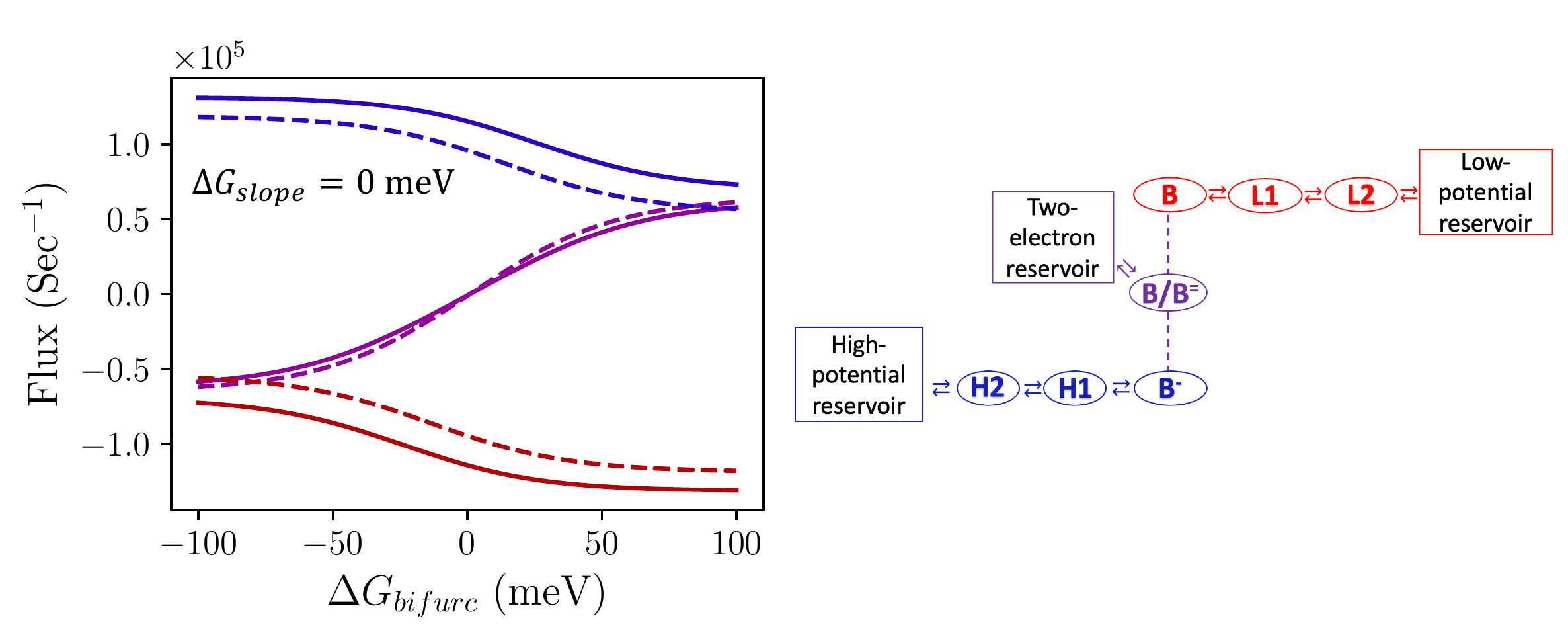}
         \label{fig:EBscheme_0meV}
     \end{subfigure}
     \begin{subfigure}[b]{0.49\textwidth}
         \subcaption{}
         \includegraphics[width=\textwidth]{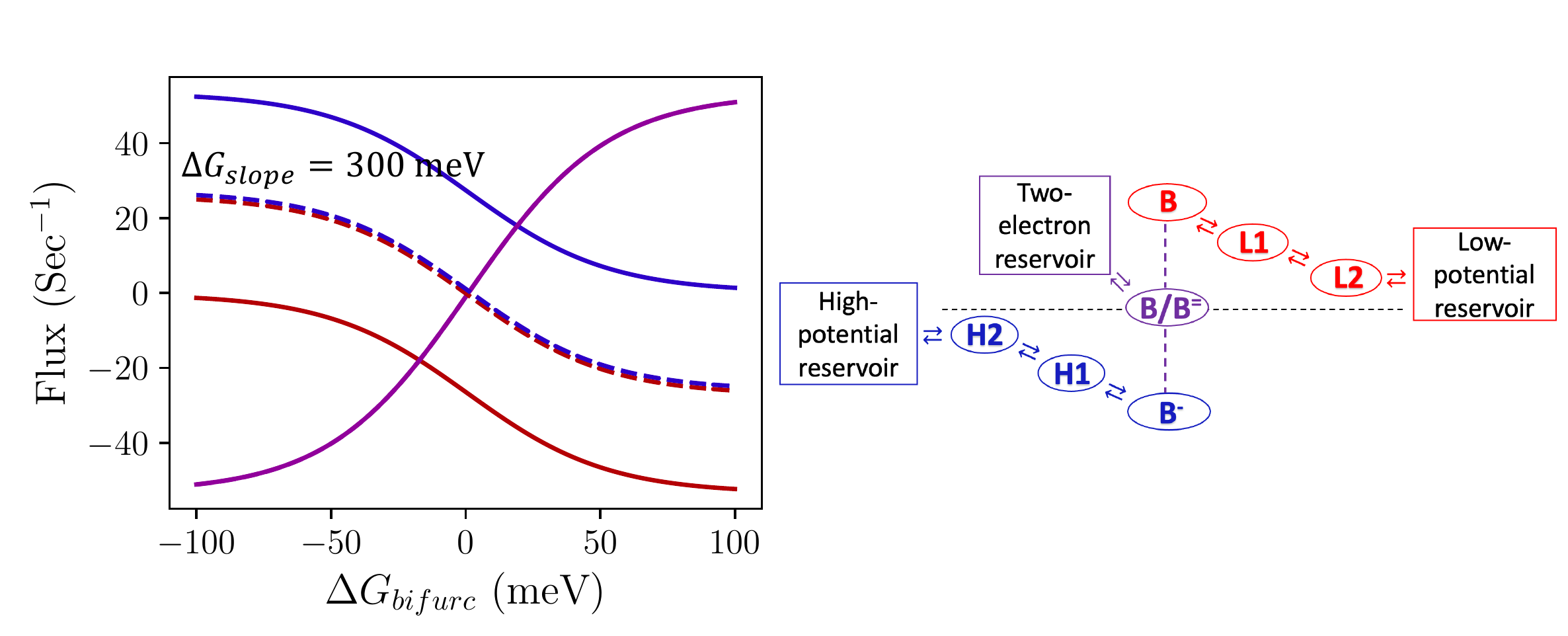}
         \label{fig:EBscheme_300meV}
     \end{subfigure}
     \begin{subfigure}[b]{0.49\textwidth}
         \subcaption{}
         \includegraphics[width=\textwidth]{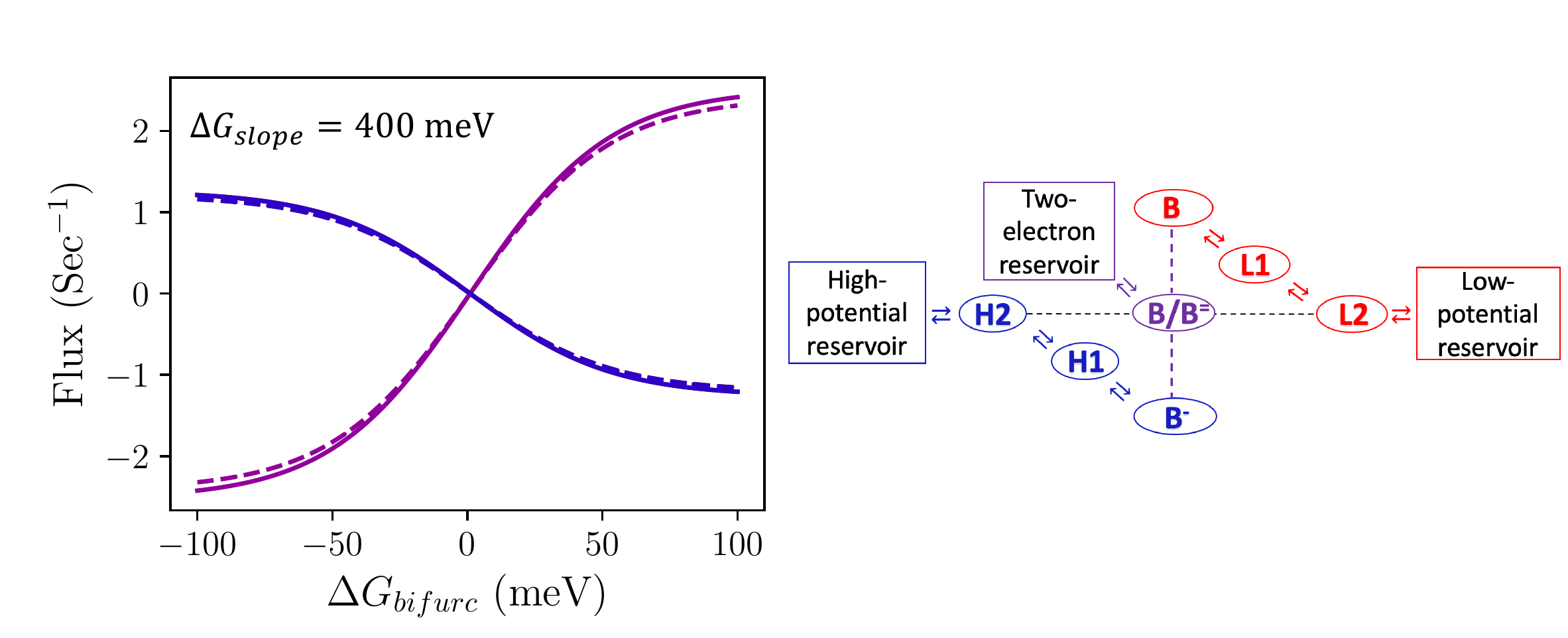}
         \label{fig:EBscheme_400meV}
     \end{subfigure}
     \begin{subfigure}[b]{0.3\textwidth}
         \includegraphics[width=\textwidth]{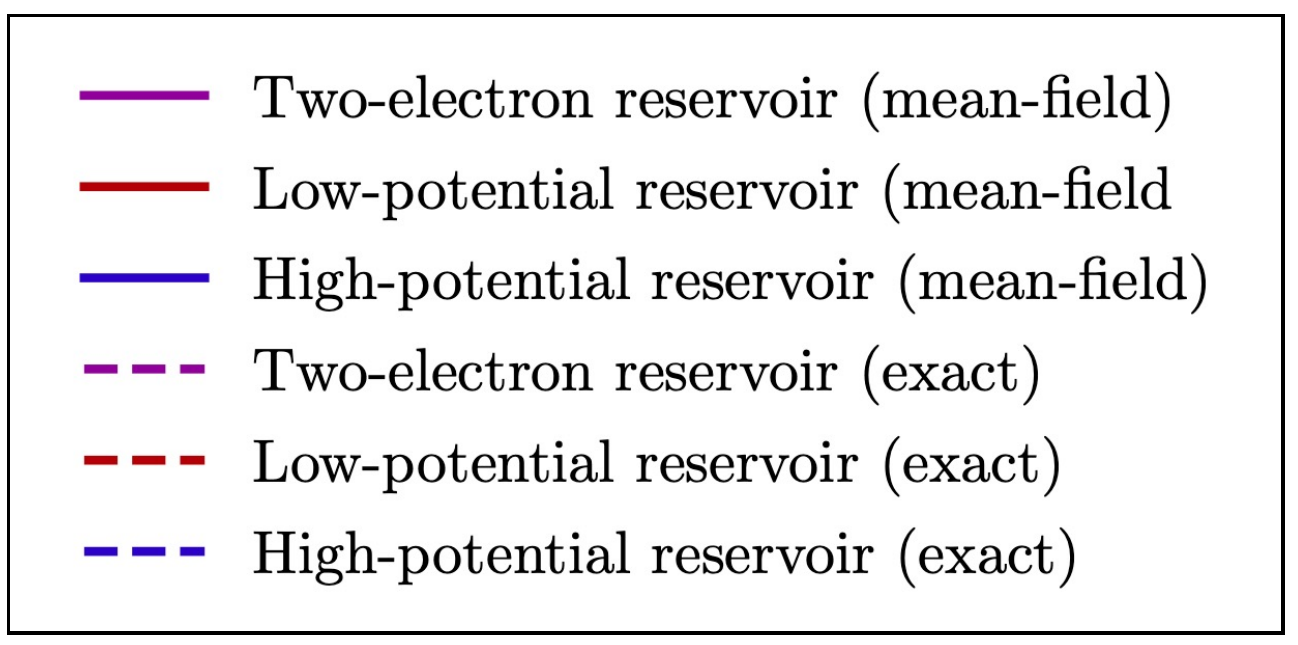}
         \vspace{10pt}
     \end{subfigure}
\caption{The net electron fluxes into the reservoirs (two-electron reservoir: purple, high-potential reservoir: blue, low-potential reservoir: red) as a function of $\Delta G_{bifurc}$ for $\Delta G_{slope}$ = (a) 0, (b) 300, and (c) 400 meV. The first and second redox potentials of $B$ are fixed at -400 and 400 meV, respectively, and the redox potentials of the low and high-potential reservoirs are equivalent to the terminal cofactor potentials. $\Delta G_{HL}$ (the chemical potential difference between the low and high-potential reservoirs) is 800, 200, and 0 meV for $\Delta G_{slope}$ = 0, 300, and 400 meV, respectively. The solid curves are electron fluxes approximated by mean-field. The dashed curves are the exact fluxes computed using the full master equation (Eq. \ref{master eq}). The blue and red curves almost overlap in panel (c), indicating efficient 1-1 electron partitioning between the high- and low-potential branches. Electron conservation (and the assumption of steady-state conditions) guarantees that the sum of the fluxes represented by the blue and red curves matches the values indicated by the purple curve. The efficiency of free-energy transduction by near reversible electron bifurcation is evaluated by the difference between the blue and red curves, since any difference between the two is caused by short-circuiting. There is efficient free energy transduction when there are approximately equal electron fluxes flowing to the high- and low-potential reservoirs~\cite{RN197}. In contrast, slippage is significant when gaps between the blue and red curves increase~\cite{RN197}. Panel (b) indicates that the mean-field approximation predicts significant short-circuiting fluxes while the exact master equation predicts that short-circuiting fluxes are suppressed. In panels (a) and (c), which are regimes that do not produce free-energy transduction (in the regime of panel (a), short circuiting is rampant, in the regime of panel (c), the high- and low-potential electron reservoirs have the same potential, so uphill electron flow is impossible), and the mean-field treatment  agrees qualitatively with the exact master equation.}
\label{fig:EBfluxresults}
\end{figure}

We simulated the kinetics of the EB scheme (Fig. \ref{fig:EB scheme}) using the exact master equation (dashed lines) and the mean-field model (solid lines) with $\Delta G_{slope}$ varied from -200 to 400 meV (see SI Appendix for parameters used,  drawn from previous studies~\cite{RN197}). We compared the steady state electron fluxes  into the two-electron (purple curve), high-potential (blue curve), and low-potential reservoirs (red curve) using the full master equation and its mean-field approximation (Fig. \ref{fig:EBfluxresults}). The slopes of the energy gradients in the branches are varied as indicated in the figure caption.
\par
For energy transduction by electron bifurcation to occur in this model~\cite{RN197, RN371}, the energy slopes in the branches ($\Delta G_{slope}$) are crucial.  Proper slopes promote electron buildup (near the bifurcating cofactor) on the high-potential branch and hole buildup on the low potential branch, suppressing microstates that short circuit. These proper slopes favor microstates that cannot short circuit, since short circuiting requires an electron on the low-potential branch and a nearby hole on the high-potential branch~\cite{RN197}. 
\par
The comparison in Fig. \ref{fig:EBfluxresults} shows that the mean-field approximation fails in regimes where the network performs energy transduction. In particular, when $\Delta G_{slope}$ = 300 meV (Fig. \ref{fig:EBscheme_300meV}), the blockade effect is present, $\Delta G_{HP} < 0$, and $\Delta G_{LP} > 0$, so the system is energy transducing.  However, the mean-field approximation fails to capture the efficient partitioning of electrons. As the slopes increase,  the system will eventually fail to transduce energy as the high and low potential reservoirs approach the same chemical potential ($\Delta G_{HL} \rightarrow 0$): electrons cannot flow uphill as required for energy transduction. In this regime, the mean-field approximation performs well (Fig. \ref{fig:EBscheme_400meV}), even though blockade effects are present in the two branches. If the slopes in the branches are absent or have the wrong sign ($\Delta G_{slope} \leq 0$), the exact master equation predicts rampant short circuiting because of the lack of appropriate electron and hole buildup, and the system behaves like a linear chain with a reservoir in the center. In this regime, the mean-field approximation is once again qualitatively accurate (Fig. \ref{fig:EBscheme_0meV}). The occupancy blockade effect is underestimated in the mean-field treatment, and this error originates in an inadequate description of statistical correlations among site occupancies. Note that there are cases of mean-field failure even when the system is not energy transducing (Fig. S4 in SI Appendix). Energy transduction is thus a sufficient but not a necessary condition for the mean-field approximation to fail.
\par
The mean-field approximation's failure to capture the electron and hole blockade effects in the EB scheme is particularly puzzling since electron and hole buildup in the high and low potential branches is not obviously  associated with particle motion correlations.
Mean-field failure to describe electron bifurcation is also interesting because the only elements that cause the mean-field approximation to fail are site-site interactions (which are not present in the electron bifurcation model~\cite{RN197}) and disequilibrium (see section \ref{Introduction}), but reversible electron bifurcation has local equilibrium in each branch. The source of mean-field failure in reversible electron bifurcation is discussed further in section \ref{Discussion}.

\section{Discussion}    \label{Discussion}
We explored the viability of the mean-field approximation to describe three canonical examples of transport networks in bioenergetics. While qualitatively describing the flow through the linear transport chain, the mean-field approximation fails to capture the energy transduction function in the redox-coupled proton pump and near reversible electron bifurcation models.
\par
The qualitative failure of the mean-field approximation is especially interesting for near reversible electron bifurcation. Near reversible electron bifurcation ($\Delta G_{bifurc} \approx 0$) allows near equilibrium conditions to persist in each redox  branch~\cite{osyczka2004reversible}, so short-circuiting fluxes at $\Delta G_{bifurc} = 0$ may be calculated using Boltzmann weights for moving electrons and holes in each branch to short circuit~\cite{RN197,RN371} (See SI appendix Section 6 for details). In addition, our model for electron bifurcation lacks interactions between the redox sites, as these interactions are not required \cite{RN197}. Since disequilibrium and site-site interactions are the only elements that cause correlations in the system, the only possible source of correlations is the disequilibrium generated by the high- and low-potential reservoirs, which must be out of equilibrium for energy transduction to occur (i.e., some electrons move uphill, others downhill).
\par
In short, if any energy transducing system can be described by a mean-field approximation, near reversible electron bifurcation without interactions would be a strong candidate. Even in this case, however, we find that the correlations between the cofactors' occupancy must be accounted for in order to describe the electron flux through the system correctly. We also find that the computed mean-field short circuiting flux is close to the geometric mean of the fluxes through the dominant short-circuiting channels computed using the exact master equation (Fig. S5 in SI Appendix). This finding may be interpreted as arising from a difference in effective energy barriers for short-circuiting between the mean-field  and the exact master equation treatments, despite the fact that the physical landscapes are the same for both calculations (See section 6
in SI Appendix for details). This result suggests that the mean-field approximation may unnaturally extend the near-equilibrium statistics of the separate branches in a  way that washes out information about the activation energy for short-circuiting.  The mean-field approximation is unable to distinguish the short-circuiting channels, instead only capturing the geometric mean of their contributions to the short-circuiting flux (the arithmetic mean of their activation free energies). We were surprised by the catastrophic failure of the mean-field approximation to describe the electron flow through the electron bifurcating system, and this finding motivated us to generalize and explore the broader applicability of the mean-field approximation to other energy transducing systems.
\par
Previous studies found that the full description of many-particle transport is necessary to describe the kinetics of  complex III~\cite{ransac2008loneliness}. Our findings suggest that the result of Ref \cite{ransac2008loneliness} is not a peculiarity of complex III, but is indeed general to any energy transduction process. Our study also extends this thinking about complex III to the broader context of particle transport theories and energy transduction. We  suggest that using the  exact master equation (Eq. \ref{master eq}) to describe transport kinetics may be unnecessary in some circumstances, and identifying which correlations are essential (and the corresponding approximate methods) to describe machine-like function  remains an outstanding challenge for future research.
\par
The failure of the mean-field approximation to describe energy transduction raises two related and compelling questions. What information, precisely, is lost when using the mean-field approximation to analyze kinetic networks, information that ultimately determines how energy transduction occurs?  Equivalently, what is the minimal additional information that must be reintroduced beyond the mean-field level to describe how slippage flux is suppressed? For example, why did the mean-field approximation fail to distinguish the various short-circuiting channels in the electron bifurcation model (See section 6 in SI Appendix)? 
\par
The study of biological transport networks with a large number of redox cofactors (such as bacterial nanowires, cable bacteria~\cite{pirbadian2014shewanella,RN241},  and large electron transfer/bifurcating enzymes~\cite{huwiler2019one}) are hampered by the exponential growth in the  computational cost of solving the exact master equation as the system size grows. The mean-field approximation provides a tempting strategy to address these large systems, but our analysis urges caution in the context of energy transduction systems, or systems with an energy transducing element (other strategies, such as use of the Gillespie algorithm~\cite{gillespie1977exact,gillespie1992rigorous} and similar approaches may capture energy transduction function at lower computational cost than directly solving Eq \ref{master eq}). For large energy transducing systems (e.g., electron bifurcating systems with many redox cofactors~\cite{huwiler2019one}), improvements beyond the mean-field limit with advantageous size scalings are needed.
\par
The failure of the mean-field approximation to describe energy transduction at the nanoscale has a corresponding principle with respect to the nature of macroscopic machines. The intrinsic coupling between machine components required for macroscopic machine function is the source of mean-field failure. It is essential that components of a molecular machine influence one other to perform free energy transduction. For example, a macroscopic piston and drive shaft must be mechanically coupled to transduce motion. This means that the piston position also determines the drive shaft's position; that is, the state of the piston and drive shaft are interdependent and their joint probability distribution function is not separable. At the nanoscale, this determinism  is weakened by the central role of thermal fluctuations. This thermal buffeting disrupts deterministic linkage between degrees of freedom in a nanoscale machine, so the essential coupling between nanoscale components is achieved via statistical correlations among the fluctuating nanoscale components, rather than among macroscale elements that that are linked deterministically. Our results do not merely indicate that a corresponding kind of coupling is  essential at the nanoscale.  Rather, we find that this coupling among components cannot be described adequately at the mean-field level. The mean-field approximation allows the state of nanoscale machine components to depend on one another, but only through their average states. We find that the mean-field approximation does not capture the interdependence of components in  nanoscale energy transducing systems, which apparently requires explicit treatment of at least some correlations.
\par
These findings suggest that principles of biological function (e.g., electron bifurcation vs. linear transport) may constrain the kinds of models that may be used successfully to simulate the structures. In other words, there may be ``top down'' approaches to build models for biological function, where the function dictates the form of the model, instead of the model being constructed only using microscopic details of the biochemical components. This concept of functional constraints on physical models for biological function has been explored in many biophysical contexts~\cite{jolivet2008benchmark,y2003computation,hartwell1999molecular}, and we find it to be essential for modeling and understanding free energy transduction.
\par
\vspace{10 pt}

This material is based upon work supported by the US Department of Energy, Office of Science, Office of Basic Energy Sciences under award number DE-SC0021144. J.Y. acknowledges partial support from the Lewis-Sigler Institute for Integrative Genomics.
\par
\vspace{10 pt}
Author contributions: K.T., J.L.Y., P.Z., and D.N.B. designed the research; K.T. performed the research; K.T. and J.L.Y. contributed new analytical tools; K.T., J.L.Y., P.Z., and D.N.B. analyzed the data; and K.T., J.L.Y., P.Z., and D.N.B. wrote the paper. \\
\par
The authors declare no competing interests. \\

\printbibliography

\end{document}


\author[a]{Kiriko Terai}
\author[b,1]{Jonathon L. Yuly}
\author[a]{Peng Zhang}
\author[a,c,d,1]{David N. Beratan}
\affil[a]{Department of Chemistry, Duke University, Durham, NC, 27708}
\affil[b]{Lewis-Sigler Institute for Integrative Genomics, Princeton University, Princeton, NJ, 08540}
\affil[c]{Department of Physics, Duke University, Durham, NC, 27708}
\affil[d]{Department of Biochemistry, Duke University, Durham, NC, 27710}
\affil[1]{\small Corresponding authors. Email: jonathon.yuly@princeton.edu or david.beratan@duke.edu}

\maketitle

\section{Full master equation and mean-field kinetics for nanoscale particle transport}  \label{Full master equation and mean-field kinetics for nanoscale particle transport}
Each microstate for a transport network with $N$  sites can be described using
\begin{equation}  \label{microstates}
\boldsymbol{\sigma} = [\sigma_{1} \hspace{0.8mm} \sigma_{2} ... \hspace{0.8mm} \sigma_{i} \hspace{0.8mm} ... \hspace{0.8mm} \sigma_{j} \hspace{0.8mm} ...\hspace{0.8mm} \sigma_{N}]
\end{equation}
where $\sigma_{i}$ and $\sigma_{j}$ are non-negative integers that indicate the number of particles on sites $i$ and $j$ in microstate $\boldsymbol{\sigma}$.
The probability $P_{\boldsymbol{\sigma}}$ that the network is in microstate $\boldsymbol{\sigma}$ as a function of time is described by the master equation~\cite{RN152}:
 \begin{equation}  \label{master eq}
 \frac{dP_{\boldsymbol{\sigma}}(t)}{dt} = -\sum_{\boldsymbol{\sigma'}} (K_{\boldsymbol{\sigma}\boldsymbol{\sigma'}}P_{\boldsymbol{\sigma}}(t)-K_{\boldsymbol{\sigma'}\boldsymbol{\sigma}}P_{\boldsymbol{\sigma'}}(t))
 \end{equation}
where $K_{\boldsymbol{\sigma}\boldsymbol{\sigma'}}$ is the rate constant for the transition from microstate $\boldsymbol{\sigma}$ to $\boldsymbol{{\sigma'}}$. $P_{\boldsymbol{\sigma}}$ can also be written as an $N$-point correlation function:
\begin{equation}    \label{N-point correlation function}
    P_{\boldsymbol{\sigma}} = p_{1\sigma_1,2\sigma_2,...,i\sigma_i,...,j\sigma_j,...,N\sigma_N}
\end{equation}
which is the probability of simultaneously finding sites 1, 2, ..., $i$, ..., $j$, ..., and $N$ in occupation states $\sigma_1$, $\sigma_2$, ..., $\sigma_i$, ..., $\sigma_j$, ..., and $\sigma_N$, respectively.
\par
Site occupancy constraints are enforced by the values of the rate constants $K_{\boldsymbol{\sigma}\boldsymbol{\sigma'}}$, such that a rate constant $K_{\boldsymbol{\sigma}\boldsymbol{\sigma'}}$ is zero if either microstate $\boldsymbol{\sigma}$ or $\boldsymbol{\sigma'}$ violates the occupancy constraints for any site in the network. The following indicates a transition between microstate $\boldsymbol{\sigma}$ and $\boldsymbol{\sigma'}$ associated with a typical $n$-particle transport process within the network:
\begin{equation}    \label{site-site}
\boldsymbol{\sigma}: [\sigma_{1} \hspace{0.8mm} \sigma_{2} ... \hspace{0.8mm} \sigma_{i} \hspace{0.8mm} ... \hspace{0.8mm} \sigma_{j} \hspace{0.8mm} ...] \rightleftharpoons \boldsymbol{\sigma'}: [\sigma_{1} \hspace{0.8mm} \sigma_{2} ... \hspace{0.8mm} (\sigma_{i}-n) \hspace{0.8mm} ... \hspace{0.8mm} (\sigma_{j}+n) \hspace{0.8mm} ...]
\end{equation}
The rate constant for an $n$-particle transfer (beginning in  microstate $\boldsymbol{\sigma}$) from site $i$, initially in occupation state $\sigma_i$, to site $j$, initially in occupation state $\sigma_j$, is denoted by $k^{\boldsymbol{\sigma}}_{i,\sigma_i \underset{n}\rightarrow j,\sigma_j}$. 
\par
In addition to $n$-particle transport between sites, particles are allowed to enter or leave the system through interactions with reservoirs (Eq \ref{site-reservoir}), $n$ particles at a time. The following is a transition between microstate $\boldsymbol{\sigma}$ and $\boldsymbol{\sigma'}$ associated with a typical $n$-particle transfer with a reservoir at site $i$:
\begin{equation}       \label{site-reservoir}
\boldsymbol{\sigma}: [\sigma_{1} \hspace{0.8mm} \sigma_{2} ... \hspace{0.8mm} \sigma_{i} \hspace{0.8mm} ... \hspace{0.8mm} \sigma_{j} \hspace{0.8mm} ...] \rightleftharpoons \boldsymbol{\sigma'}: [\sigma_{1} \hspace{0.8mm} \sigma_{2} ... \hspace{0.8mm} (\sigma_{i}+n) \hspace{0.8mm} ... \hspace{0.8mm} \sigma_{j} \hspace{0.8mm} ...]
\end{equation}
The rate constant of an $n$-particle transfer from site $i$ (in microstate $\boldsymbol{\sigma}$) , with occupancy $\sigma_i$, to a reservoir (or from a reservoir to site $i$, initially in occupation state $\sigma_i$) is denoted  $k^{\boldsymbol{\sigma}}_{i,\sigma_i \underset{n}\rightarrow \text{out}}$ ($k^{\boldsymbol{\sigma}}_{\text{in} \underset{n}\rightarrow i,\sigma_i}$). 
\par
$K_{\boldsymbol{\sigma}\boldsymbol{\sigma'}}$, written in terms of $k^{\boldsymbol{\sigma}}_{i,\sigma_i \underset{n}\rightarrow j,\sigma_j}$, $k^{\boldsymbol{\sigma}}_{i,\sigma_i \underset{n}\rightarrow \text{out}}$, and $k^{\boldsymbol{\sigma}}_{\text{in} \underset{n}\rightarrow i,\sigma_i}$, is:
\begin{equation}    \label{microstate-microstate rate}
\begin{split}
K_{\boldsymbol{\sigma}\boldsymbol{\sigma'}} &= \sum_{\substack{i \\ j\neq i \\n}} k^{\boldsymbol{\sigma}}_{i,\sigma_i \underset{n}\rightarrow j,\sigma_j} \hspace{0.8mm} \delta_{\sigma_{i},\sigma'_{i}+n} \delta_{\sigma_{j}+n,\sigma'_{j}} \prod_{l \neq i,j} \delta_{\sigma_{l}\sigma'_{l}} \\
&+ \sum_{i,n} k^{\boldsymbol{\sigma}}_{i,\sigma_i \underset{n}\rightarrow \text{out}} \hspace{0.8mm} \delta_{\sigma_i-n,\sigma'_{i}} \prod_{l \neq i} \delta_{\sigma_{l}\sigma'_{l}} \\
&+ \sum_{i,n} k^{\boldsymbol{\sigma}}_{\text{in} \underset{n}\rightarrow i,\sigma_i} \hspace{0.8mm} \delta_{\sigma_i+n,\sigma'_{i}} \prod_{l \neq i} \delta_{\sigma_{l}\sigma'_{l}}
\end{split}
\end{equation}
\par
The probability of finding site $i$ with occupancy   $q$ is denoted  $p_{iq}$. This average quantity is:
\begin{equation}    \label{probability}
p_{iq} = \sum_{\boldsymbol{\sigma}} \delta_{\sigma_{i},q} P_{\boldsymbol{\sigma}}
\end{equation}
where $\delta$ is the Kronecker delta. Therefore, the exact master equation (Eq \ref{master eq}) gives the exact time derivative of $p_{iq}$ as~\cite{RN152}:
\begin{equation}   \label{exact rate eq} 
\begin{split}
    \frac{dp_{iq}}{dt} &= \sum_{\boldsymbol{\sigma}} \delta_{\sigma_{i},q} \frac{dP_{\boldsymbol{\sigma}}(t)}{dt} \\
    &= -\sum_{\boldsymbol{\sigma}} \delta_{\sigma_{i},q} \{ \sum_{\boldsymbol{\sigma'}} (K_{\boldsymbol{\sigma}\boldsymbol{\sigma'}}P_{\boldsymbol{\sigma}}-K_{\boldsymbol{\sigma'}\boldsymbol{\sigma}}P_{\boldsymbol{\sigma'}}) \}
\end{split}
\end{equation}

\subsection{An exact formula for the time derivative of $p_{iq}$ from the master equation}   \label{An exact formula for the time derivative of $p_iq$ from the master equation}
When site-site interactions are present, the rate of $n$ particle transfer from site $i$ to site $j$ depends on the occupancy of all the $n$-particle transfer rate constant depends on the initial and the final microstates. 
\par
Beginning with Eq \ref{exact rate eq}, the exact time derivative of $p_{iq}$ when site-site interactions are present is:
\begin{equation}    \label{exact p_iq - interaction}
\begin{split}
    \frac{dp_{iq}}{dt} &= -\sum_{\boldsymbol{\sigma}} \delta_{\sigma_{i},q} \{ \sum_{\boldsymbol{\sigma'}} (K_{\boldsymbol{\sigma}\boldsymbol{\sigma'}}P_{\boldsymbol{\sigma}}-K_{\boldsymbol{\sigma'}\boldsymbol{\sigma}}P_{\boldsymbol{\sigma'}}) \} \\
    &= -\sum_{\boldsymbol{\sigma}} \delta_{\sigma_{i},q} \\
    & \indent \{ \sum_{\substack{j\neq i \\n}} [\sum_{\boldsymbol{\sigma'}}(k^{\boldsymbol{\sigma}}_{i,\sigma_i \underset{n}\rightarrow j,\sigma_j}P_{\boldsymbol{\sigma}}-k^{\boldsymbol{\sigma'}}_{j,\sigma_j+n \underset{n}\rightarrow i,\sigma_i-n}P_{\boldsymbol{\sigma'}})\delta_{\sigma_i-n,\sigma'_i}\delta_{\sigma_j+n,\sigma'_j}\prod_{l \neq i,j}\delta_{\sigma_l,\sigma'_l}] \\
    & \indent + \sum_{n}[\sum_{\boldsymbol{\sigma'}}(k^{\boldsymbol{\sigma}}_{i,\sigma_i \underset{n}\rightarrow \text{out}}P_{\boldsymbol{\sigma}}-k^{\boldsymbol{\sigma'}}_{\text{in} \underset{n}\rightarrow i,\sigma_i-n}P_{\boldsymbol{\sigma'}})\delta_{\sigma_i-n,\sigma'_i}\prod_{l \neq i}\delta_{\sigma_l,\sigma'_l}] \}  \\
    &= -\sum_{\boldsymbol{\sigma}} \delta_{\sigma_{i},q} \\
    & \indent \{ \sum_{\substack{j\neq i \\n}}[\sum_{\boldsymbol{\sigma'}}(k^{\boldsymbol{\sigma}}_{i,\sigma_i \underset{n}\rightarrow j,\sigma_j}p_{1\sigma_1, ..., i\sigma_i,..., j\sigma_j,...,N\sigma_N}-k^{\boldsymbol{\sigma'}}_{j,\sigma_j+n \underset{n}\rightarrow i,\sigma_i-n}p_{1\sigma'_1, ...,i\sigma'_i, ..., j\sigma'_j,...,N\sigma'_N}) \delta_{\sigma_i-n,\sigma'_i}\delta_{\sigma_j+n,\sigma'_j}\prod_{l \neq i,j}\delta_{\sigma_l,\sigma'_l}] \\
    & \indent + \sum_{n}[\sum_{\boldsymbol{\sigma'}}(k^{\boldsymbol{\sigma}}_{i,\sigma_i \underset{n}\rightarrow \text{out}}p_{1\sigma_1, ..., i\sigma_i,..., j\sigma_j,...,N\sigma_N}-k^{\boldsymbol{\sigma'}}_{\text{in} \underset{n}\rightarrow i,\sigma_i-n}p_{1\sigma'_1, ...,i\sigma'_i, ..., j\sigma'_j,...,N\sigma'_N})\delta_{\sigma_i-n,\sigma'_i}\prod_{l \neq i}\delta_{\sigma_l,\sigma'_l}] \}, \\
\end{split}
\end{equation}
where the first step uses Eq \ref{microstate-microstate rate} and last step uses Eq \ref{N-point correlation function}.

\subsection{Derivation of Eq. 8 of the main text}   \label{Deriving Eq. 8 of the main text}
In the absence of site-site interactions, the rate for $n$ particle transfer from site $i$ to site $j$ does not depend on the occupancy of other sites in the transport network. The $n$-particle transfer rate constant is independent of the initial and the final microstates. In the absence of site interactions, the rate constant for $n$-particle transfer from site $i$, initially in occupation state $\sigma_i$, to site $j$, initially in occupation state $\sigma_j$ is $k_{i,\sigma_i \underset{n}\rightarrow j,\sigma_j}$. Beginning with Eq \ref{exact p_iq - interaction}, the exact time derivative of $p_{iq}$ in the absence of site-site interactions is:
\begin{equation*} 
\begin{split}
\frac{dp_{iq}}{dt} &= -\sum_{\boldsymbol{\sigma}} \delta_{\sigma_{i},q}\\
& \indent \{ \sum_{\substack{j\neq i \\n}}[\sum_{\boldsymbol{\sigma'}}(k_{i,\sigma_i \underset{n}\rightarrow j,\sigma_j}p_{1\sigma_1, ...,i\sigma_i, ..., j\sigma_j,...,N\sigma_N}-k_{j,\sigma_j+n \underset{n}\rightarrow i,\sigma_i-n}p_{1\sigma'_1, ...,i\sigma'_i, ..., j\sigma'_j,...,N\sigma'_N}) \delta_{\sigma_i-n,\sigma'_i}\delta_{\sigma_j+n,\sigma'_j}\prod_{l \neq i,j}\delta_{\sigma_l,\sigma'_l}] \\
& \indent + \sum_{n}[\sum_{\boldsymbol{\sigma'}}(k_{i,\sigma_i \underset{n}\rightarrow \text{out}}p_{1\sigma_1, ...,i\sigma_i, ..., j\sigma_j,...,N\sigma_N}-k_{\text{in} \underset{n}\rightarrow i,\sigma_i-n}p_{1\sigma'_1, ...,i\sigma'_i, ..., j\sigma'_j,...,N\sigma'_N})\delta_{\sigma_i-n,\sigma'_i}\prod_{l \neq i}\delta_{\sigma_l,\sigma'_l}] \} \\
\end{split}
\end{equation*}
Summing over the $\boldsymbol{ \sigma}'$ variables (applying the Kronecker deltas) and using $\sum_{\boldsymbol{\sigma}}=\sum_{\sigma_1}\cdot \cdot \cdot \sum_{\sigma_i}\cdot \cdot \cdot \sum_{\sigma_j} \cdot \cdot \cdot\sum_{\sigma_N}$ gives
\begin{equation*}
\begin{split}
\frac{dp_{iq}}{dt} &= -\sum_{\sigma_i} \delta_{\sigma_{i},q} \\
& \indent \{ \sum_{\substack{j\neq i \\n}}  \sum_{\sigma_j} \prod_{l\neq i, j}\sum_{\sigma_l} (k_{i,\sigma_i \underset{n}\rightarrow j,\sigma_j}p_{1\sigma_1,...,i\sigma_i,...,j\sigma_j,...,N\sigma_N}-k_{j,\sigma_j+n \underset{n}\rightarrow i,\sigma_i-n}p_{1\sigma_1,...,i\sigma_i-n,...,j\sigma_j+n,...,N\sigma_N})  \\
& \indent + \sum_{\substack{j\neq i \\n}} \sum_{\sigma_j} \prod_{l\neq i,j}\sum_{\sigma_l} (k_{i,\sigma_i \underset{n}\rightarrow \text{out}}p_{1\sigma_1,...,i\sigma_i,...,j\sigma_j,...,N\sigma_N}-k_{\text{in} \underset{n}\rightarrow i,\sigma_i-n}p_{i\sigma_i-n,j\sigma_j,l\sigma_l,...,N\sigma_N}) \} \\
\end{split}
\end{equation*}
Using the identities of Eq. \ref{probability identity} below gives
\begin{equation*}
\begin{split}
\frac{dp_{iq}}{dt} &= -\sum_{\sigma_i} \delta_{\sigma_{i},q} \\
& \indent \sum_{\substack{j\neq i \\n}}  \sum_{\sigma_j} \left(k_{i,\sigma_i \underset{n}\rightarrow j,\sigma_j}p_{i\sigma_i,j\sigma_j}-k_{j,\sigma_j+n \underset{n}\rightarrow i,\sigma_i-n}p_{i\sigma_i-n,j\sigma_j+n} + (k_{i,\sigma_i \underset{n}\rightarrow \text{out}}p_{i\sigma_i, j\sigma_j}-k_{\text{in} \underset{n}\rightarrow i,\sigma_i-n}p_{i\sigma_i-n, j\sigma_j}) \right).
\end{split}
\end{equation*}
The second term in the double sum above depends only on the probabilities $p_{i\sigma_i}$ and $p_{i \sigma_i - n}$ when the sum over $\sigma_j$ is carried out (because of the identity $\sum_{\sigma_j} p_{i\sigma_i, j\sigma_j} = p_{i\sigma_i}$). However, we choose to keep the sum in the form of Eq \ref{exact p_iq - no interaction} to obtain $w_{i,j, n, s}^{r,u}$ of Eq 8 in the main text. Performing the sum over $\sigma_i$ and substituting $\sigma_j$ with $s$ gives
\begin{equation}     \label{exact p_iq - no interaction}
\begin{split}
\frac{dp_{iq}}{dt} &= -\sum_{\substack{j\neq i\\n}}  \sum_{s} \left(k_{i,q \underset{n}\rightarrow j,s}p_{iq,js}-k_{j,s+n \underset{n}\rightarrow i,q-n}p_{iq-n,js+n} + (k_{i,q \underset{n}\rightarrow \text{out}}p_{iq, js}-k_{\text{in} \underset{n}\rightarrow i,q-n}p_{iq-n, js}) \right)\\
&= -\sum_{\substack{j\neq i\\n,s}} \sum_{r, u} w_{i,j, n, s}^{r,u} \hspace{3pt} p_{ir,ju} \hspace{10 pt}
\end{split}
\end{equation}
where $w_{i,j, n, s}^{r,u}$ is defined by
\begin{equation}    \label{def of w}
\begin{split}
    w_{i,j,n,s}^{r,u} &=  \delta_{r,q} \delta_{u,s} k_{i,r \underset{n}\rightarrow j,u} - \delta_{r,q-n}\delta_{u,s+n}k_{j,u \underset{n}\rightarrow i,r} + \delta_{r,q} \delta_{u,s} k_{i,r \underset{n}\rightarrow \text{out}} - \delta_{r, q-n} \delta_{u,s} k_{\text{in} \underset{n}\rightarrow i,q-n}.
\end{split}
\end{equation}
The derivation of Eq. \ref{exact p_iq - no interaction} uses the identities
\begin{equation}   \label{probability identity}
\begin{split}
    \prod_{l\neq i,j} \sum_{\sigma_l} p_{1\sigma_1,...,i\sigma_i,...,j\sigma_j,...,N\sigma_N} &= p_{i\sigma_i,j\sigma_j}  \\
    \sum_{\sigma_j}\prod_{l\neq i,j} \sum_{\sigma_l} p_{1\sigma_1,...,i\sigma_i,...,j\sigma_j,...,N\sigma_N} &= p_{i\sigma_i}   \\
\end{split}
\end{equation}
The mean-field kinetics in the absence of interactions (Eq. 9 in the main text)  follows immediately from Eq. \ref{exact p_iq - no interaction} on using the mean-field approximation $p_{ir,ju} \approx p_{ir} p_{ju}$ (Eq. \ref{mean-field approximation}), as shown in Section \ref{The mean-field time derivative of p_iq without site-site interactions}.

\subsection{Mean-field kinetics with site-site interactions}   \label{The mean-field time derivative of p_iq in the presence of site-site interactions}
For a transport network with $N$ sites for the particles, the mean-field approximation neglects all statistical correlations among site occupancies. That is, any $X$-point correlation function (Eq \ref{N-point correlation function}) with $X \leq N$, can be factored into individual probabilities~\cite{RN152,RN250,RN290}:
\begin{equation}    \label{mean-field approximation}
    p_{i_{1}q_{1},i_{2}q_{2},...,i_{X}q_{X}} \approx \prod_{x=1}^{X}p_{i_{x}q_{x}}   \hspace{1cm}    (X \leq N)
\end{equation}
Eq \ref{mean-field approximation} defines the mean-field approximation.
\par
Beginning with Eq. \ref{exact p_iq - interaction} and using the mean-field approximation, the mean-field time derivative of $p_{iq}$ including site-site interactions is:
\begin{equation*}
\begin{split}
\frac{dp_{iq}}{dt} &= -\sum_{\boldsymbol{\sigma}} \delta_{\sigma_{i},q} \\
& \indent \{ \sum_{\substack{j\neq i \\n}}[\sum_{\boldsymbol{\sigma'}}(k^{\boldsymbol{\sigma}}_{i,\sigma_i \underset{n}\rightarrow j,\sigma_j}p_{1\sigma_1}...p_{i\sigma_i}...p_{j\sigma_j}...p_{N\sigma_N} \\
& \indent -k^{\boldsymbol{\sigma'}}_{j,\sigma_j+n \underset{n}\rightarrow i,\sigma_i-n}p_{1 \sigma'_1}...p_{i\sigma_i'}...p_{j\sigma'_j}...p_{N\sigma'_N})\delta_{\sigma_i-n,\sigma'_i}\delta_{\sigma_j+n,\sigma'_j}\prod_{l \neq i,j}\delta_{\sigma_l,\sigma'_l}] \\
& \indent + \sum_{n}[\sum_{\boldsymbol{\sigma'}}(k^{\boldsymbol{\sigma}}_{i,\sigma_i \underset{n}\rightarrow \text{out}}p_{1 \sigma_1}...p_{i\sigma_i}...p_{j\sigma_j}...p_{N\sigma_N}-k^{\boldsymbol{\sigma'}}_{\text{in} \underset{n}\rightarrow i,\sigma_i-n}p_{1 \sigma'_1}...p_{i\sigma'_i}...p_{j\sigma'_j}...p_{N\sigma'_N})\delta_{\sigma_i-n,\sigma'_i} \prod_{l \neq i}\delta_{\sigma_l,\sigma'_l}] \} .\\
\end{split}
\end{equation*}
Summing over the $\boldsymbol{ \sigma}'$ variables (applying the Kronecker deltas) and using $\sum_{\boldsymbol{\sigma}}=\sum_{\sigma_1}\cdot \cdot \cdot \sum_{\sigma_i}\cdot \cdot \cdot \sum_{\sigma_j} \cdot \cdot \cdot\sum_{\sigma_N}$ gives
\begin{equation*}
\begin{split}
\frac{dp_{iq}}{dt} &= -\sum_{\sigma_i} \delta_{\sigma_{i},q} \\
& \indent  \sum_{\substack{j\neq i \\n}}\sum_{\sigma_j} \{ \prod_{l\neq i,j} [\sum_{\sigma_l} k^{\boldsymbol{\sigma}}_{i,\sigma_i \underset{n}\rightarrow j,\sigma_j} \prod_{u \neq i,j}p_{u\sigma_u}]p_{i\sigma_i}p_{j\sigma_j}-\prod_{l\neq i,j} [\sum_{\sigma_l}k^{\boldsymbol{\sigma}^f}_{j,\sigma_j+n \underset{n}\rightarrow i,\sigma_i-n}\prod_{u \neq i,j}p_{u\sigma_u}]p_{i\hspace{0.5mm}\sigma_i-n}p_{j\hspace{0.5mm}\sigma_j+n} \} \\
& \indent + \sum_{\substack{j\neq i \\n}}\sum_{\sigma_j}\{\prod_{l\neq i,j}[\sum_{\sigma_l}k^{\boldsymbol{\sigma}}_{i,\sigma_i \underset{n}\rightarrow \text{out}} \prod_{u \neq i,j}p_{u\sigma_u}]p_{i\sigma_i}p_{j\sigma_j}-\prod_{l\neq i,j}[\sum_{\sigma_l}k^{\boldsymbol{\sigma}^{f'}}_{\text{in} \underset{n}\rightarrow i,\sigma_i-n}\prod_{u \neq i,j}p_{u\sigma_u}]p_{i\hspace{0.5mm}\sigma_i-n}p_{j\sigma_j} \}, \\
\end{split}
\end{equation*}
where $\boldsymbol{\sigma}^{f}$ and $\boldsymbol{\sigma}^{f'}$ are given by
\begin{equation}
\begin{split}
\boldsymbol{\sigma}^f &= [\sigma_1, ..., \sigma_i-n,...,\sigma_j +n,...,\sigma_N] \\
\boldsymbol{\sigma}^{f'} &= [\sigma_1, ..., \sigma_i-n,...,\sigma_j,...,\sigma_N].
\end{split}
\end{equation}
Performing the sum over $\sigma_i$ and substituting $\sigma_j$ with $s$ gives
\begin{equation}       \label{mean-field p_iq - interaction}
\begin{split}
\frac{dp_{iq}}{dt} &=
-\sum_{\substack{j\neq i \\n,s}} \{ \prod_{l\neq i,j} [\sum_{\sigma_l} k^{\boldsymbol{\sigma}}_{i,q \underset{n}\rightarrow j,s} \prod_{u \neq i,j}p_{u\sigma_u}]p_{iq}p_{js}-\prod_{l\neq i,j} [\sum_{\sigma_l}k^{\boldsymbol{\sigma}^f}_{j,s+n \underset{n}\rightarrow i,q-n}\prod_{u \neq i,j}p_{u\sigma_u}]p_{i\hspace{0.5mm}q-n}p_{j\hspace{0.5mm}s+n} \} \\
& \indent + \sum_{\substack{j\neq i \\n,s}} \{\prod_{l\neq i,j}[\sum_{\sigma_l}k^{\boldsymbol{\sigma}}_{i,q \underset{n}\rightarrow \text{out}} \prod_{u \neq i,j}p_{u\sigma_u}]p_{iq}p_{js}-\prod_{l\neq i,j}[\sum_{\sigma_l}k^{\boldsymbol{\sigma}^{f'}}_{\text{in} \underset{n}\rightarrow i,q-n}\prod_{u \neq i,j}p_{u\sigma_u}]p_{i\hspace{0.5mm}q-n}p_{js} \} \\
&= -\sum_{\substack{j\neq i \\n,s}} [ k^{\text{MF}}_{i,q \underset{n}\rightarrow j,s} p_{iq}p_{js}- k^{\text{MF}}_{j,s+n \underset{n}\rightarrow i,q-n}p_{i\hspace{0.5mm}q-n}p_{j\hspace{0.5mm}s+n} ] \\
& \indent + \sum_{\substack{j\neq i \\n,s}} [k^{\text{MF}}_{i,q \underset{n}\rightarrow \text{out}} p_{iq}p_{js}-k^{\text{MF}}_{\text{in} \underset{n}\rightarrow i,q-n}p_{i\hspace{0.5mm}q-n}p_{js} ] \\
&= -\sum_{\substack{j\neq i \\n,s}} \sum_{r,u} w^{r,u(\text{MF})}_{i,j,n,s} p_{ir}p_{ju}
\end{split}
\end{equation}
where the mean-field approximated $n$-particle transfer rate constants are:
\begin{equation}    \label{mean-field rate constants}
\begin{split}
k^{\text{MF}}_{i,q \underset{n}\rightarrow j,s} &= \prod_{l\neq i,j} \sum_{\sigma_l} k^{\boldsymbol{\sigma}}_{i,q \underset{n}\rightarrow j,s} \prod_{u \neq i,j}p_{u\sigma_u}  \\
k^{\text{MF}}_{j,s+n \underset{n}\rightarrow i,q-n} &= \prod_{l\neq i,j} \sum_{\sigma_l}k^{\boldsymbol{\sigma}^f}_{j,s+n \underset{n}\rightarrow i,q-n}\prod_{u \neq i,j}p_{u\sigma_u} \\
k^{\text{MF}}_{i,q \underset{n}\rightarrow \text{out}} &= \prod_{l\neq i,j} \sum_{\sigma_l}k^{\boldsymbol{\sigma}}_{i,q \underset{n}\rightarrow \text{out}} \prod_{u \neq i,j}p_{u\sigma_u} \\
k^{\text{MF}}_{\text{in} \underset{n}\rightarrow i,q-n} &= \prod_{l\neq i,j} \sum_{\sigma_l}k^{\boldsymbol{\sigma}^{f'}}_{\text{in} \underset{n}\rightarrow i,q-n}\prod_{u \neq i,j}p_{u\sigma_u}
\end{split}
\end{equation}
and $w^{r,u(\text{MF})}_{i,j,n,s}$ is defined as
\begin{equation}    \label{def of w - mean-field}
\begin{split}
w^{r,u(\text{MF})}_{i,j,n,s} = \delta_{r,q} \delta_{u,s} k^{\text{MF}}_{i,r \underset{n}\rightarrow j,u} - \delta_{r,q-n}\delta_{u,s+n}k^{\text{MF}}_{j,u \underset{n}\rightarrow i,r} + \delta_{r,q} \delta_{u,s} k^{\text{MF}}_{i,r \underset{n}\rightarrow \text{out}} - \delta_{r, q-n} \delta_{u,s} k^{\text{MF}}_{\text{in} \underset{n}\rightarrow i,q-n} \hspace{10 pt}.
\end{split}
\end{equation}

\subsection{The mean-field kinetics without site-site interactions}   \label{The mean-field time derivative of p_iq without site-site interactions}
Using the mean-field approximation (Eq \ref{mean-field approximation}) with Eq \ref{exact p_iq - no interaction} gives the mean-field time derivative of $p_{iq}$ when site-site interactions are absent:
\begin{equation}    \label{mean-field p_iq - no interaction}
\begin{split}
\frac{dp_{iq}}{dt} &= -\sum_{\substack{j\neq i\\n,s}} \left(k_{i,q \underset{n}\rightarrow j,s}p_{iq}p_{js}-k_{j,s+n \underset{n}\rightarrow i,q-n}p_{iq-n}p_{js+n} + (k_{i,q \underset{n}\rightarrow \text{out}}p_{iq}p_{js}-k_{\text{in} \underset{n}\rightarrow i,q-n}p_{iq-n}p_{js}) \right)\\
&= -\sum_{\substack{j\neq i \\n,s}} \sum_{r,u} w^{r,u(\text{MF})}_{i,j,n,s} p_{ir}p_{ju} = -\sum_{\substack{j\neq i \\n,s}} \sum_{r,u} w^{r,u}_{i,j,n,s} p_{ir}p_{ju}
\end{split}
\end{equation}
Thus, in the absence of site-site interactions, 
\begin{equation}
\begin{split}
k^{\text{MF}}_{i,q \underset{n}\rightarrow j,s}&\rightarrow k_{i,q \underset{n}\rightarrow j,s} \\
k^{\text{MF}}_{j,s+n \underset{n}\rightarrow i,q-n}&\rightarrow k_{j,s+n \underset{n}\rightarrow i,q-n}\\
k^{\text{MF}}_{i,q \underset{n}\rightarrow \text{out}}&\rightarrow k_{i,q \underset{n}\rightarrow \text{out}} \\
k^{\text{MF}}_{\text{in} \underset{n}\rightarrow i,q-n}&\rightarrow k_{\text{in} \underset{n}\rightarrow i,q-n}
\end{split}
\end{equation}
so $w^{r,u(\text{MF})}_{i,j,n,s} \rightarrow w^{r,u}_{i,j,n,s}$, and one obtains Eq 9 in the main text, that is the mean-field kinetics in the absence of site-site interactions.

\section{The mean-field approximation is exact at equilibrium without site-site interactions} \label{The mean-field approximation at equilibrium without site-site interactions}
The mean-field approximation for transport kinetics is exact at thermal equilibrium, provided there are no site-site interactions. This can be understood by noting that the microstates $\boldsymbol{\sigma}$, and their total energies, map directly onto those of a non-interacting spin chain. In particular, the  state occupancies $s_j$ may be mapped onto the magnetic quantum numbers $m_s^j$ of spins $j$. Non-interacting spins are known to be uncorrelated at thermal equilibrium~\cite{chandler1987introduction}. We provide a mathematical proof of this well-known result~\cite{richards1977theory,ambegaokar1971hopping}. 
\par
Each microstate of a transport network is described using
\begin{equation}  \label{microstates  - SI}
\boldsymbol{\sigma} = [\sigma_{1} \hspace{0.8mm} \sigma_{2} ... \hspace{0.8mm} \sigma_{i} \hspace{0.8mm} ... \hspace{0.8mm} \sigma_{j} \hspace{0.8mm} ...]
\end{equation}
where $\sigma_{i}$ and $\sigma_{j}$ are non-negative integers that indicate the number of particles on sites $i$ and $j$. Therefore, microstate $\boldsymbol{\sigma}$ thus tracks the occupation states of all sites. At thermodynamic equilibrium in the grand canonical ensemble, the probability to occupy microstate $\boldsymbol{\sigma}$ ($P_{\boldsymbol{\sigma}}$) and the partition function ($Z$) of a many-site transport chain are:
\begin{equation}    \label{equilibriumprobsigma}
P_{\boldsymbol{\sigma}}= \frac{\exp({-\beta G_{\boldsymbol{\sigma}})}}{Z} =\frac{\exp{(-\beta\sum_{k}G^{\boldsymbol{\sigma}}_{k})}}{Z} = \frac{\prod_{k}\exp{(-\beta G^{\boldsymbol{\sigma}}_{k})}}{Z}
\end{equation}
\begin{equation}    \label{partitionfunction}
Z = \sum_{\boldsymbol{\sigma}}\exp{(-\beta\sum_{k}G^{\boldsymbol{\sigma}}_{k})}
= \sum_{\boldsymbol{\sigma}}\prod_{k}\exp{(-\beta G^{\boldsymbol{\sigma}}_{k})}
= \prod_{k}\sum_{\boldsymbol{\sigma}}\exp{(-\beta G^{\boldsymbol{\sigma}}_{k})}
= \prod_{k}Z_{k}
\end{equation}
where
\begin{equation}    \label{Zk}
Z_{k}=\sum_{\boldsymbol{\sigma}}\exp{(-\beta G^{\boldsymbol{\sigma}}_{k})}
\end{equation}
is the partition function associated with site $k$. The quantity $G_{\boldsymbol{\sigma}}$ is the free energy associated with microstate $\boldsymbol{\sigma}$ and $G^{\boldsymbol{\sigma}}_{k}$ is the free energy associated with site $k$ in any microstate $\boldsymbol{\sigma}$.
Thus, at equilibrium, the probability of finding site $i$ with occupancy $q$ ($p_{iq}$) is
\begin{equation}
\begin{split}
p_{iq} &= \sum_{\boldsymbol{\sigma}}\delta_{\sigma_{i},q}P_{\boldsymbol{\sigma}} \\
& = \frac{1}{Z}\sum_{\boldsymbol{\sigma}}\delta_{\sigma _{i},q}\prod_{k}\exp{(-\beta G^{\boldsymbol{\sigma}}_{k})} \\
& = \frac{1}{\sum_{\boldsymbol{\sigma}}\prod_{k}\exp{(-\beta G^{\boldsymbol{\sigma}}_{k})}}\sum_{\boldsymbol{\sigma}}\delta_{\sigma _{i},q}\prod_{k}\exp{(-\beta G^{\boldsymbol{\sigma}}_{k})} \\
& = \frac{\sum_{\boldsymbol{\sigma}}\delta_{\sigma_{i},q}\exp{(-\beta G^{\boldsymbol{\sigma}}_{i})}\prod_{k \neq i}\exp{(-\beta G^{\boldsymbol{\sigma}}_{k})}}{\sum_{\boldsymbol{\sigma}}\exp{(-\beta G^{\boldsymbol{\sigma}}_{i})}\prod_{k \neq i}\exp{(-\beta G^{\boldsymbol{\sigma}}_{k})}} \\
& = \frac{\sum_{\boldsymbol{\sigma}}\delta_{\sigma_{i},q}\exp{(-\beta G^{\boldsymbol{\sigma}}_{i})}}{Z_{i}}
\end{split}
\end{equation}
The joint probability to find sites $i_1, ..., i_X$ with occupancies $q_1,..., q_X$ (i.e., the $X$-point correlation function $p_{i_1q_1,...,i_Xq_X}$) at equilibrium is
\begin{equation}
\begin{split}
p_{i_1q_1,...,i_Xq_X} & = \sum_{\boldsymbol{\sigma}}\delta_{\sigma_{i_1},q_1} \cdot \cdot \cdot \delta_{\sigma_{i_X},q_X}P_{\boldsymbol{\sigma}} \\
& = \sum_{\boldsymbol{\sigma}}\delta_{\sigma_{i_1},q_1} \cdot \cdot \cdot \delta_{\sigma_{i_X},q_X}\frac{\prod_{k}\exp{(-\beta G^{\boldsymbol{\sigma}}_{k})}}{Z}\\
& = \frac{1}{\prod_{k}Z_{k}}\sum_{\boldsymbol{\sigma}}\delta_{\sigma_{i_1},q_1} \cdot \cdot \cdot \delta_{\sigma_{i_X},q_X}\prod_{k}\exp{(-\beta G^{\boldsymbol{\sigma}}_{k})} \\
& = \frac{1}{\prod_{k}Z_{k}}\sum_{\boldsymbol{\sigma}}\delta_{\sigma_{i_1},q_1} \cdot \cdot \cdot \delta_{\sigma_{i_X},q_X}\exp{(-\beta G^{\boldsymbol{\sigma}}_{i_1})} \cdot \cdot \cdot \exp{(-\beta G^{\boldsymbol{\sigma}}_{i_X})}\prod_{k \neq \{i_1,...,i_X\}}Z_{k} \\
& = \left(\frac{1}{Z_{i_1}}\sum_{\boldsymbol{\sigma}}\delta_{\sigma_{i_1},q_1}\exp{(-\beta G^{\boldsymbol{\sigma}}_{i_1})} \right) \cdot \cdot \cdot \left(\frac{1}{Z_{i_X}} \sum_{\boldsymbol{\sigma}}\delta_{\sigma_{i_X},q_X}\exp{(-\beta G^{\boldsymbol{\sigma}}_{i_X})}\right) \\
& = p_{i_1q_1} \cdot \cdot \cdot p_{i_X q_X}\\
& = \prod_{x = 1}^{X} p_{i_x q_x}.
\end{split}
\end{equation}
Thus, the mean-field approximation (Eq. \ref{mean-field approximation}) is exact at equilibrium for a classical transport chain without site-site interactions.

\section{Hill's model for biological energy transduction}  \label{Hill's model of biological energy transduction}
Free energy transduction occurs when one or more exergonic reactions ($\Delta G < 0$ or ``downhill'')  are leveraged to drive one or more endergonic reactions ($\Delta G > 0$ or ``uphill'')~\cite{hill2013free}.
Reaction cycles that only perform exergonic processes without driving endergonic cycles are known as slippage events, dissipating otherwise available free energy as heat~\cite{hill2013free}. Slippage  reduces the turnover rates of the endergonic processes, so efficient free energy transduction occurs when slippage is suppressed.
\par
\begin{figure}[bt!]
    \centering
     \begin{subfigure}{0.28\textwidth}
         \includegraphics[width=\textwidth]{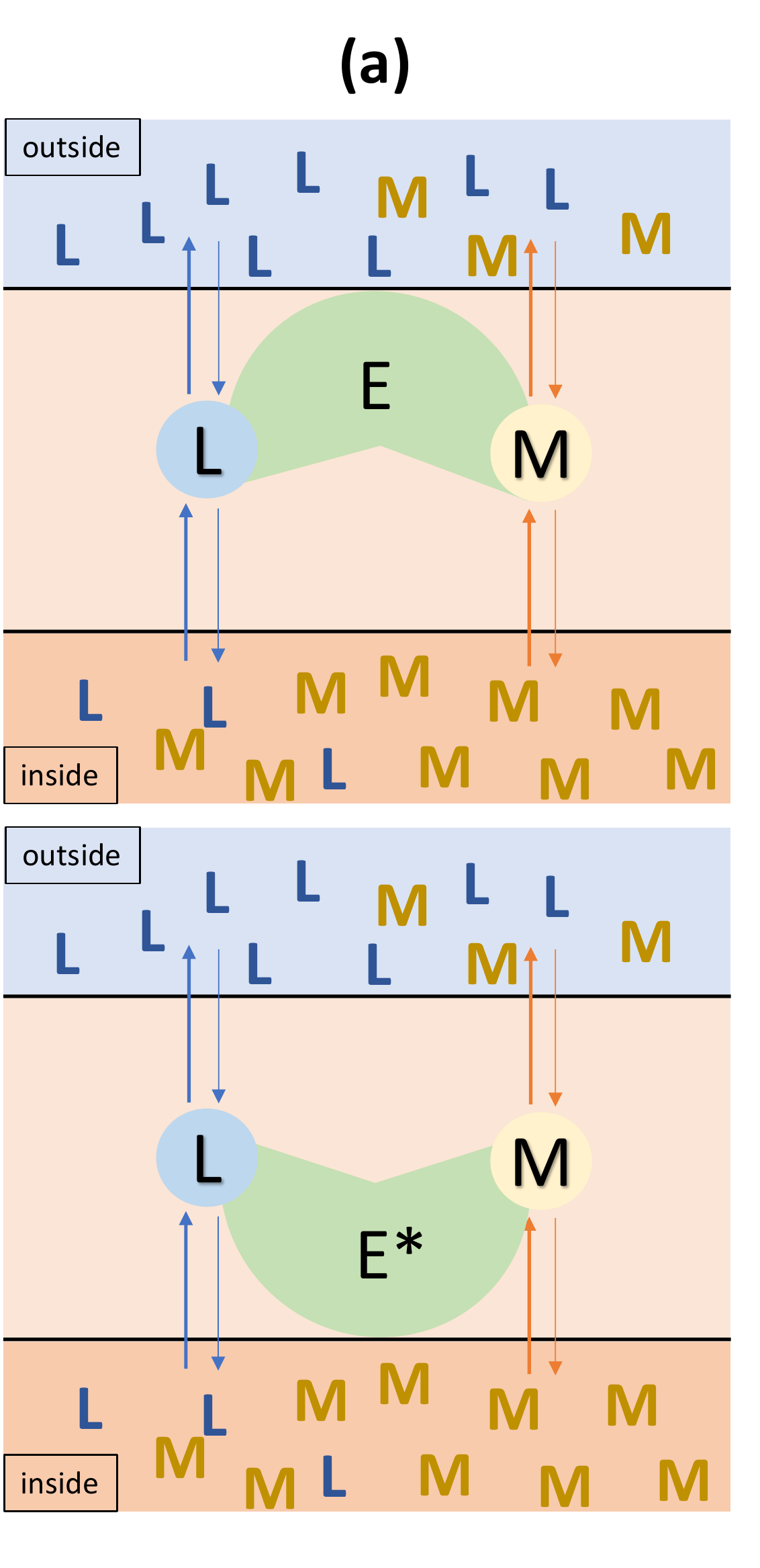}
         \label{fig:Hill's model in picture}
     \end{subfigure}
     \begin{subfigure}{0.45\textwidth}
         \includegraphics[width=\textwidth]{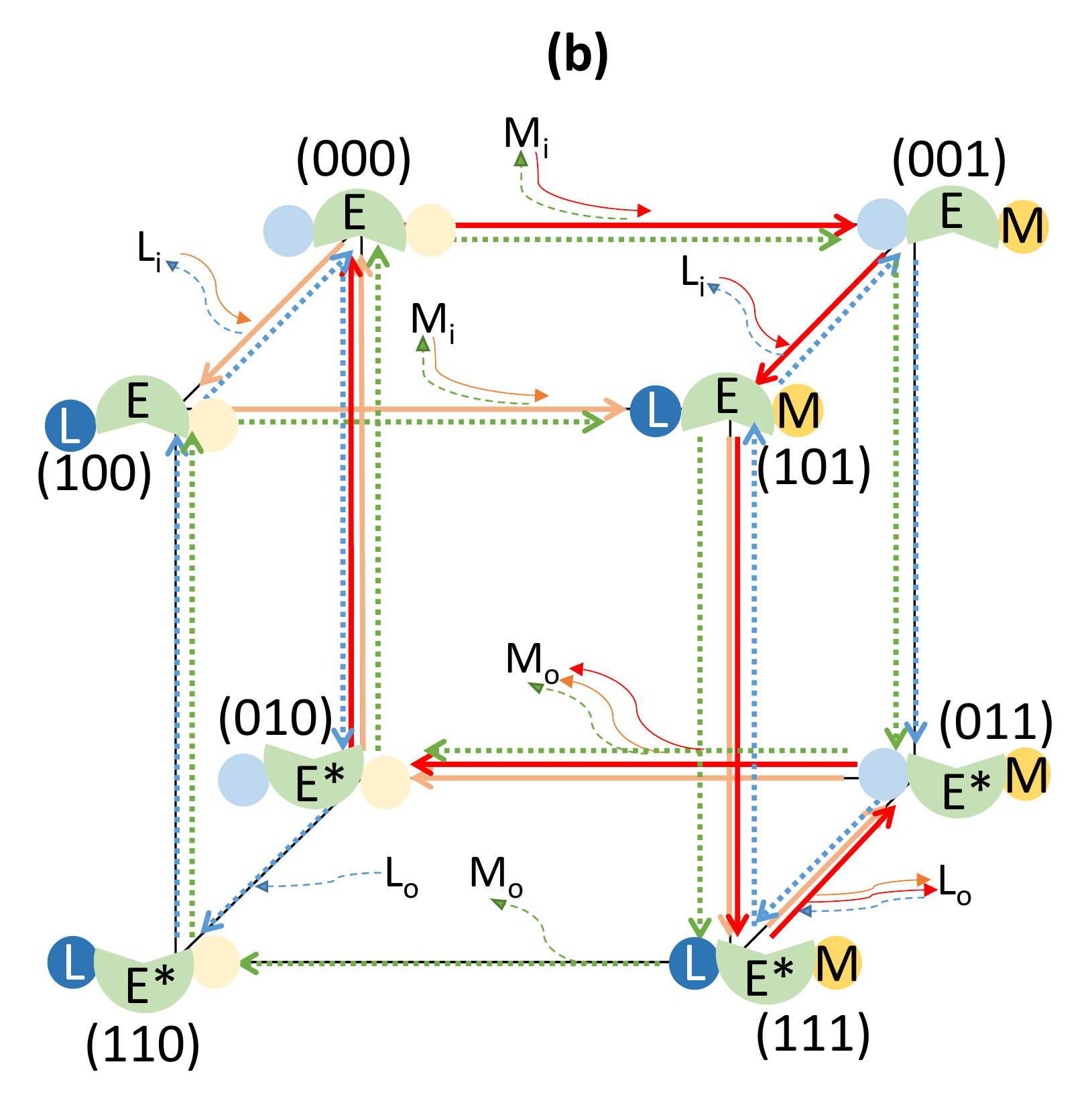}
         \label{fig:Hill's model in microstates}
     \end{subfigure}
\caption{Hill's model for free energy transduction. (a) Schematic representation of the model. Two substrates, L and M, have different concentrations (chemical potentials) on the two sides of a membrane.  The concentration of molecule M inside ($c_{M_{i}}$) is greater than outside ($c_{M_{o}}$), while molecule L has greater concentration outside~\cite{hill2013free}. An energy transducing enzyme, E, with  binding sites for L and M, is bound to the membrane. The enzyme pumps L from the inside to the outside against its concentration gradient (``L-cycle'': uphill transport of L), using the energy released by moving M from inside to outside down its concentration gradient (``M-cycle'': downhill transport of M)~\cite{hill2013free}. E may exist in two  conformations (E and E*)~\cite{hill2013free}. In conformation E (E*), the binding sites are accessible only for L and M that are inside (outside)~\cite{hill2013free}. The thicker arrows show pumping of L against a concentration gradient and transport of M down the concentration gradient. (b) Hill's model mapped onto particle occupancy microstates. The microstates are ($\sigma_{L}$, $\sigma_{E}$, $\sigma_{M}$). $\sigma_{L}$ and $\sigma_{M}$ indicate the site occupancies of the L and M binding sites. Different enzyme conformations are denoted by the fictitious E site, using $\sigma_{E}=0$ and E*:$\sigma_{E}=1$. The schematics adjacent to the microstates correspond to the binding sites and enzyme shown in panel (a) and illustrate occupied sites and conformations of E for each microstate. $\text{M}_{i}$ ($\text{L}_{i}$) and $\text{M}_{o}$ ($\text{L}_{o}$) represent reservoirs of molecule M (L) that are inside and outside, respectively. The red and orange arrows represent the free energy transduction cycle, which transports one M and one L species~\cite{hill2013free}. The green dashed arrow represents slippage, namely the transport of M from the inside to the outside without coupling to L transport, and the blue dashed arrow represents slippage associated with the flow of L from the outside to the inside~\cite{hill2013free}.}
\label{fig:Hill's model}
\end{figure}

A useful model to understand biological free energy transduction was described by T.L. Hill~\cite{hill2013free}, and Fig. \ref{fig:Hill's model} summarizes his model. This model maps onto an equivalent model for many-particle transport, by assigning the microstates of the enzyme and binding sites to a particle transport network with isomorphic kinetics. We denote the states of the sites in this mapping by ($\sigma_{L}$, $\sigma_{E}$, $\sigma_{M}$). The occupancy of the L and M binding sites,  $\sigma_{L}$ and $\sigma_{M}$ respectively, indicate whether L and M are bound to E (0 if no ligand is bound, 1 if a ligand is bound). An additional site is used to denote the state of the enzyme $E$ in its two available conformations (E:$\sigma_{E}=0$ and E*:$\sigma_{E}=1$). Importantly, particles cannot move between the sites in this mapping, as such processes are unphysical (that is, M is eliminated and converted into bound species L). However, the particle sites in the mapping do influence each other through interactions. The L and M concentration gradients across the membrane, and the energy of the E$\rightarrow$E* conformational change, are reflected in the energies of the reservoirs that are coupled to the three sites. There are eight microstates in this model. The redox-coupled proton pump described in Figure 3 of the main text is analogous to Hill's model (Fig \ref{fig:Hill's model}), albeit with additional transitions between the states (diagonals of the cube faces).

\section{Electron transfer through redox cofactor chains}   \label{Near reversible electron bifurcation}
Electron transport through proteins is often mediated by a chain of multiple redox cofactors between two or more substrates that interact with the cofactor chain. We model the cofactors as hopping sites and the substrates as electron reservoirs. In studies of  electron transport, the electron hopping rate from site $i$ to site $j$ may be approximated by~\cite{RN152,RN222,RN207,RN198,RN312}
\begin{equation} \label{Marcus ET}
k_{ij}=\frac{2\pi}{\hbar} \vert H_{ij}\vert ^{2} \frac{1}{\sqrt{4\pi \lambda_{ij} k_{B}T}}\exp(-\frac{((\epsilon_{j}-\epsilon_{i})+\lambda_{ij})^{2}}{4\lambda_{ij} k_{B}T})
\end{equation}
where $H_{ij}$ ({\AA}) and $\lambda_{ij}$ (eV) are the distance dependent electronic coupling and reorganization energy for electron transfer between sites $i$ and $j$.  $k_B T$ is the thermal energy. The electronic coupling is approximated as $\vert H_{ij}\vert^{2} \approx V^{2}\exp(-\beta R)$ ($\beta$ is the tunneling decay factor) and $V$ is the electronic coupling at contact between cofactors $i$ and $j$. The parameters were approximated as $\beta$ = 1 {\AA}$^{-1}$, $V$ = 0.1 eV, $\lambda$ = 1.0 eV, and $\sfrac{1}{k_{B}T}$ = 39.06 $\text{eV}^{-1}$, which are typical for protein electron transfer~\cite{RN312,RN182,RN132}. The distances between cofactors in the electron bifurcation model were set to  typical values for electron bifurcating enzymes.  We set $R$ = 10 {\AA} for nearest-neighbor cofactors, $R$ = 20 {\AA} for second nearest-neighbors, and $R$ = 30 {\AA} for third nearest-neighbors, as in a previous model~\cite{RN197}.
\par
In the electron bifurcation model (Fig. 5 of the main text), the high- and low-potential substrates $A_H$ and $A_L$, respectively, were modeled as one electron exchange  reservoirs, and the two-electron substrate $D$ was modeled as a two-electron exchange reservoir ($n=2$ in Eq \ref{microstate-microstate rate}). The rate constants for electron transfer from the system to the reservoirs were set to be fast ($k_{i,q \underset{n}\rightarrow \text{out}}= 10^7 s^{-1}$) so they are not not rate-limiting, and the rate constants for electron transfer into the system from these reservoirs were determined using Eq \ref{reservoir ET}.
The rate constant for $n$ electrons to transfer into a reservoir from site $i$ in redox state $q$, $k_{\text{in} \underset{n}\rightarrow i,q-n}$, is~\cite{RN197}:
\begin{equation}    \label{reservoir ET}
k_{\text{in} \underset{n}\rightarrow i,q-n} = k_{i,q \underset{n}\rightarrow \text{out}} \exp\left[\frac{n\times(\mu_{\text{res}}+|e^{-}|E_{i}^{\circ \text{mid}})}{k_{B}T}\right]
\end{equation}
where $\mu_{\text{res}}$ is the chemical potential of electrons in the reservoir, and $E_{i}^{\circ \text{mid}}$ is the average midpoint redox potential of cofactor $i$ (our model with rate constants given by Eq \ref{Marcus ET} and Eq \ref{reservoir ET} satisfy local detailed balance~\cite{bauer2014local}) over all $n$-electron transfers with the reservoir. 
\par
The exact kinetics within this theoretical framework is described by the master equation (Eq \ref{master eq} in main text), using rate constants defined by Eq \ref{Marcus ET} and \ref{reservoir ET}. The mean-field approximated kinetics is described by the approximate rate equation (Eq \ref{mean-field p_iq - no interaction}) with rate constants defined by equations \ref{Marcus ET} and \ref{reservoir ET}, particularly noting that no interactions between redox cofactors are included in this model. To simulate the linear transport chain of Fig. 1 of the main text, the parameters of Eq \ref{Marcus ET} above were taken from calculations~\cite{RN222} and experiment~\cite{RN215}, summarized in Table \ref{tab:psuedo_STC_parameters}. \\

\section{Summary of the kinetics model of Jiang {\textbf {\textit{et al.}}}} \label{Summary of the kinetics approach of Jiang $et$ $al.$}
\begin{figure}
    \centering
    \begin{subfigure}{0.48\textwidth}
    \includegraphics[width=\textwidth]{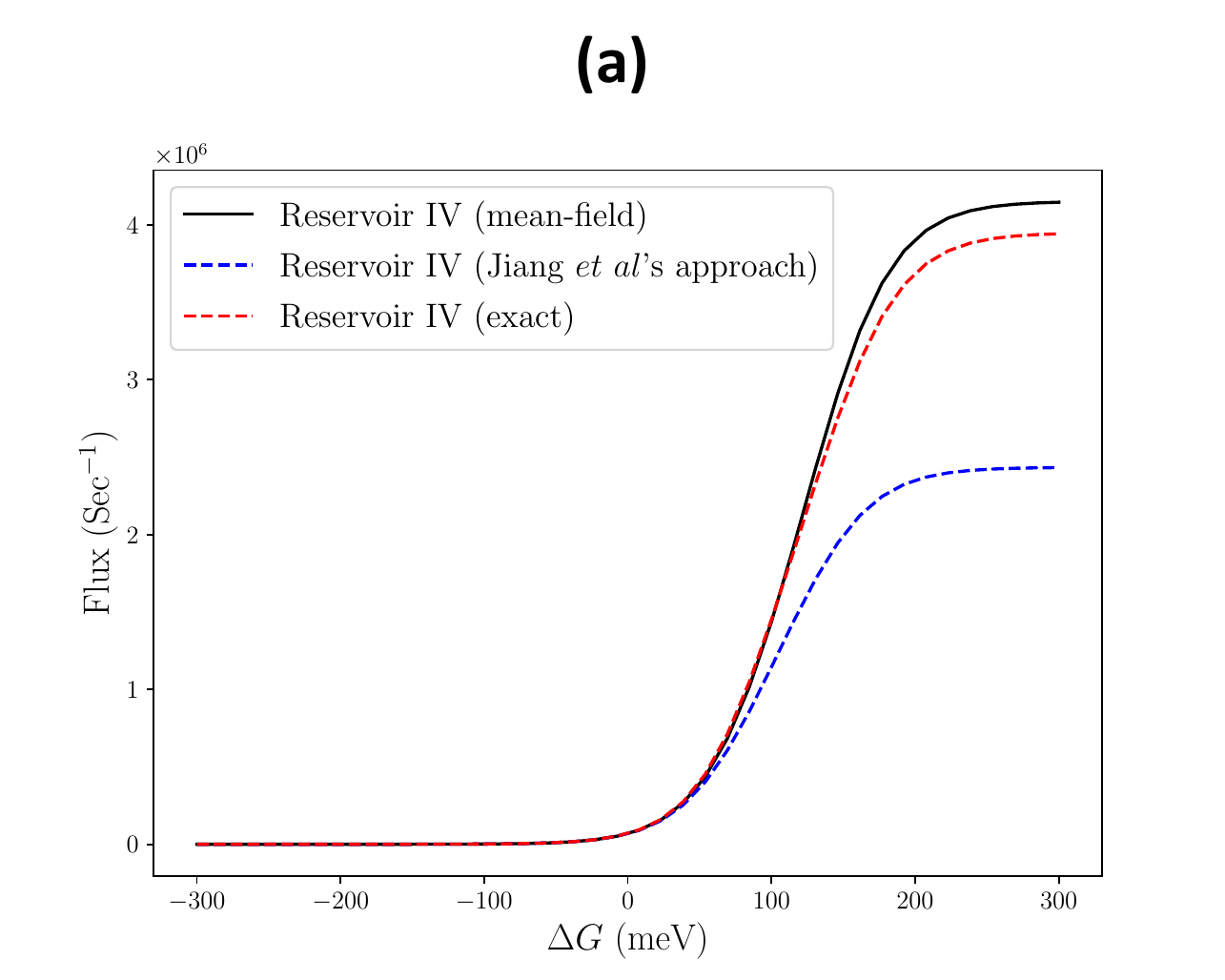}
    \end{subfigure}
    \begin{subfigure}{0.48\textwidth}
    \includegraphics[width=\textwidth]{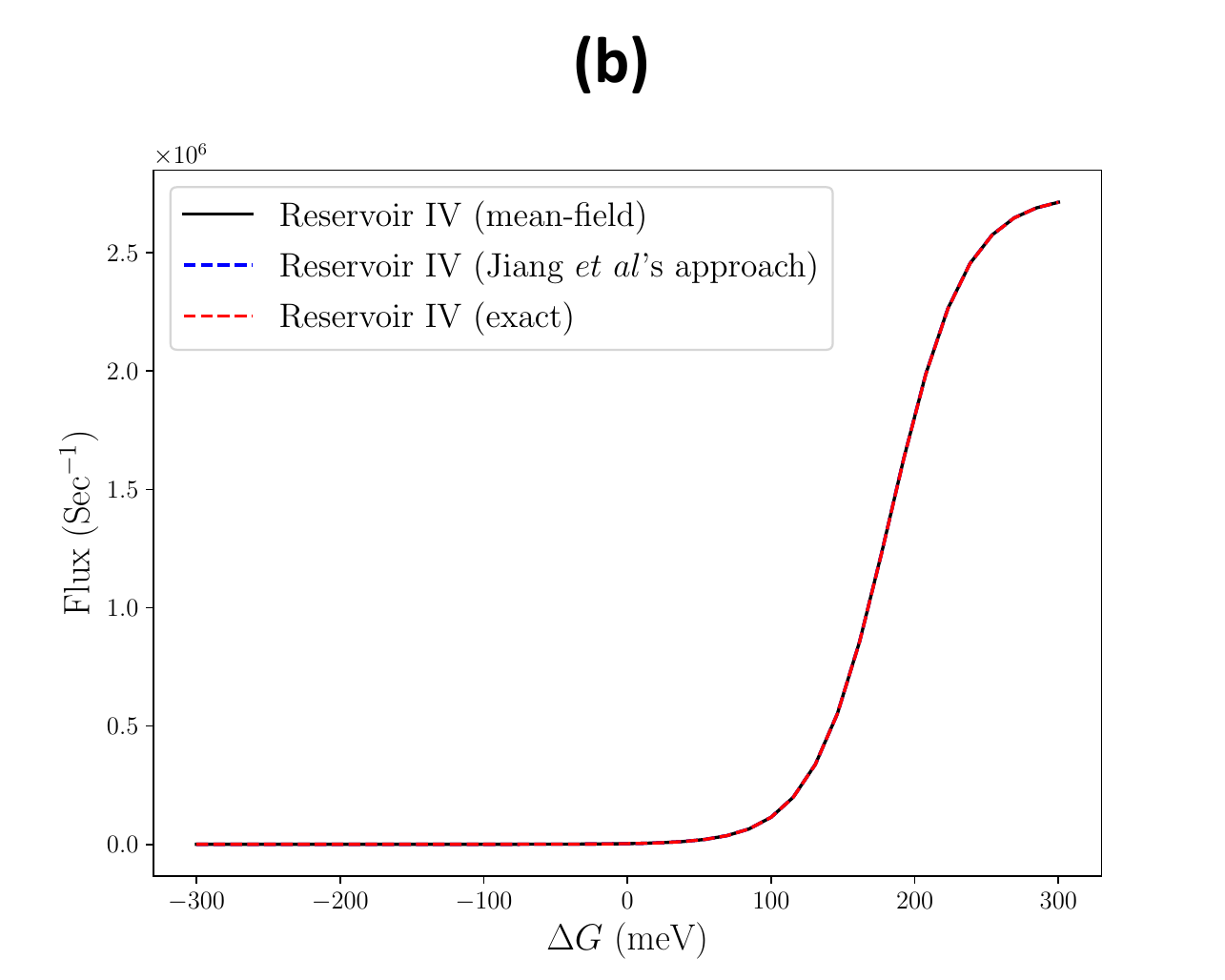}
    \end{subfigure}
\caption{Kinetics of a linear transport chain inspired by the STC protein. The plot shows the steady-state net electron fluxes into reservoir IV as a function of $\Delta G$ varied from -300 to 300 meV, computed using the mean-field approximation (black solid line), the approach of Jiang $et$ $al.$ (blue dashed line), and exact master equation (red dashed line). Panel (a) shows the case with site-site interactions and panel (b) shows the case without site-site interactions (i.e., $\Delta\epsilon_{ij}$ = 0, where $i$ and $j$ are labels of the redox cofactors shown in main text Fig. 1). The mean-field approximation, the approach of Jiang {\it et al.} and the exact master equation all agree (in both regimes) within an order of magnitude. In (b), the mean-field limit approximates the kinetics without site-site interactions (error $<1\%$) very well. Without site-site interactions, the mean-field approximation and the approach of Jiang {\it et al.} are equivalent.}
\label{fig:flux_psuedoSTC_MEK_trueMF_Blum}
\end{figure}

This section describes the kinetic approach used by Jiang {\it et al.}~\cite{RN222} and its differences from the mean-field approximation in the presence of interactions between particle binding sites. Jiang {\it et al.}  included interactions between sites by making the driving force for electron transfer depende on
 site occupancies. Jiang {\it et al.} made $\epsilon_{i}$ (and $\epsilon_{j}$) in Eq. \ref{Marcus ET} dependent on the site occupations using:
\begin{equation}    \label{Jiang et al site energy - STC}
\epsilon^{\circ}_{i} = \epsilon_{i}^{r} + \sum_{j\neq i}(\Delta\epsilon_{ij})(1-p_{j1})
\end{equation}
where $\epsilon_{i}^{r}$ is the redox potential of site $i$ when all sites (including site $i$) are occupied, $\Delta\epsilon_{ij}$ is the change in redox potential of site $i$ as site $j$ changes from occupied to unoccupied, and $p_{j1}$ is the probability of site $j$ being occupied. Table \ref{tab:psuedo_STC_parameters} shows $\Delta \epsilon_{ij}$ values taken from experiments~\cite{RN215} (and used for both the mean-field model, and the model of Jiang {\it et al.}~\cite{RN222}) for the STC protein. The approach of Jiang {\it et al.}  uses the equations
\begin{equation} \label{blumberger}
    \begin{split}
    \frac{dp_{i1}}{dt}&= -\sum_{j\neq i} [k^{\circ}_{i,1 \underset{1}\rightarrow j,0} p_{i1}p_{j0}- k^{\circ}_{j,1 \underset{1}\rightarrow i,0}p_{i0}p_{j1} ] \\
    & \indent +k^{\circ}_{i,1 \underset{1}\rightarrow \text{out}} p_{i1}-k^{\circ}_{\text{in} \underset{1}\rightarrow i,0}p_{i0}
    \end{split}
\end{equation}
where $ k^{\circ}_{i,q \underset{n}\rightarrow j,s}$ is the electron transfer rate  (Eq. \ref{Marcus ET}) based on the modified reduction potentials $\epsilon^{\circ}_{i}$ and $\epsilon^{\circ}_{j}$ values of Eq. \ref{Jiang et al site energy - STC} (the reservoir rates $k^{\circ}_{i,q \underset{1}\rightarrow \text{out}}$ and $k^{\circ}_{\text{in} \underset{1}\rightarrow i,q-1}$ were modeled as irreversible rates that are independent of the site occupancies, which we reproduce in the limit of strong driving forces, infinite $\Delta G$ in Fig. \ref{fig:flux_psuedoSTC_MEK_trueMF_Blum}). The kinetics of Eq \ref{blumberger} is clearly distinct from the mean-field kinetics (Eq. \ref{mean-field p_iq - interaction} and \ref{mean-field rate constants}), except when there are no interactions (Eq. \ref{mean-field p_iq - no interaction}). This case is relevant for our electron bifurcation model, where the mean-field approximation and the approach of Jiang {\it et al.} are identical. The results for the linear transport chain that are derived from the exact master equation and the mean-field approximation are shown in Fig. 2 of the main text, and the results using the approach of Jiang {\it et al.} are shown in Fig. \ref{fig:flux_psuedoSTC_MEK_trueMF_Blum}.
\par
\begin{figure}[bt!]
    \centering
     \begin{subfigure}{0.45\textwidth}
         \includegraphics[width=\textwidth]{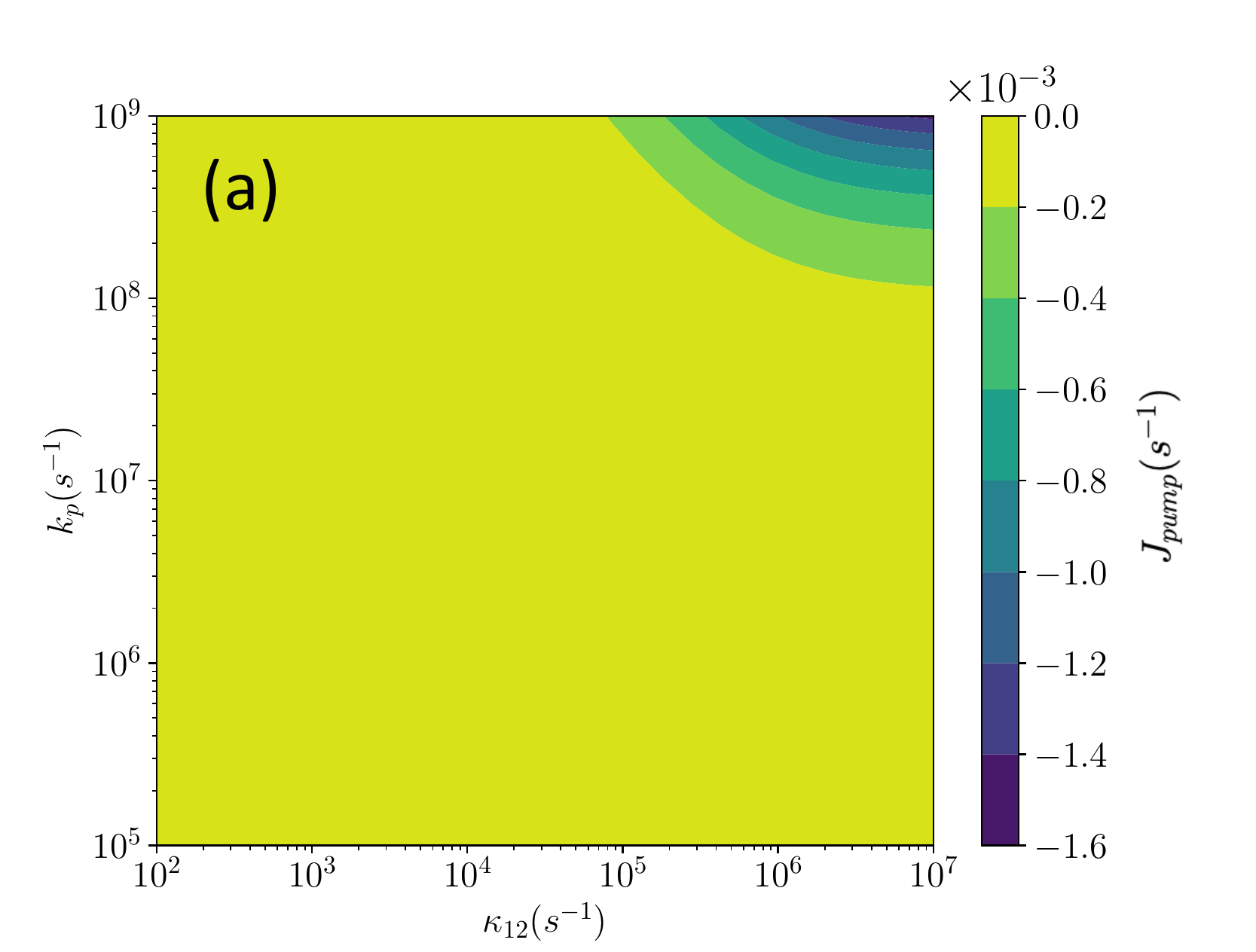}
     \end{subfigure}
     \begin{subfigure}{0.45\textwidth}
         \includegraphics[width=\textwidth]{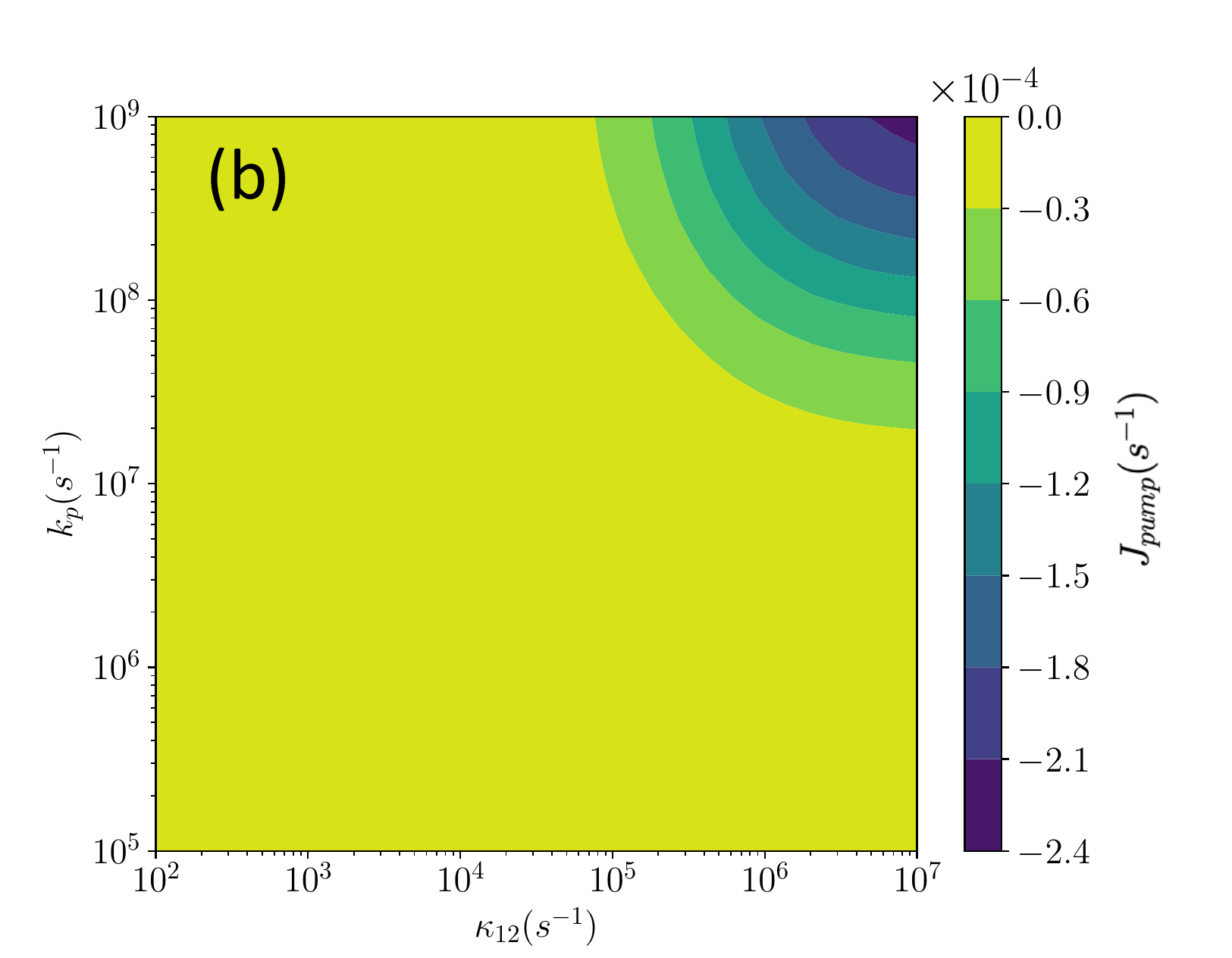}
     \end{subfigure}
     \begin{subfigure}{0.45\textwidth}
         \includegraphics[width=\textwidth]{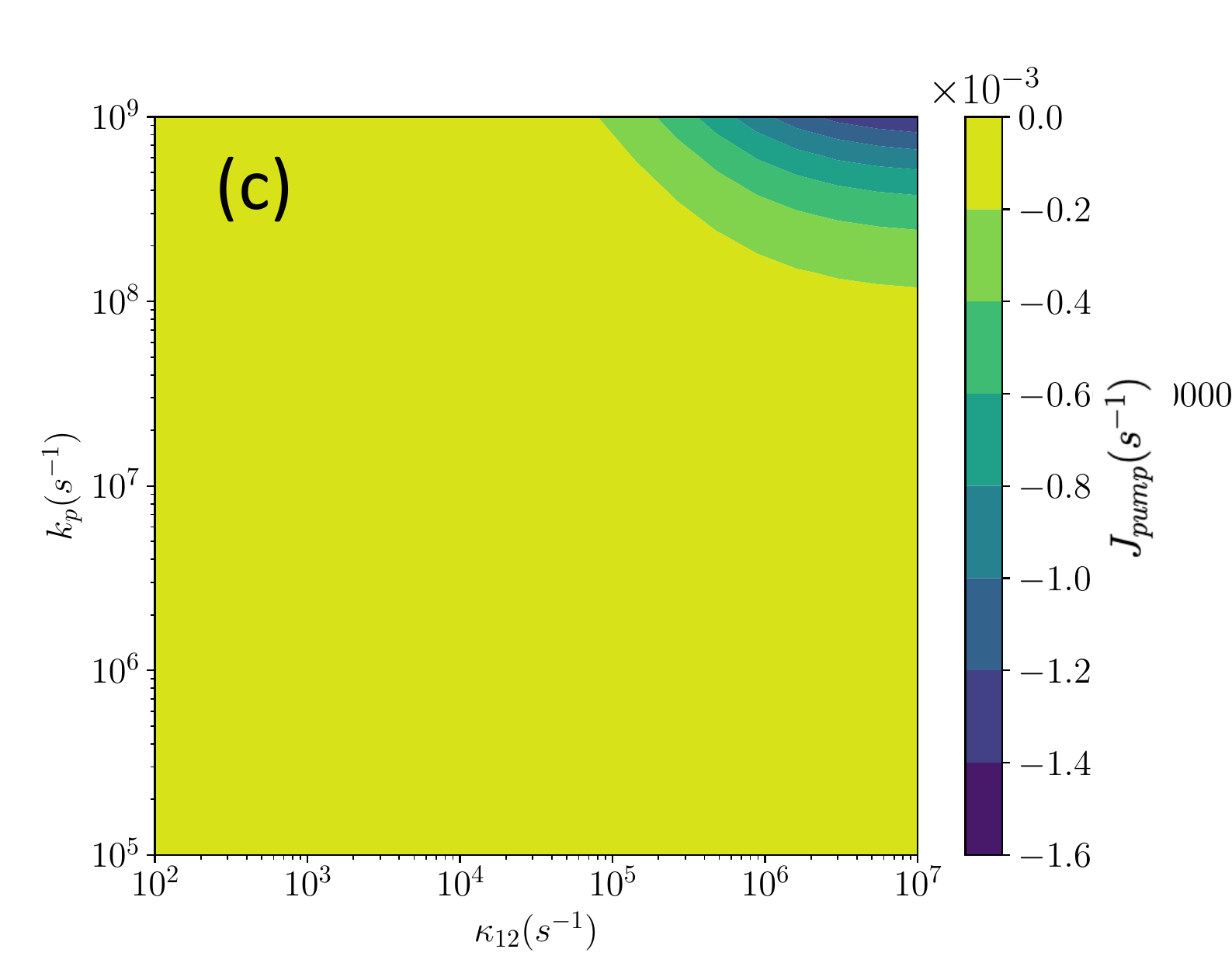}
     \end{subfigure}
     \begin{subfigure}{0.45\textwidth}
         \includegraphics[width=\textwidth]{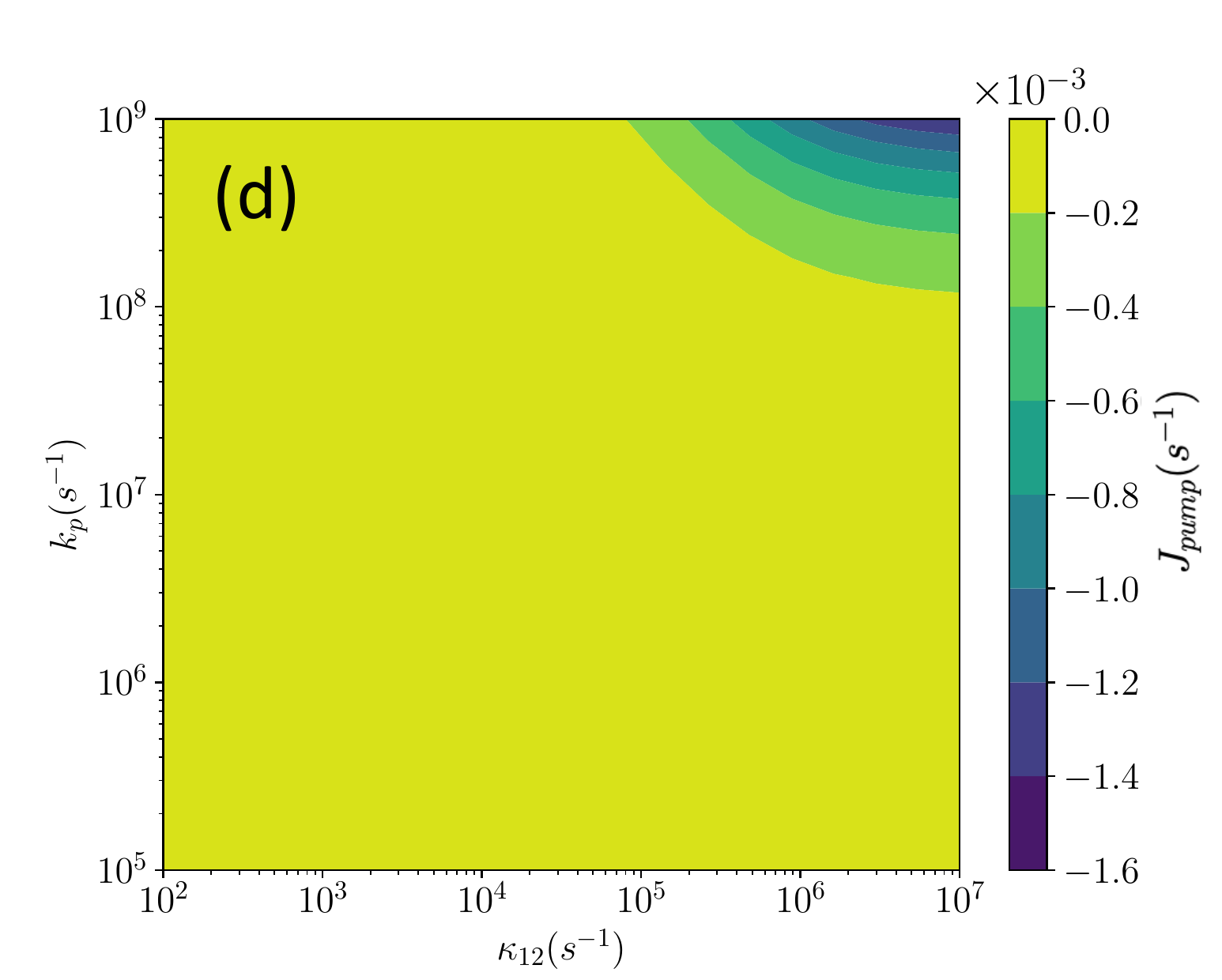}
     \end{subfigure}
\caption{Steady-state flux of protons pumped to the P-side ($J_{pump} (s^{-1})$), calculated using (a) Jiang $et$ $al.$'s approach including site-site interactions, (b) the exact master equation without site-site interactions, (c) the mean-field model without site-site interactions, and (d) Jiang $et$ $al.$'s approach without site-site interactions. Positive $J_{pump}$ values indicate energy transduction since $J_{pump} > 0$ indicates spontaneous proton flow uphill from the N- to the P-side of the membrane (against the the membrane potential or proton motive force $V_{m} =$ 100 mV). Negative $J_{pump}$ indicates that slippage reactions dominate, causing proton leakage from the P-side to the N-side. Comparing panel (a) here with panel (a) of Fig. 4 in the main text, we find that in regimes ($k_{p} \gtrsim 10^{8}$ sec$^{-1}$) where the exact master equation predicts no free energy transduction, the approach of Jiang $et$ $al.$ also predicts no free energy transduction. However, in regimes with $k_{p} \lesssim 10^{8}$ sec$^{-1}$, the exact master equation predicts free energy transduction, and the approach of Jiang $et$ $al.$  does not predict free energy transduction. Thus, the approach of Jiang $et$ $al.$  is not able to describe free energy transduction for this  redox-coupled proton pump model, similar to the failure of the mean-field kinetics to describe the energy transduction. We observe that site-site interactions are required for free energy transduction in this proton-pump model, consistent with conclusions from earlier  studies~\cite{kim2007kinetic,kim2012proton}. Panels (c) and (d) are exactly the same, as expected because the mean-field kinetics and the approach of Jiang {\it et al.}~\cite{RN222} are equivalent in the absence of site-site interactions.}
\label{fig:Pump flux - SI}
\end{figure}

We extended the approach of Jiang {\it et al.} to study the redox-coupled proton pump. In this model (Fig. 3 of the main text), the rate constant for a proton to transfer from site $i$ to site $j$ is~\cite{kim2007kinetic,kim2012proton}
\begin{equation}    \label{rate constant - proton pump}
k_{i,1 \underset{1}\rightarrow j,0} = \kappa_{ij}\exp{(-\frac{E_{j}-E_{i}}{2k_{B}T})}
\end{equation}
The reverse rate constant is
\begin{equation}    \label{reverse rate constant - proton pump}
k_{j,1 \underset{1}\rightarrow i,0} = \kappa_{ij}\exp{(-\frac{E_{i}-E_{j}}{2k_{B}T})},
\end{equation}
and the rate constant for a single proton or electron transfer from a reservoir to site $i$ is
\begin{equation}    \label{reservoir rate constant - protonpump}
k_{\text{in} \underset{1}\rightarrow i,0} = \kappa_{ii}\exp{(\frac{E_{i}}{2k_{B}T})}
\end{equation}
$\kappa_{ij}$ is the rate constant between sites $i$ and $j$ when all sites, other than site $i$, are unoccupied. $\kappa_{ii}$ is the rate constant for a single proton or electron transfer from a reservoir to site $i$ when all sites are unoccupied, and $E_{i}$($E_{j}$) is the energy of site $i$($j$). As in the approach of Jiang {\it et al.}, the electrostatic interactions between sites are included by making the driving force for a proton or electron transfer dependent on the site occupancies (i.e. making $E_i$ (and $E_j$) in Eq. \ref{rate constant - proton pump} and \ref{reservoir rate constant - protonpump} dependent on site occupancies). Studies by Kim {\it et al.}~\cite{kim2007kinetic,kim2012proton} modified $E_i$ to be
\begin{equation}    \label{proton pump energy}
E^{\circ}_{i} = E_{i}^{*} + \sum_{j\neq i}\epsilon_{ij}p_{j1} + \frac{q_{i}Z_{i}V_{m}}{L}
\end{equation}
where $E_{i}^{*}$ is the free energy to place a particle on site $i$ while all other sites are unoccupied and the membrane potential $V_m$ is zero, $\epsilon_{ij}$ is the electrostatic interaction between sites $i$ and $j$, $p_{j1}$ is the probability to find site $j$ occupied, $q_{i}$ is the charge of site $i$, $Z_{i}$ is the distance from the N-side to site $i$, and $L$ is the membrane thickness. Table \ref{tab:Proton pump parameters} shows the relevant parameters. The kinetics are computed using Eq. \ref{blumberger} with the modified site energies $E^{\circ}_{i}$ and $E^{\circ}_{j}$ of Eq. \ref{proton pump energy}. The results for the analysis of the redox-coupled proton pump based on the exact master equation and the mean-field approximation are shown in Fig. 4 of the main text, and the results using the approach of Jiang {\it et al.} are shown in Fig. \ref{fig:Pump flux - SI}.

\section{Effective activation energies for short-circuiting in the mean-field approximation} \label{Effective activation energies for short-circuiting in the mean-field kinetics}
This section explores the short-circuit flux in the electron bifurcation network of Fig. 5 when $\Delta G_{bifurc} = 0$ (i.e., ``idle short-circuiting"~\cite{RN197}). We find that mean-field kinetic analysis fails to produce two separate short-circuiting regimes as a function of temperature (see Fig. \ref{fig:T-dependent_SCfluxes} below). These regimes are determined by two separate short-circuit channels (illustrated in Fig. \ref{fig:EBshortcircuits}a and \ref{fig:EBshortcircuits}b) that have different activation energies and hence dominate the kinetics at different temperatures. The electron transfer steps that occur in three possible short-circuit channels are shown in Fig. \ref{fig:EBshortcircuits}. In contrast, the mean-field approximation to the kinetics predicts one regime with an effective activation energy that is the average of the activation energies of the regimes from the exact master equation (purple curve in Fig. \ref{fig:T-dependent_SCfluxes}).
\par
Earlier studies found that the short-circuiting flux predicted by the exact master equation agrees well with the estimated flux through the dominant short-circuit channel at a given temperature, given by ~\cite{RN197, RN371}
\begin{equation}   \label{quasi-eq flux}
\text{flux} \sim k_{rl}\exp(-\frac{\Delta G^{\ddagger}}{k_{B}T}),
\end{equation}
where $\Delta G^{\ddagger} = \Delta G^{e^{-}} + \Delta G^{h^{+}}$ is the free energy that moves an electron and hole to the cofactors required for short-circuiting to occur, and $k_{rl}$ is the rate-limiting electron transfer rate constant of the short-circuiting channel(Fig. \ref{fig:EBshortcircuits}). Thus, short-circuits are thermally-activated processes~\cite{RN197}. We define $J1$ and $J2$ as the short-circuiting electron fluxes through the orange and green channels indicated in Fig. 5 of the main text, respectively. The factor of $2$ in the equation for $J1$ arises from the fact that there are two-short circuits $B^- \rightarrow H_1$ and $L_1^- \rightarrow B^-$ that both contribute equally to $J1$. Using Eq. \ref{quasi-eq flux}, the flux $J1$  is approximated as
\begin{equation}   \label{J1}
\begin{split}
J1 &= \text{flux}_{B^{-}\rightarrow H1} + \text{flux}_{L1^{-}\rightarrow B^{-}} \\
&\approx k_{L1,1 \rightarrow B,0} \exp(-\frac{\Delta G_{B^{-} \rightarrow H1}^{e^{-}}}{k_{B}T}) \exp(-\frac{\Delta G_{B^{-} \rightarrow H1}^{h^{+}}}{k_{B}T}) \\
& \hspace{15 pt} + k_{B,2 \rightarrow H1,0} \exp(-\frac{\Delta G_{L1^{-} \rightarrow B^{-}}^{e^{-}}}{k_{B}T}) \exp(-\frac{\Delta G_{L1^{-} \rightarrow B^{-}}^{h^{+}}}{k_{B}T}) \\
&= k_{L1,1 \rightarrow B,0}\exp(-\frac{\Delta G_{B^{-}\rightarrow H1}^{\ddagger}}{k_{B}T}) + k_{B,2 \rightarrow H1,0}\exp(-\frac{\Delta G_{L1^{-} \rightarrow B^{-}}^{\ddagger}}{k_{B}T}) \\
&= 2k^{J1}_{rl}\exp(-\frac{\Delta G^{\ddagger}_{J1}}{k_{B}T})
\end{split}
\end{equation}
where $\Delta G^{\ddagger}_{SC}=\Delta G^{e^-}_{SC}+\Delta G^{h^+}_{SC}$ $(SC=B^- \rightarrow H1, L1^- \rightarrow B^-)$ is the energy required to move a hole from $A_H$ and an electron from $A_L$ to realize the precursor state from which the rate-limiting process with rate constant $k_{rl}$ initiates the short-circuiting  (see Fig \ref{fig:EBshortcircuits}). $\Delta G^{\ddagger}_{SC}$ is a product of Boltzmann weights for particle occupancies in each branch. There are two short-circuits that contribute equally to $J1$ (see Fig. 5 of the main text), and their contributions are identical for the parameters of the electron bifurcation model used (see Fig. 5 of the main text). Thus, $k^{J1}_{rl} = k_{L1,1 \rightarrow B,0} = k_{B,2 \rightarrow H1,0}$ and $\Delta G^{\ddagger}_{J1} = \Delta G_{B^{-}\rightarrow H1}^{\ddagger} = \Delta G_{L1^{-} \rightarrow B^{-}}^{\ddagger}$
Similarly, the $J2$ flux is approximated as
\begin{equation}    \label{J2}
\begin{split}
J2 &= \text{flux}_{L1^{-}\rightarrow H1} \\
&\approx k_{L1,1 \rightarrow H1,0} \exp(-\frac{\Delta G_{L1^{-} \rightarrow H1}^{e^{-}}}{k_{B}T}) \exp(-\frac{\Delta G_{L1^{-} \rightarrow H1}^{h^{+}}}{k_{B}T}) \\
&= k_{L1,1 \rightarrow H1,0}\exp(-\frac{\Delta G_{L1^{-} \rightarrow H1}^{\ddagger}}{k_{B}T}) \\
&= k_{L1,1 \rightarrow H1,0}\exp(-\frac{\Delta G^{\ddagger}_{J2}}{k_{B}T})
\end{split}
\end{equation}
These expressions for the $J1$ and $J2$ fluxes are plotted as orange and green dashed lines, respectively, in Fig. \ref{fig:T-dependent_SCfluxes}. The activation energies $\Delta G^{\ddagger}_{J1}$ and $\Delta G^{\ddagger}_{J2}$ are estimated by analyzing the energy landscape for the electrons in the system, again as illustrated in Fig. \ref{fig:EBshortcircuits}. Since the activation energy for $J2$ is lower than that for $J1$, the $J2$ flux dominates at low temperature (large $1/k_BT$).
\par
We were intrigued to find that the mean-field approximated short-circuiting flux has an effective activation energy (slope of the purple line in Fig. \ref{fig:T-dependent_SCfluxes}) that is approximately the mean of the slopes of $J1$ and $J2$ on a log scale (supplementary Fig. \ref{fig:T-dependent_SCfluxes}). Defining $J3$ as the geometric mean of $J1$ and $J2$, 
\begin{equation}    \label{J3}
J3 = (J1 J2)^{1/2} = \sqrt{2k_{L1,1 \rightarrow D,0}k_{L1,1 \rightarrow H1,0}}\exp(-\frac{\Delta G^{\ddagger}_{J1}+\Delta G^{\ddagger}_{J2}}{2k_{B}T})
\end{equation}
we find that $J3$ has the activation energy $\frac{\Delta G^{\ddagger}_{J1}+\Delta G^{\ddagger}_{J2}}{2k_{B}T}$ that is the average of the activation energies of $J1$ and $J2$ ($J3$ is indicated with the pink line in Fig. \ref{fig:T-dependent_SCfluxes})
Since the purple curve (mean-field flux) is nearly parallel to $J3$ (supplementary Fig. \ref{fig:T-dependent_SCfluxes}), the effective activation energy of the flux calculated with the mean-field approximation is approximately the geometric mean of $J1$ and $J2$, but the mean-field flux is not equal to $J3$, having a different $Y$-intercept in Fig. \ref{fig:T-dependent_SCfluxes}. 
\par
The underlying cause of the mean-field approximation to fail to capture the different short-circuiting regimes as a function of temperature remains unclear. As mentioned in the main text, mean-field failure may involve extending the near-equilibrium statistics within each branch unnaturally across the two branches. Understanding why the effective activation energy of the mean-field flux is approximately the average of the activation energies for  the orange and green short-circuiting channels in Fig. \ref{fig:T-dependent_SCfluxes} and is of interest for future analysis.

\clearpage
\section{Supplementary tables and figures}
\begin{table}[hbt!]
\centering
\caption{\label{tab:psuedo_STC_parameters}  \textbf{Parameters for the STC linear transport chain taken from previous studies} ~\cite{RN222,fonseca2009tetraheme}. $\lambda_{ij}$ is the reorganization energy (meV), and $H_{ij}$ is the electronic coupling (meV) between cofactors $i$ and $j$. $\epsilon_{i}^{r}$ is the redox potential (meV) of cofactor $i$ when all other sites are occupied, and $\Delta\epsilon_{ij}$ is the change in redox potential (meV) of cofactor $i(j)$ when $j(i)$ is unoccupied.}
\begin{subtable}{0.3\textwidth}
\centering
\begin{tabular}[t]{cc}
\hline
$\lambda_{12}$ &  1080 meV \\
$\lambda_{23}$ &  760 meV \\
$\lambda_{34}$ &  880 meV \\
\hline
$H_{12}$ & 2.17 meV \\
$H_{23}$ & 3.08 meV \\
$H_{34}$ & 2.08 meV \\
\hline
\end{tabular}
\end{subtable}
\begin{subtable}{0.3\textwidth}
\begin{tabular}[t]{cc}
\hline
$\epsilon_{1}^{r}$ & -243 meV \\
$\epsilon_{2}^{r}$ & -222 meV \\
$\epsilon_{3}^{r}$ & -189 meV \\
$\epsilon_{4}^{r}$ & -171 meV \\
\hline
$\Delta\epsilon_{12}$ & 28 meV \\
$\Delta\epsilon_{13}$ & 21 meV \\
$\Delta\epsilon_{14}$ & 11 meV \\
$\Delta\epsilon_{23}$ & 72 meV \\
$\Delta\epsilon_{24}$ & 11 meV \\
$\Delta\epsilon_{34}$ & 29 meV \\
\hline
\end{tabular}
\end{subtable}
\end{table}

\clearpage
\begin{table}
\centering
\caption{\label{tab:Proton pump parameters} \textbf{Parameters for the redox-coupled proton pump model used in earlier studies} ~\cite{kim2012proton,kim2007kinetic,kim2009kinetic}. $E_{i}^{*}$ is the free energy associated with site $i$ when site $i$ is occupied while all other sites are unoccupied and the membrane potential $V_m$ is zero. $\epsilon_{ij}$ is the electrostatic interaction between sites $i$ and $j$. $\kappa_{ii}$ is the rate constant for a single proton or electron transfer from a reservoir to site $i$ when all other sites are unoccupied. $V_{m}$ is the membrane potential. $q_{i}$ is the elementary charge of site $i$. $Z_{i}$ is the distance from the N-side to site $i$. $L$ is the membrane thickness.}
\begin{subtable}{0.3\textwidth}
\centering
\begin{tabular}[t]{cc} 
\hline
$E_{1}^{*}$ &  $98.4$ meV \\
$E_{2}^{*}$ &  $229$ meV \\
$E_{3}^{*}$ &  $386$ meV \\
\hline
$\epsilon_{12}$ &  $319$ meV \\
$\epsilon_{13}$ &  $-386$ meV \\
$\epsilon_{23}$ &  $-578$ meV \\
\hline
$\kappa_{11}$ &  $7.58 \times 10^{6}$ sec$^{-1}$ \\
$\kappa_{22}$ &  $2.84 \times 10^{4}$ sec$^{-1}$\\
$\kappa_{33}$ &  $2.77 \times 10^{4}$ sec$^{-1}$\\
\hline
\end{tabular}
\end{subtable}
\begin{subtable}{0.3\textwidth}
\centering
\begin{tabular}[t]{cc}
\hline
$V_{m}$ & 100 \\
\hline
$q_{1}$ & 1 $|e^-|$ \\
$q_{2}$ & 1 $|e^-|$ \\
$q_{3}$ & -1 $|e^-|$ \\
\hline
$Z_{1}$ &  15 {\AA} \\
$Z_{2}$ &  25 {\AA} \\
$Z_{3}$ &  20 {\AA} \\
\hline
$L$ & 30 {\AA} \\
\hline
\end{tabular}
\end{subtable}
\end{table}

\clearpage

\clearpage

\clearpage

\clearpage
\begin{figure}
    \centering
     \begin{subfigure}{0.7\textwidth}
         \includegraphics[width=\textwidth]{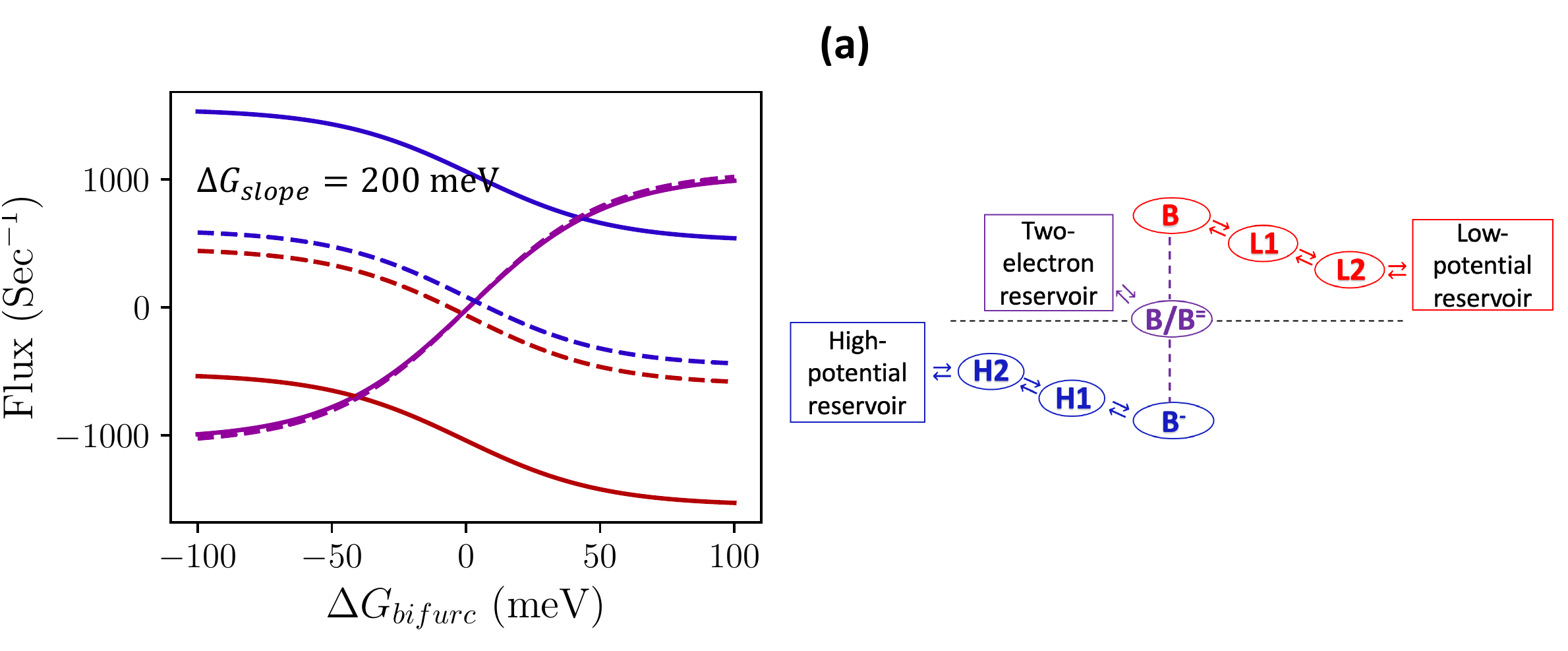}
         \label{fig:EBscheme_200meV}
     \end{subfigure}
     \begin{subfigure}{0.7\textwidth}
         \includegraphics[width=\textwidth]{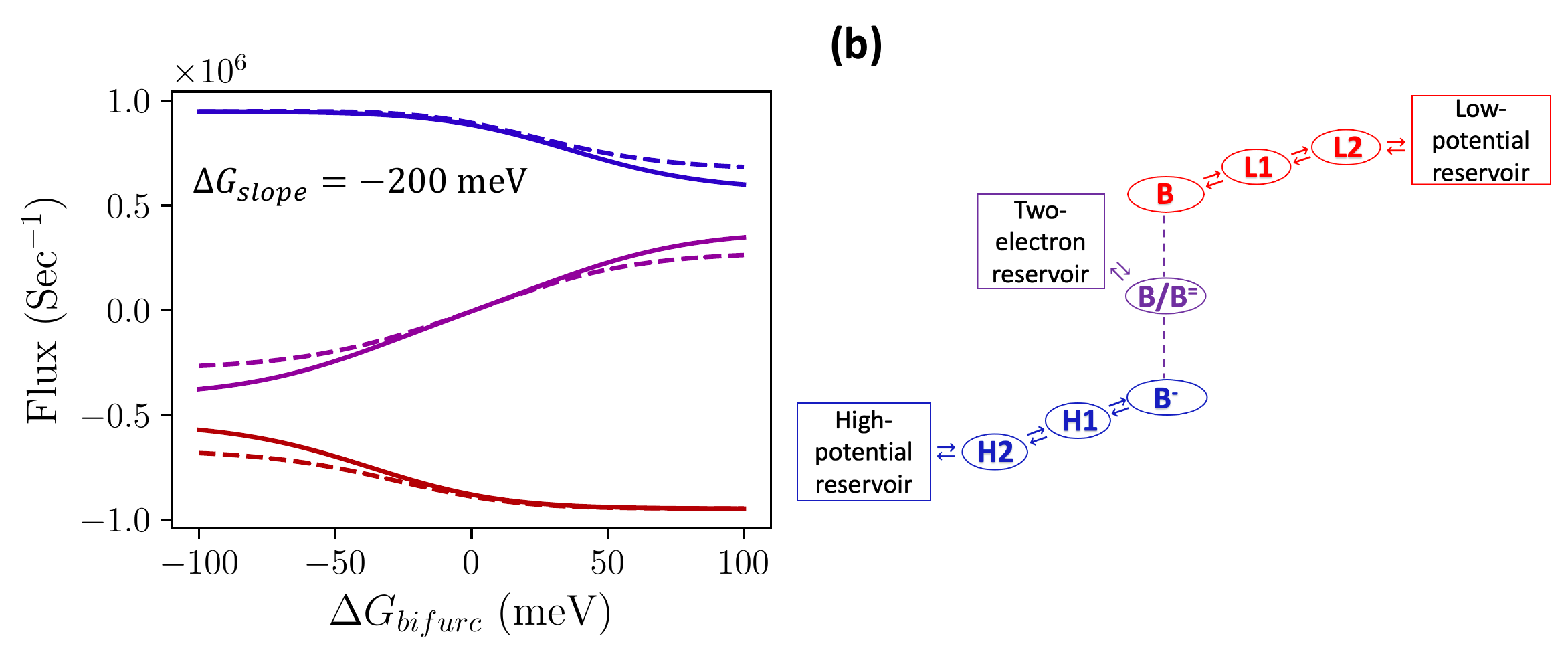}
         \label{fig:EBscheme_-200meV}
     \end{subfigure}
     \begin{subfigure}{0.7\textwidth}
         \includegraphics[width=\textwidth]{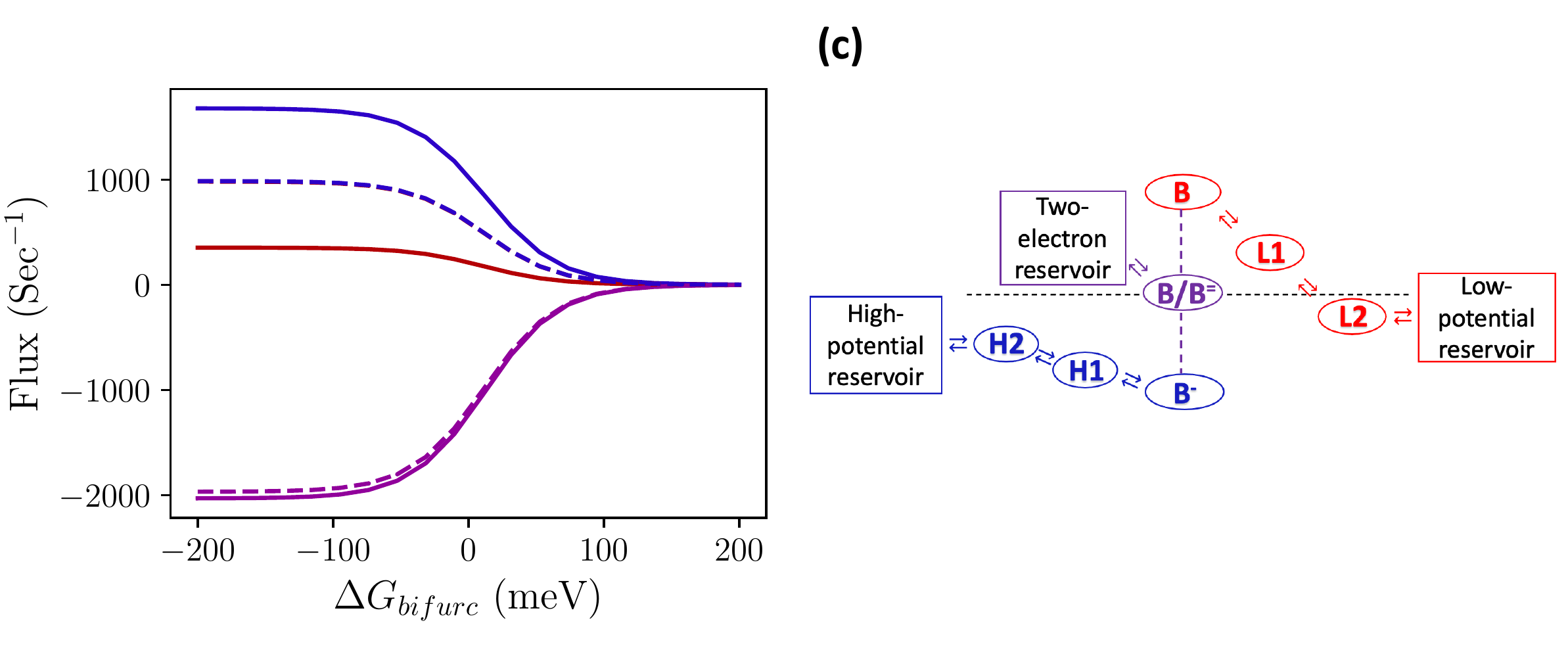}
         \label{fig:EBflux_L5_H2}
     \end{subfigure}     
     \begin{subfigure}{0.3\textwidth}
         \includegraphics[width=\textwidth]{Figures/EBflux_legend.pdf}
     \end{subfigure}
\caption{Kinetic simulations of the electron bifurcation network in regimes not explored in Fig. 6 of the main text. The solid curves indicate the electron fluxes computed with the mean-field approximation, and the dashed curves are the fluxes computed using the exact master equation. The sum of the values represented by the blue and red curves matches the values indicated by the purple curve, consistent with the  conservation of electrons at steady-state. The efficiency of electron bifurcation is indicated by the gap between the blue and red curves, which arises from short-circuiting. Panel (c) shows the case when the high and low-potential reservoirs have different chemical potentials ($\Delta G_{slope}$=250 meV for the low-potential branch and $\Delta G_{slope}$=100 meV for the high-potential branch), but have lower chemical potentials than the two-electron reservoir ($\Delta G_{HP} < 0, \Delta G_{LP} < 0$). Therefore, case (c) does not produce energy transduction because energy transduction requires an uphill and downhill process, or $\Delta G_{HP} < 0$ and $\Delta G_{LP} > 0$. A significant discrepancy remains between the blue curves, indicating that the mean-field approximation of the kinetics may fail in systems that do not transduce energy.}
\label{fig:EBfluxresults - SI}
\end{figure}

\clearpage
\begin{figure}
\centering
\includegraphics[width=0.6\textwidth]{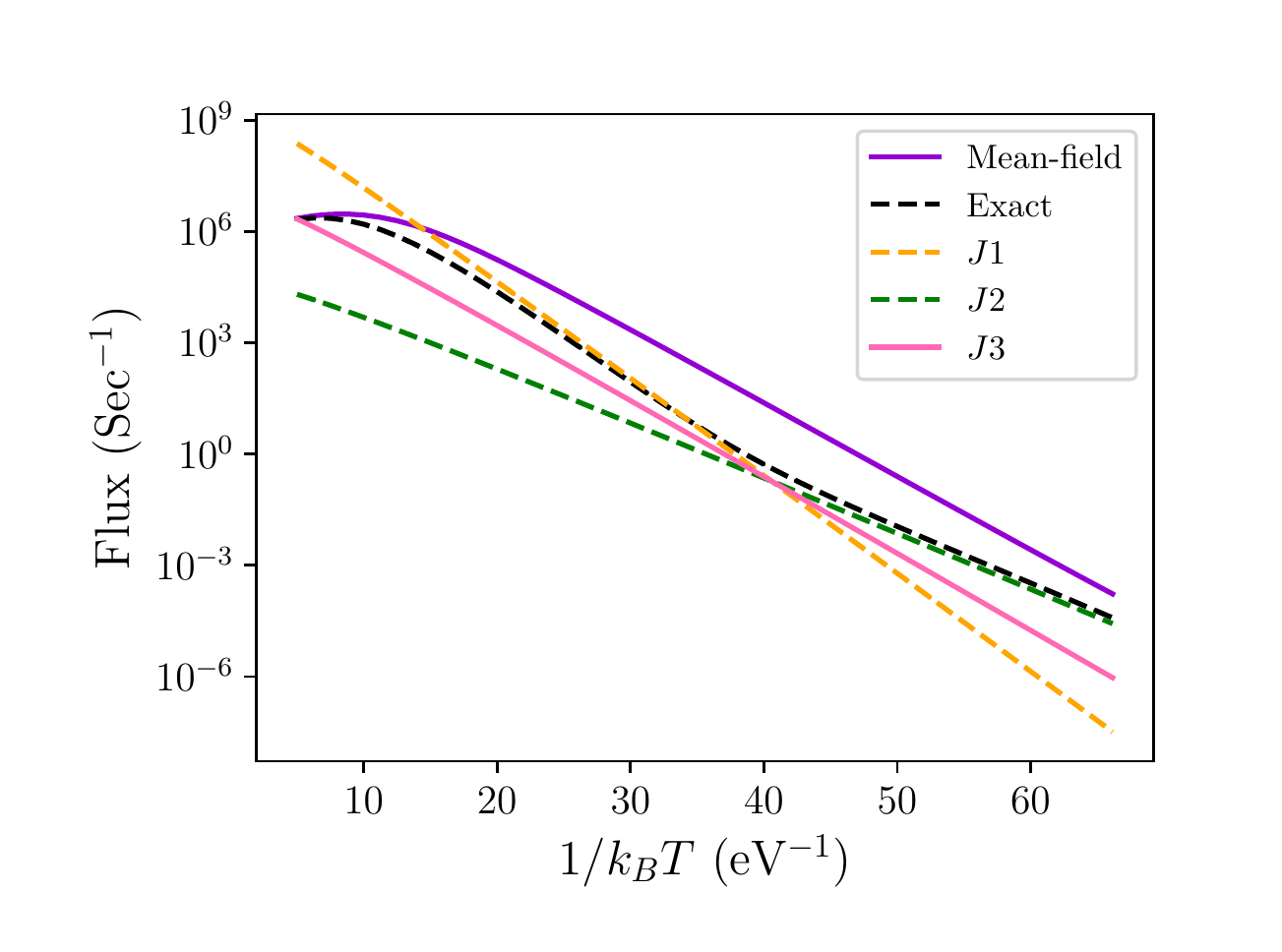}
\caption{Temperature-dependent short-circuiting fluxes at zero driving force $\Delta G_{bifurc} = 0$ meV plotted on a log scale. The exact master equation calculated flux (black), mean-field approximated flux (purple), $J1$ calculated using Eq. \ref{J1} (orange), $J2$ calculated using Eq. \ref{J2} (green), and $J3$ calculated using Eq. \ref{J3} (pink). These data were computed for the network with ($\Delta G_{slope}, \Delta G_{HL}$) = (300 meV, 200 meV), as shown in Fig. 5 of the main text. The slope of the purple line that represents the mean-field calculated flux is approximately the same as the slope of the line representing $J3$. Since the plot is on a log scale, this indicates that the mean-field calculated flux has an effective activation energy that is the average of the activation energies of the short circuiting channels $J1$ and $J2$, as described in Section \ref{Effective activation energies for short-circuiting in the mean-field kinetics} of the SI appendix.}
\label{fig:T-dependent_SCfluxes}
\end{figure}

\clearpage
\begin{figure}[h]
 \centering
 \begin{subfigure}{\textwidth}
     \centering
     \includegraphics[width=0.6\textwidth]{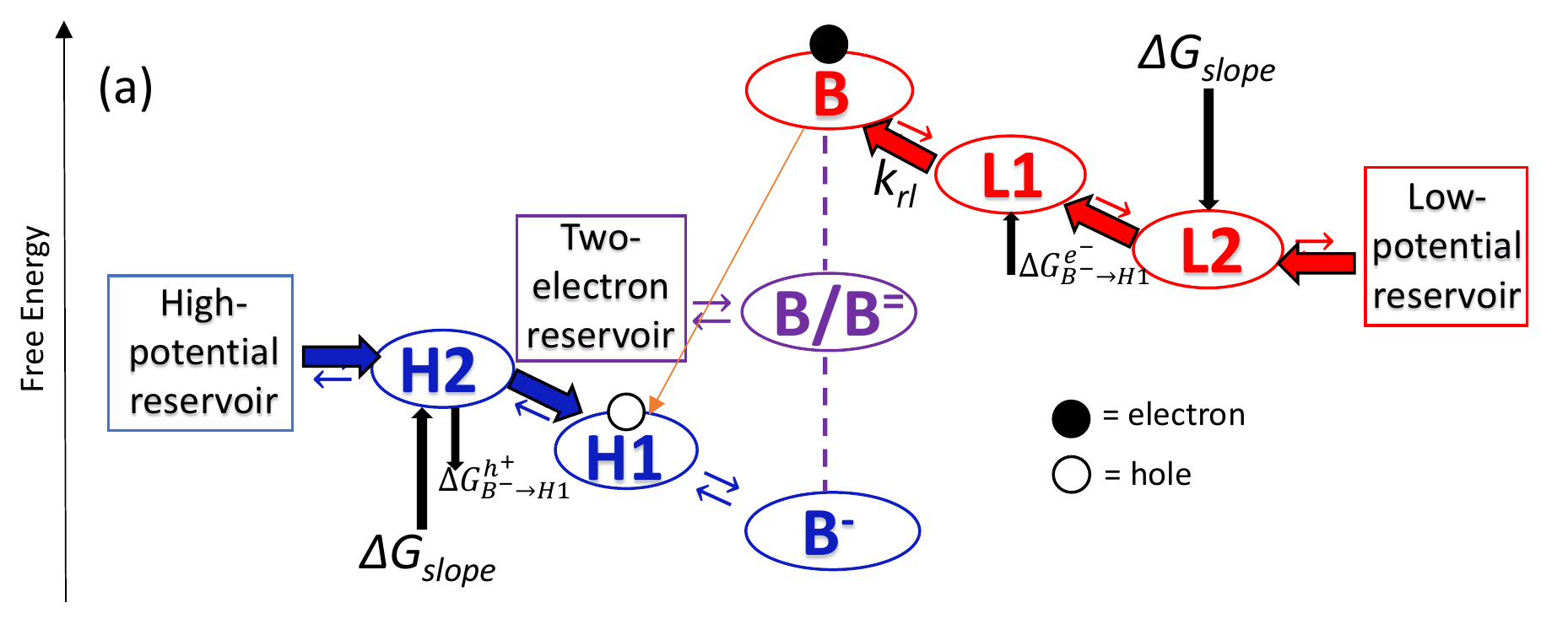}
     \label{fig:B1_H10}
 \end{subfigure}
 \begin{subfigure}{\textwidth}
     \centering
     \includegraphics[width=0.6\textwidth]{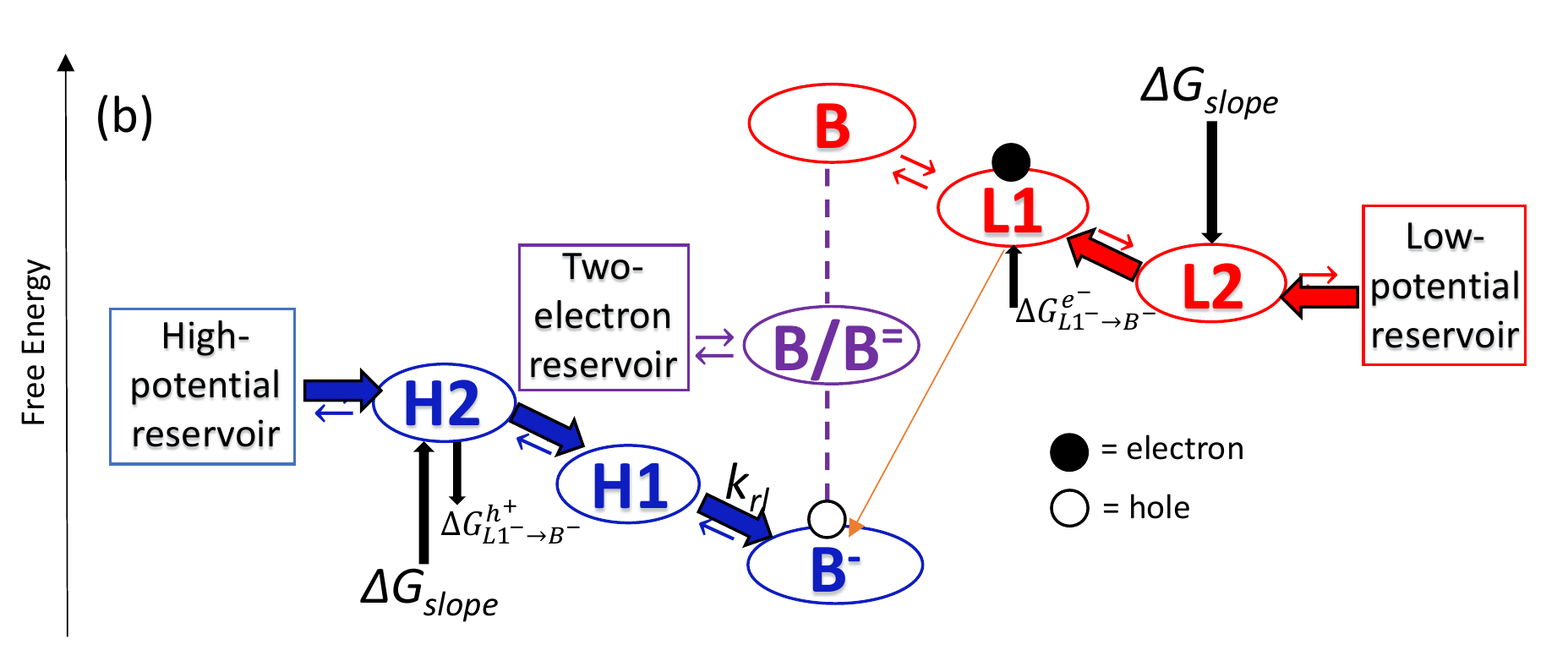}
     \label{fig:L11_B1}
 \end{subfigure}
 \begin{subfigure}{\textwidth}
     \centering
     \includegraphics[width=0.6\textwidth]{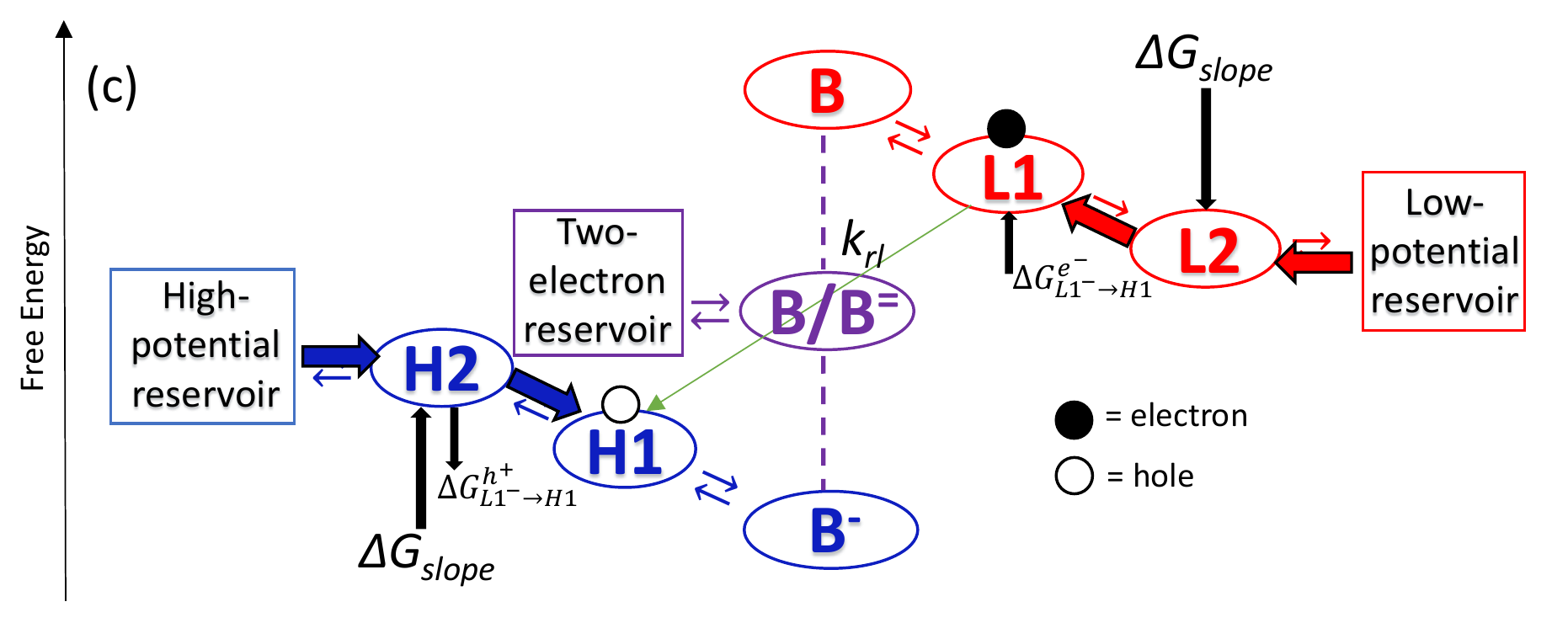}
     \label{fig:L11_H10}
 \end{subfigure}
    \caption{Illustration of electron transfer steps associated with three different short-circuiting channels: (a) $B^{-} \rightarrow H1$ (orange arrow), (b) $L1^{-} \rightarrow B^{-}$ (orange arrow) and (c) $L1^{-} \rightarrow H1$ (green arrow)~\cite{RN197,RN371}. When $\Delta G_{bifurc} = 0$, there is no energy cost to refill the two-electron cofactor $B$. $k_{rl}$ labels the rate-limiting electron-transfer rate constant in each short-circuiting channel. $\Delta G^{e^{-}}_{SC}$ ($\Delta G^{h^{+}}_{SC}$) is the free energy to bring an electron (hole) to the site required for the short-circuiting channel shown. (a) An electron is transferred from the low-potential reservoir with energy cost $\Delta G_{B^{-} \rightarrow H1}^{e^{-}}$, and a hole is transferred from the high-potential reservoir to $H_1$ with energy  $\Delta G_{B^{-} \rightarrow H1}^{h^{+}}$, followed by rate-limiting electron transfer from $L_1^-$ to $B$. (b) A hole is transferred from the high-potential reservoir to $B^=$, forming $B^-$, costing energy $\Delta G_{L1^{-} \rightarrow B^{-}}^{h^{+}}$. Next, an electron is transferred from the low-potential reservoir to $L_1$ with energy cost $\Delta G_{L1^{-} \rightarrow B^{-}}^{e^{-}}$, followed by a rate-limiting electron transfer to $B^{-}$ from $L_1^-$. (c) An electron (hole) is transferred from the low-potential (high-potential) reservoir with $\Delta G_{L1^{-} \rightarrow H1}^{e^{-}}$ ($\Delta G_{L1^{-} \rightarrow H1}^{h^{+}}$) to $L1$ ($H1$) followed by the short-circuit electron transfer from $L_1^-$ to $H_1$ that is the rate-limiting electron transfer for the short-circuit channel shown in (c).}
    \label{fig:EBshortcircuits}
\end{figure}

\clearpage

\printbibliography